\documentclass[]{JHEP3}
\usepackage{amsmath,bm}
\usepackage{graphicx}

\newcommand{\la}{\langle}
\newcommand{\ra}{\rangle}

\newcommand{\as}{\alpha_{\rm s}}

\def\ket#1{\big|{#1}\big\rangle}
\def\bra#1{\big\langle{#1}\big|}

\title{Matching parton showers to NLO computations}

\author{
Zolt\'an Nagy\\
Institute for Theoretical Physics\\
University of Z\"urich\\
Winterthurerstrasse 190\\
CH-8057 Z\"urich, Switzerland\\ 
E-mail: \email{nagyz@physik.unizh.ch}
}
\author{
Davison E. Soper\\
Institute of Theoretical Science\\
University of Oregon\\
Eugene, OR  97403-5203, USA\\
E-mail: \email{soper@physics.uoregon.edu}
}

\abstract{
We give a prescription for attaching parton showers to
next-to-leading order (NLO) partonic jet cross sections in
electron-positron annihilation. Our method effectively extends to
NLO the scheme of Catani, Krauss, Kuhn, and Webber for
matching between $m$ hard jets and $(m+1)$ hard jets. The matching between
parton splitting as part of a shower and parton splitting as part of
NLO matrix elements is based on the Catani-Seymour
dipole subtraction method that is commonly used for removing the
singularities from the NLO matrix elements.}

\keywords{perturbative QCD, NLO calculation, parton shower}
\preprint{
ZH-TH 03/05\\
hep-ph/0503053
}

\begin{document}

\section{Introduction}
\label{sec:Introduction}

One often uses perturbation theory to produce predictions for the results
of particle physics experiments in which the strong interaction is
involved. Let us suppose that the measurement to be made is cast in the
form of the cross section weighted by a function $F(f)$ that assigns
a number to each possible final state $f$. The object then is to predict
$\sigma[F]$. Perturbation theory can be relevant if a high
momentum transfer $Q$ is involved in the reaction. In that case, one
arranges the calculation so that one is performing an expansion in powers
of $\alpha_{\rm s}(Q)$, which is small when $Q$ is large even though
$\alpha_{\rm s}(1\ {\rm GeV})$ is not small. If there are hadrons in the
initial state, then low scale effects related to the initial state must
be factored into parton distribution functions. Low scale effects related
to the final state can be avoided if $F$ is ``infrared safe,'' as
described in Sec.~\ref{sec:notations} and if its definition does not
involve any small parameters. 

It will prove helpful to have an example, taken from $e^+e^- \to {\it
hadrons}$. Let $F(f)$ be $(1-t_f)^4$ where $t_f$ is
the thrust of the particles in $f$. If we call this observable $T_4$, we
have
\begin{equation}
\sigma[T_4] = \int_0^1\! dt\,(1-t)^4\, \frac{d\sigma}{dt}\;\;.
\label{T4def}
\end{equation}
A related example is obtained by collecting the particles in $f$ into
three jets using a suitable jet algorithm and defining $m_f$ to be the
mass of the most massive of the three jets. Then we can take $F(f)$ to be
$dT_4/dM^2 = (1-t_f)^4\delta(m_f^2 - M^2)$, so that
\begin{equation}
\sigma[dT_4/dM^2] \equiv d\sigma[T_4]/dM^2 =
\int_0^1\! dt\,(1-t)^4\, \frac{d\sigma}{dt\,dM^2}\;\;.
\end{equation}
For fixed $M^2/s$, this is an infrared safe observable, but if $M^2/s \ll
1$, its perturbative expansion contains large logarithms, $\log(M^2/s)$.
Quantities like this need a treatment beyond a fixed order of
perturbation theory. In practical applications one often needs not only a
treatment of the large logarithms but also a good model for how partons
turn into hadrons. That is, to make predictions for $\sigma[T_4]$, we need
just short distance physics, but to understand $d\sigma[T_4]/dM^2$ for all
$M^2$, we need to model also long distance physics.
 
The most widely used tools available for perturbative calculations can be
classified into two groups. The first is parton shower Monte Carlo
(MC) event generators, such as  H{\small ERWIG} \cite{Herwig} and P{\small
YTHIA} \cite{Pythia}. The second is next-to-leading order (NLO) programs
such as N{\small LOJET++} \cite{Nagy} and M{\small CFM} \cite{MCFM}. The
NLO programs have NLO accuracy for hard cross sections like $\sigma[T_4]$,
but do a bad job of predicting quantities like $d\sigma[T_4]/dM^2$ that
represent the inner structure of jets. The parton shower Monte Carlo
programs do not have NLO accuracy for hard cross sections but do a good
job of predicting quantities like $d\sigma[T_4]/dM^2$. Additionally, they
have the advantage that their final state particles are hadrons rather
than partons. There are also some NLO-MC hybrids.  One example is the
program of Frixione, Nason, and Webber
\cite{FrixioneWebberI, FrixioneWebberII, FrixioneWebberIII}, which so far
has been applied to cases with massless incoming partons but not to cases
with massless final state partons at the Born level of calculation. The
other example is that of \cite{nloshowersI, nloshowersII, nloshowersIII},
which concerns three-jet observables in electron-positron annihilation
and thus addresses massless final state partons but not massless initial
state partons. This paper concerns a method for constructing NLO-MC
hybrid programs.

The shower Monte Carlo event generators can be made more powerful by
incorporating the $k_T$-jet matching scheme of Catani, Kuhn, Krauss and
Webber \cite{CKKW}. This scheme is defined by first considering the cross
section $\sigma[F]$ to be divided into partial cross sections
\begin{equation}
\sigma[F] = \sum_{m=2}^\infty \sigma_m[F]\;\;.
\end{equation}
Here $\sigma_m[F]$ consists of the contribution to $\sigma[F]$ from
final states that are classified as consisting of $m$ jets according to a
slightly modified version of the $k_T$ jet algorithm (as described in
Sec.~\ref{sec:PartialCrossSections} of this paper). This jet algorithm
depends on the choice of a resolution scale for the jets, which we call
$d_{\rm ini}$. 

A parton shower Monte Carlo program based on the $k_T$-jet matching
scheme is based on the Born squared matrix element for $\sigma_m$
multiplied by a reweighting factor that we call $W$, which contains
Sudakov factors of the sort found in parton shower Monte Carlo programs.
This produces a realistic structure for jets at a scale above $d_{\rm
ini}$.  The base calculation for $\sigma_m[F]$ produces $m$ partons. For
each of these partons, the full program then produces a complete shower
with $d_{\rm ini}$ as a maximum scale. This fills in the details of jet
structure at scales below $d_{\rm ini}$. Both the factor $W$ and the
Sudakov factors that control the shower development at scales below
$d_{\rm ini}$ depend on $d_{\rm ini}$, but the product of these two
factors is approximately independent of $d_{\rm ini}$.

Suppose that we use a program of this sort to calculate $\sigma[F]$ for
an infrared safe $N$-jet observable as defined in
Sec.~\ref{sec:notations}. We assume that the definition $F$ is such that
the perturbative expansion of $\sigma[F]$ does not involve large
logarithms. Thus $F$ measures large momentum scale features of events
with $N$ or more jets. It vanishes if there are fewer than $N$ narrow
jets. As long as the jet resolution parameter $d_{\rm ini}$ is
appropriately chosen, the main contribution to $\sigma[F]$ comes from
$\sigma_m[F]$ with $m=N$. This contribution is of order
$\alpha_{\rm s}^{B_m}$, where $B_m = m - 2$ for $e^+e^-\to {\it
hadrons}$.  For instance, in our example, $T_4$ is a three jet
observable. A calculation using P{\small YTHIA} with $\sqrt s = M_{\rm
Z}$ shows that if we pick $d_{\rm ini}
\approx 0.03$, then $\sigma_3[T_4]$ gives the dominant contribution to
$\sigma[T_4]$, with  $\sigma_2[T_4] \approx \sigma_4[T_4] \approx 0.1\,
\sigma_3[T_4]$. We discuss this further in
Sec.~\ref{sec:PartialCrossSections}.

The perturbative expansion of $\sigma_m[F]$ has the form
\begin{equation}
\sigma_m[F]  =  
C_{m,0}[F]\, \alpha_{\rm s}^{B_m}(Q) 
+ C_{m,1}[F]\, \alpha_{\rm s}^{B_m+1}(Q)
+ C_{m,2}[F]\, \alpha_{\rm s}^{B_m+2}(Q) +\cdots\;\;.
\label{pertexpansion}
\end{equation}
(This applies for any $m$, not just $m=N$, but some of the coefficients
are zero if $m<N$.) A leading order parton shower Monte Carlo program
based on the $k_T$-jet matching scheme will get the Born term,
$C_{m,0}[F]\, \alpha_{\rm s}^{B_m}(Q)$, in this expansion exactly right.
Higher order contributions will be present, but getting the higher order
perturbative contributions correctly is beyond the order of approximation
intended in such a program.

If we wanted more perturbative accuracy than described above, we might
use an NLO calculation. There are, in fact, NLO calculations available for
a wide variety of important processes. Modern calculations are in the form
of computer programs, each of which is designed to work for a certain
class of observables $F$. Such programs are very important for
producing accurate predictions for the class of measurement functions $F$
for which they were designed. Unfortunately ordinary NLO programs have
significant limitations. For instance, an NLO program can calculate
$\sigma[T_4]$, but if one were to use the same program to calculate
$d\sigma[T_4]/dM^2$ for $M^2/s \ll 1$, the result would not be even
qualitatively right. For $\sqrt s = M_{\rm Z}$, the physical distribution
peaks at a jet mass of a few GeV and tends to zero for $M\to 0$. In
contrast, the perturbative program gives for $d\sigma[T_4]/dM^2$ a result
that increases without bound as $M^2 \to 0$ and additionally contains a
delta function at zero mass, $\delta(M^2)$, with a coefficient that is
negative and infinite (see \cite{nloshowersIII} for a similar example
worked out numerically).

The purpose of this paper is to extend to next-to-leading order the
parton shower Monte Carlo idea based on the $k_T$-jet matching scheme.
That is, we want to keep the feature that each outgoing hard parton
generates a full parton shower and hadronization. We also want to utilize
the decomposition of $\sigma[F]$ into $\sum\sigma_m[F]$ so that we can
include calculations for different numbers of hard jets in the same
program. However, we now want the perturbative expansion of the calculated
$\sigma_m[F]$ to be correct with respect to the first {\it two}
coefficients in Eq.~(\ref{pertexpansion}), namely $C_{m,0}[F]$ and
$C_{m,1}[F]$.\footnote{Of course, we can have NLO accuracy only for those
values of $m$ for which NLO calculations exist. For $e^+ e^- \to {\it
hadrons}$, that would currently be $m = 2,3,$ and $4$.}

The general idea of the algorithm that we present applies,
we believe, to lepton-lepton collisions, lepton-hadron collisions, and
hadron-hadron collisions. This is because the dipole subtraction scheme
that we use applies to all of these cases. However, we have elected to
work out only the case of lepton-lepton collisions here and to leave the
other cases for future publications.

Before we launch into a construction that must be rather involved if it
is to be precise, it may be useful to begin with an informal preview.
Consider as an example the calculation $\sigma[T_4]$. The dominant
contribution to $\sigma[T_4]$ comes from $\sigma_3[T_4]$. This
quantity has a perturbative expansion of the form~(\ref{pertexpansion})
with $B_3 = 1$. We seek a calculation that includes parton showering and
hadronization and gets the two coefficients $C_{3,0}[T_4]$ and
$C_{3,1}[T_4]$ right. At the Born level ($\alpha_{\rm s}^{B_3}$) there
are three partons in the final state. At order $\alpha_{\rm s}^{B_3+1}$
there can be four partons and we might think of the four partons as
arising from three partons by parton splitting, say $q \to q + g$, even
though the actual calculation involves the exact $\alpha_{\rm s}^{B_3+1}$
matrix element. If we now add parton showering to one of the Born graphs
we are in big trouble. The showering includes splittings $q \to q + g$,
which we had already included. To include parton shower splittings of the
Born level partons without double counting, we must expand their effect in
perturbation theory and subtract the order $\alpha_{\rm s}^{B_3+1}$
contributions from the $\alpha_{\rm s}^{B_3+1}$ graphs. 

We see that there are subtractions from the order $\alpha_{\rm s}^{B_3+1}$
graphs that are derived from the splitting functions that generate the
initial step of the parton shower. That is good, because NLO calculations
inevitably involve subtractions. In this paper, we want to have an
algorithm that can be used by NLO practitioners in a reasonably
straightforward manner. For this reason, we design the parton splitting
so that the subtractions are (with some minor changes) those of the
dipole subtraction scheme of Catani and Seymour \cite{CataniSeymour}.
This scheme is expressed in a Lorentz covariant style and nicely
expresses the complete available phase space for parton splitting. It is
quite widely used for NLO calculations (for example in the programs
N{\small LOJET++} \cite{Nagy} and M{\small CFM} \cite{MCFM}).

The one parton splitting just described represents the hardest splitting
in the eventual shower. After this comes secondary showering that can be
performed according to an existing shower Monte Carlo program suitably
modified to account for the starting condition that the hardest splitting
has already occurred.

Now, the main contribution to a calculation of $\sigma[T_4]$ comes from
final states that have three jets. The first splitting and the secondary
showering generates the inner structure of these jets, which would be
reflected in $d\sigma[T_4]/dM^2$. Experience has shown that quite a
variety of showering algorithms are capable of doing a good job of
representing this inner structure. In this paper, we leave open the
choice of methods for the secondary showers. As long as the secondary
showering has certain simple properties, the coefficients $C_{3,0}[T_4]$
and $C_{3,1}[T_4]$ will not be affected.

With this structure, we hope to largely decouple the part of the
calculation that needs NLO computations from the bulk of the parton
showering and hadronization. We have in mind that NLO practitioners 
could then write code that could be used in conjunction with a variety of
Monte Carlo event generators, even with Monte Carlo event generators that
are written after the NLO code. 

We seek to be quite flexible not only with respect to the Monte Carlo
event generator but also with respect to several of the function choices
internal to the algorithm. The reason for this flexibility is to allow for
improvements in the methods we suggest.

The complete algorithm calculates $\sigma[F]$ in the form
\begin{equation}
\sigma[F] \approx \frac{1}{N}\sum_{n=1}^N
w_n\,F(f_{\!n})\;\;.
\label{average}
\end{equation}
That is, it is in the form of a Monte Carlo event generator with weights
$w_n$. As is natural for a description that includes quantum interference,
the weights for individual simulated events can be positive or
negative.\footnote{In NLO programs, there are also weights $w_n$, but the
weights can be very large, with an event with a large positive weight
being followed by a counter-event with a large negative weight. Here,
the weights are not large.}

The algorithm presented in this paper combines two separate ideas.
First, we use the $k_T$-jet matching scheme to match calculations
involving different numbers of hard jets \cite{CKKW}. One might
call this $m$-jet/$(m+1)$-jet matching. Second, we use the
dipole subtraction scheme \cite{CataniSeymour} to match between the
hardest splitting in a parton shower and the next-to-leading order
contributions to the $m$-jet matrix elements. One might call this PS/NLO
matching. Most of the paper concerns PS/NLO matching. The reader might
want to work on this part first. For this purpose, one can simply take
$d_{\rm ini}$ to be infinitesimally less than 1. In that case,
the partial cross sections $\sigma_m[F]$ all vanish except for
$\sigma_2[F]$ and the reweighting factors $W$ that appear in the paper
can be taken to be $W=1$. Of course such a calculation would be of only
limited phenomenological usefulness since it would apply only to two-jet
infrared-safe observables $F$. To get to a method that applies to
three-jet observables, we need to do something about the two-jet region
and then we need $m$-jet/$(m+1)$-jet matching.

In the following sections, we review the dipole algorithm
\cite{CataniSeymour} for doing NLO calculations in the case of
electron-positron annihilation. For readers who are familiar with this
algorithm, our aim is to set up the notation that we use, which differs
in some instances from that used in Ref.~\cite{CataniSeymour}. For other
readers, our aim is to provide a compact introduction to the algorithm. 
We provide no proofs that the algorithm works. Furthermore, there are
several functions that must be defined in order to fully specify the
algorithm, but in the introductory sections that follow, we skip the
formulas for these functions and merely state the important properties of
the functions that follow from these formulas. The formulas are then
summarized later on, in Secs.~\ref{sec:splittingkinematics},
\ref{sec:splittingfctns} and \ref{sec:virtualfunctions}.  Once the dipole
subtraction formalism has been set up, we define in
Sec.~\ref{sec:PartialCrossSections} how to break up the cross section
into partial cross sections involving different numbers of resolved jets,
along the lines of Ref.~\cite{CKKW}. Then in
Sec.~\ref{sec:PartialCrossSectionsandDipoles} we modify the subtraction
scheme just a little to accommodate this division. The main subject of the
paper begins in Sec.~\ref{sec:PartialCrossSectionswithShowers} with an
outline of the general structure for adding showers to the partial cross
sections. This is followed by an exposition in
Sec.~\ref{sec:ShowerConstruction}, with several subsections, of how the
showers are to be added while keeping track of the next-to-leading order
correction terms needed to keep from changing the coefficient
$C_{m,1}[F]$ in  Eq.~(\ref{pertexpansion}). The following sections
contain details on the functions used in the dipole subtraction and
showering constructions. We present some conclusions in
Sec.~\ref{sec:Conclusion}.

\section{Notations}
\label{sec:notations}

In this section, we introduce some of the notations that we will use
throughout the paper.

We wish to describe the process $e^+ + e^- \to {\it hadrons}$. In a
calculation at a finite order of perturbation theory, we consider
final states consisting of $n$ partons, with $n \ge 2$. We denote the
momenta of these partons by $\{p\}_n = \{p_1, p_2, \dots, p_n\}$. We
represent the phase space integration for the final state partons as
\begin{equation}
d\varGamma(\{p\}_{n})
=
\prod_{l = 1}^n\biggl( (2\pi)^{-3} d^4p_l\ \delta_+(p_l^2)\biggr)\,
(2 \pi)^4
\delta^4\!\left(\sum_{l = 1}^n p_l - P_0\right)\;\;,
\label{Gammadef}
\end{equation}
where $P_0$ is the momentum of the initial $e^+ e^-$ pair, $P_0 =
(\sqrt s,0,0,0)$ if we use the c.m.\ frame.

We denote the flavors of the partons by labels $\{f\}_n$ with $f_i \in
\{{\rm g}, {\rm u}, \bar {\rm u}, {\rm d}, \bar {\rm d}, \dots\}$. Then
the complete description of the final state momenta and flavors is
specified by the list $\{p,f\}_n = \{(p_1,f_1),(p_2,f_2), \dots,
(p_n,f_n)\}$. We can define ``addition'' on flavors by saying that $f_1 +
f_2 = f_3$ if there is a QCD vertex for $f_1 + f_2 \to f_3$. Thus, for
instance, $\bar {\rm d} + {\rm g} = \bar {\rm d}$ and ${\rm u} + \bar {\rm
u} = {\rm g}$, while ${\rm d} +  \bar {\rm u}$ is not defined. The
splitting of a final state parton with flavor $f_l$ can be represented by
giving the pair of daughter flavors $\{\hat f_{l,1}, \hat f_{l,2}\}$ in
the set of flavor pairs such that $\hat f_{l,1} + \hat f_{l,2} = f_l$.

The matrix element ${\cal M}$ for $a + b \to n\ {\it partons}$ depends
on the spin and color indices of the $n$ partons. In order to avoid
writing these indices, we follow the notation of Catani and Seymour and
write the matrix element as a vector
\begin{equation}
\ket{{\cal M}(\{p,f\}_n)}
\end{equation}
in color $\otimes$ spin space. The squared matrix element, summed over
colors and spins, is then
\begin{equation}
\big\langle{\cal M}(\{p,f\}_n)\big|{\cal M}(\{p,f\}_n)\big\rangle\;\;.
\label{squaredM}
\end{equation}

We consider a perturbative calculation that is designed to predict an
infrared safe $N$-jet observable. We need to be careful about
what we mean by an infrared safe observable. There should be a
function $F_n(\{p_1, p_2,\dots, p_n\})$ defined for any number $n$ of
massless partons with momenta $p_i$. The functions $F_n$ do not depend
on the flavors of the partons and should be invariant under
permutations of the momentum arguments $p_1, p_2, \dots, p_n$. In a
calculation at ``all'' orders of perturbation theory, the cross section
for the observable would be
\begin{equation}
\sigma[F] = \sum_n \frac{1}{n!} \int
d\vec p_1\,\cdots \, d\vec p_n\
\frac{d\sigma}{d\vec p_1\,\cdots \, d\vec p_n}\
F_n(\{p\}_n)\;\;.
\end{equation}
Here $p_i^0 = |\vec p_i|$ in $F$ and $d\sigma$ contains the delta
functions for momentum and energy conservation. The cross section
needs to contain a regulator to control infrared divergences, which
cancel between terms with different numbers $n$ of partons. To ensure
that these cancellations work, the $F_n$ functions for different
values of $n$ need to be related. Specifically, if two partons become
collinear or one becomes soft ($p_i \to 0$), the outcome of the
measurement should be unaffected:
\begin{equation}
F_n(\{p_1, p_2, \dots, (1-\lambda) p_{n-1}, \lambda p_{n-1}\})
=
F_{n-1}(\{p_1, p_2, \dots, p_{n-1}\})
\label{irsafe}
\end{equation}
for $0 \le \lambda < 1$. (The case $\lambda = 0$ covers the
possibility that parton $n$ is soft and not necessarily collinear
with any other parton.) Cancellation of infrared
divergences also requires that the measurement is unaffected when many
partons become soft or group themselves into collinear groups (jets).
To cover this possibility, we also require that the functions
$F_n(\{p\}_n)$ be smooth functions of their arguments.\footnote{A
  smooth function has an infinite number of derivatives, and thus a
  Taylor expansion to any order. This property is stronger than one
  really needs, but successively weaker requirements become
  successively more unwieldy.}

Ultimately, one wants to apply the measurement function to hadrons.
Indeed, we want to do that in this paper after supplementing the
perturbative calculation with a simulation of parton showering and
hadronization. For this purpose, one uses the same functions $F_n$,
but now applied to hadron momenta $p_i$ with $p_i^2 > 0$. The
extensions of the $F_n$ to $p_i^2>0$ should then have smooth limits as
any or all of the $p_i^2$ approach 0.

We have defined what we mean by an infrared safe observable. In this
paper, we consider an infrared-safe ``$N$-jet'' observable. This means
that additionally
\begin{equation}
F_{m}(\{p_1, p_2, \dots, p_{m}\}) = 0\;\;,
\hskip 1 cm m < N \;\;.
\label{mjetsonly}
\end{equation}

Throughout the paper we use the standard notations $C_{\rm F} = (N_{\rm
c}^2 - 1)/2N_{\rm c}$, $C_{\rm A} = N_{\rm c}$, $T_{\rm R} = 1/2$ for color
SU($N_{\rm c}$). We assume that there are $N_{\rm f}$  flavors ( +
$N_{\rm f}$ antiflavors) of quarks in the fundamental representation. We
typically use the notation $f$ to represent a parton flavor, $f\in \{{\rm
g},{\rm u},\bar {\rm u},{\rm d},\dots\}$. Then we use coefficients $C_f$,
$K_f$, and $\gamma_f$ defined by
\begin{equation}
\begin{split}
\label{calVcoefficients}
C_f ={}& C_{\rm F}   \quad {\rm for}\ 
f\in \{{\rm u},\bar {\rm u},{\rm d},\dots\}\,,
\\
C_{\rm g} ={}& C_{\rm A} \,,
\\
\gamma_f={}& \frac{3}{2}\,C_{\rm F} 
\quad {\rm for}\  f\in \{{\rm u},\bar {\rm u},{\rm d},\dots\}\,,
\\
\gamma_{\rm g} ={}& \frac{11}{6}C_{\rm A} 
- \frac{2}{3}T_{\rm R}N_{\rm f}\,,
\\ 
K_f ={}& \left(\frac72-\frac{\pi^2}6\right)C_{\rm F} 
\quad {\rm for}\  f\in \{{\rm u},\bar {\rm u},{\rm d},\dots\}\,,
\\
K_{\rm g} ={}&  \left(\frac{67}{18}-\frac{\pi^2}6\right)C_{\rm A} -
  \frac{10}{9}T_{\rm R} N_{\rm f}
\;\;.
\end{split}
\end{equation}

The running coupling is $\alpha_{\rm s}(\mu)$ evaluated at scale $\mu$,
often a transverse momentum. When no scale is indicated, we
mean $\alpha_{\rm s} \equiv \alpha_{\rm s}(\mu_{\rm R})$ where the
$\mu_{\rm R}$ is a fixed renormalization scale, usually chosen as some
fraction of $\sqrt s$. For our NLO calculations, we would use the two
loop running coupling. In construction our matching with parton showers,
we will require the first order relation
\begin{equation}
\frac{\as\big(\mu\big)}
 	{\as\big(\mu_{\rm R}\big)}
= 1 - \beta_0\log\left(\frac{\mu^2}{\mu_{\rm R}^2}\right)
\frac{\alpha_{\rm s}(\mu_{\rm R})}{2\pi} + {\cal O}(\as^2)\;\;,
\label{runningalphas}
\end{equation}
where $\beta_0 = \gamma_{\rm g}$.

\section{Construction and deconstruction of parton splitting}
\label{sec:splittingconstruction}

The dipole algorithm of Catani and Seymour \cite{CataniSeymour} is
based on a physical picture involving parton splitting, which turns
$m$ partons into $m+1$ partons. Deconstructing the splitting turns the
$m+1$ partons back into $m$ partons. In this section, we describe the
splitting kinematics without giving the detailed formulas from
Ref.~\cite{CataniSeymour}, which are given in
Sec.~\ref{sec:splittingkinematics}.

We describe deconstruction first. Suppose that we have a list of $m+1$
parton momenta and flavors, $\{\hat p, \hat f\}_{m+1}$. One imagines that
partons $i$ and $j$ are produced by the splitting of a mother parton with
flavor $\tilde f_{ij}$ and momentum $\tilde p_{ij}$. We need one more
parton, with index $k$, to describe the splitting in the scheme of
Catani and Seymour. Parton $k$ is a ``spectator parton'' that absorbs
some momentum associated with the splitting.\footnote{The spectator
  parton also plays a role in the color flow associated with the
  splitting, as expressed in the color matrices in the definition of
  ${\cal D}_{ij,k}$ below. }

Consider first the flavors.  The mother parton has flavor $\tilde
f_{ij} = \hat f_i + \hat f_j$. The flavor of the spectator parton is
not changed: $\tilde f_k = \hat f_k$. There is less information in the
list of just one flavor $\tilde f_{ij}$ than in the list of two
flavors $\{\hat f_i,\hat f_j\}$. The missing information is the flavor
splitting choice, which can be specified by giving the pair
$\{\hat f_i,\hat f_j\}$ in the set with $\hat f_i + \hat f_j = \tilde
f_{ij}$.

We now extend this idea to the momenta. The massless momenta $\{\hat
p_i, \hat p_j, \hat p_k\}$ determine new momenta $\{\tilde p_{ij},
\tilde p_k\}$ of just two on-shell massless partons together with
three splitting variables. The structure of this transformation is
simple in the limiting case of collinear splitting. If $\hat p_i \cdot
\hat p_j = 0$, we have
\begin{equation}
\tilde p_{ij} = \hat p_i + \hat p_j  \hskip 1 cm
({\rm when}\ \hat p_i \cdot \hat p_j = 0)\;\;,
\label{limitingpsumij}
\end{equation}
while the momentum of the spectator parton remains unchanged,
\begin{equation}
\tilde p_{k} = \hat p_k  \hskip 1 cm
({\rm when}\ \hat p_i \cdot \hat p_j = 0)\;\;.
\label{limitingpk}
\end{equation}
It is not possible to retain these relations away from the collinear
limit. However, Catani and Seymour still maintain
\begin{equation}
\tilde p_{ij} + \tilde p_k= \hat p_i + \hat p_j + \hat p_k 
\label{exactpsum}
\end{equation}
while keeping all of the momenta massless.

Since one eliminates three degrees of freedom in going from $\{\hat
p_i, \hat p_j, \hat p_k\}$ to $\{\tilde p_{ij}, \tilde p_k\}$, we can
supplement $\{\tilde p_{ij}, \tilde p_k\}$ with three splitting
variables. For our purposes, it is convenient to call these
$y,z,\phi$.  The most important role in the formalism is played by the
dimensionless variable $y$ proportional to the virtuality of the
splitting, so that
\begin{equation}
y = 0 \hskip 1 cm
{\rm if\ and\ only\ if}\ \ \ \hat p_i \cdot \hat p_j = 0\;\; .
\label{yproperty}
\end{equation}
The variable $z$ is a momentum fraction representing the share of the
mother parton momentum that is carried by parton $i$ and $\phi$ is an
azimuthal angle.

The discussion above can be summarized by saying that there is a map
\begin{equation}
\{(\hat p_i,\hat f_i), (\hat p_j,\hat f_j), (\hat p_k,\hat f_k)\}
\to
\{(\tilde p_{ij},\tilde f_{ij}), (\tilde p_k,\tilde f_k),y,z,\phi,
\hat f_i,\hat f_j\}\;\;.
\label{CSantisplitting}
\end{equation}

Parton splitting is the other way around. Here we begin with a list of
the momenta and flavors $\{p,f\}_m$ of $m$ partons. We imagine that
one of these, parton $l$, splits, producing daughter partons with with
flavors $\{\hat f_{l,1},\hat f_{l,2}\}$ and momenta $\{\hat
p_{l,1},\hat p_{l,2}\}$. We again need a spectator parton, with index
$k$. To specify the splitting we need splitting parameters
$\{y,z,\phi,\hat f_{l,1},\hat f_{l,2}\}$, with $\hat f_{l,1} + \hat
f_{l,2} = f_l$.  Then Catani and Seymour specify a map
\begin{equation}
\{(p_l,f_l), (p_k,f_k),y,z,\phi,
\hat f_{l,1},\hat f_{l,2}\}
\to
\{(\hat p_{l,1},\hat f_{l,1}), 
(\hat p_{l,2},\hat f_{l,2}), (\hat p_k,\hat f_k)\}\;\;.
\label{CSsplitting}
\end{equation}
The map (\ref{CSsplitting}) is precisely the inverse of the map
(\ref{CSantisplitting}). Only the variable names are different. The
part of these maps that concerns the flavor splitting is trivial. The
part that concerns the momentum splitting is not trivial. The formulas
from Ref.~\cite{CataniSeymour} are given in
Sec.~\ref{sec:splittingkinematics}.

These maps can be rather trivially extended to include all of the
partons. Suppose that we start with a list $\{p,f\}_{m}$ of $m$ parton
momenta and flavors and that we want to split parton $l$ with the help of
spectator parton $k$ using splitting variables
$\{y,z,\phi,\hat f_{l,1},\hat f_{l,2}\}$. We need $\hat f_{l,1} + \hat
f_{l,2} = f_l$.  We can produce a list $\{\hat p, \hat f\}_{m+1}$ of
$m+1$ parton momenta and flavors by removing partons $l$ and
$k$ from the original list and adding $\hat p_{l,1},\hat f_{l,1}$, $\hat
p_{l,2},\hat f_{l,2}$ and $\hat p_k,\hat f_k$ from
Eq.~(\ref{CSsplitting}) to the end of the list. Then $\{\hat p, \hat
f\}_{m+1}$ equals the new list. It will prove useful to call the complete
transformation ${\cal R}_{l,k}$,
\begin{equation}
\{\hat p,\hat f\}_{m+1}
=
{\cal R}_{l,k}(\{p,f\}_{m},\{y,z,\phi, 
\hat f_{l,1},\hat f_{l,2}\})\;\;.
\label{Rldef}
\end{equation}

Now suppose that we start with a list $\{\hat p,\hat f\}_{m+1}$ of
$m+1$ parton momenta and flavors and that we want to combine partons
$i$ and $j$ with the help of spectator parton $k$. We can produce a
list $\{p,f\}_{m}$ of $m$ parton momenta and flavors by removing
partons $i,j$ and $k$ from the original list and adding $\tilde
p_{ij},\tilde f_{ij}$ and $\tilde p_k,\tilde f_k$ from
Eq.~(\ref{CSantisplitting}) to the end of the list. Then $\{p,f\}_{m}$
equals the new list. We also get the splitting variables
$\{y,z,\phi, \hat f_{i},\hat f_{j}\}$ with the help of
Eq.~(\ref{CSantisplitting}). It will prove useful to call the complete
transformation ${\cal Q}_{ij,k}$,
\begin{equation}
\{\{p,f\}_{m},\{y,z,\phi, \hat f_{i},\hat f_{j}\}\}
=
{\cal Q}_{ij,k}(\{\hat p,\hat f\}_{m+1})\;\;.
\label{Qijkdef}
\end{equation}

The transformations ${\cal R}$ and ${\cal Q}$ are inverses of each
other in the sense that if we supply the right permutation operators
$\Pi(i,j,k)$ and $\Pi(l,k')$ we have $\Pi(i,j,k)  {\cal R}_{l,k'} 
 \Pi(l,k') {\cal Q}_{ij,k} = 1$. The permutations will not be of
much concern to us since the functions that we use that are functions of
$\{p,f\}$ are invariant under permutations of the parton labels.

Where possible, for a final state parton $l$, we denote the complete
set of splitting variables by a single letter,
\begin{equation}
Y_l = \{y_l, z_l,\phi_l,\hat f_{l,1},\hat f_{l,2}\}\;\;.
\label{Yldef}
\end{equation}
With this notation we can abbreviate
\begin{equation}
\int_0^1 \frac{dy_l}{y_l}
\int_0^1\!dz_l\,\int_0^{2\pi}\frac{d\phi_l}{2\pi}\,
\frac{1}{2}\!\sum_{\hat f_{l,1},\hat f_{l,2}}\!
\delta_{\hat f_{l,1} + \hat f_{l,2}}^{f_l}
\equiv \int\! dY_l\;\; .
\label{dYldef}
\end{equation}

Now that we understand the parton splitting, we are ready to examine
the construction of the cross section at NLO.

\section{Structure of the NLO cross section}
\label{sec:StructureNLO}

We consider an $N$-jet cross section in electron-positron
collisions. The cross section correct to NLO is based on a tree level
cross section for the production of $m$ partons, with $m = N$,
together with one loop graphs for the production of $m$ partons and
tree graphs for the production of $m+1$ partons. The cross section is
constructed as a sum of terms
\begin{equation}
\sigma^{\rm NLO}  = \sigma^{\rm B}
+\sigma^{\rm R - A}
+\sigma^{\rm V + A}\;\;.
\label{CSstart}
\end{equation}
In the first term there is an integration over $m$-parton phase space,
while in the second term there is an integration over $m+1$-parton phase
space. The first term is the Born contribution, proportional to
$\alpha_{\rm s}^{B_m} \equiv \alpha_{\rm s}^{m-2}$.  The second term is a
correction proportional to $\alpha_{\rm s}^{B_m+1}$ associated with real
parton emission, which comes with a subtraction term that eliminates the
soft and collinear divergences. The third term is a correction
proportional to $\alpha_{\rm s}^{B_m+1}$ that is associated with a
virtual parton loop.  There are $m$ partons in the final state. There is
a corresponding subtraction term in which there is an integration over
the phase space for one parton, which is performed analytically in
$4-2\epsilon$ dimensions to produce $1/\epsilon^2$ and  $1/\epsilon$ terms
that cancel $1/\epsilon^2$ and  $1/\epsilon$ terms that would be
present without the subtraction.

We will begin with the Born contribution.

\subsection{The Born contribution}

The Born contribution takes the form
\begin{equation}
\sigma^{\rm B}
= \frac{1}{m!}
\sum_{\{f\}_m} 
\int\!d\varGamma(\{p\}_m)\,
\big\langle{\cal M}(\{p,f\}_m)\big|
{\cal M}(\{p,f\}_m)\big\rangle\,
F_m(\{p\}_m)\;\;.
\label{sigmaB}
\end{equation}
There is an integration over the final state momenta and a sum over
final state flavors $\{f\}_m$ with a symmetry factor $1/m!$. Next is
the squared matrix element for the production of the $m$ final state
partons.  Finally, there is a final state measurement function,
$F_m(\{p\}_m)$.

\subsection{The real emission contribution}

The real emission contribution $\int d\sigma^{\rm R}$ along with its
subtraction $\int d\sigma^{\rm A}$ has the form
\begin{equation}
\begin{split}
  \sigma^{\rm R - A} = {}&
\int_{m+1}\left[d\sigma^{\rm R}-d\sigma^{\rm A}\right]
  \\= {} &
  \frac{1}{(m+1)!}  \sum_{\{\hat f\}_{m+1}} \int\!d\varGamma(\{\hat
  p\}_{m+1})
  \\
  &\times \biggl[ \big\langle{\cal M}(\{\hat p,\hat f\}_{m+1})\big|
  {\cal M}(\{\hat p,\hat f\}_{m+1})\big\rangle\ F_{m+1}(\{\hat
  p\}_{m+1})
  \\
  &\qquad - \sum_{\substack{{i,j}\\{\rm pairs}}} \sum_{k \ne i,j}
  {\cal D}_{ij,k} (\{\hat p,\hat f\}_{m+1})\, F_m(\{p\}_{m}^{ij,k})
  \biggr]\;\;.
\end{split}
\label{realandsubtraction}
\end{equation}
The real emission term $d\sigma^{\rm R}$ is represented by the first term
in square brackets. Here we have the same sums and integrals as in
$\sigma^{\rm B}$ except that now there are $m+1$ final state partons. The
momentum and flavor variables are all written with hats in order to
distinguish them from the $m$-parton variables. There is the squared
amplitude to produce the $m+1$ partons. The resulting partonic cross
section is multiplied by the measurement function $F$ for $m+1$ partons.
 
The contribution $d\sigma^{\rm R}$ has a potential singularity when
any of the dot products $\hat p_i\cdot \hat p_j$ tends to zero. In the
second term in the square brackets, representing $d\sigma^{\rm A}$, we
sum over subtractions designed to reduce the strength of these
singularities. The subtractions are labeled by (unordered) pairs
$\{i,j\}$ where $i,j \in \{1,\dots, m+1\}$. For each choice of
$\{i,j\}$, there are a number of subtraction terms labeled by the
index $k$ of a spectator parton, which can be any of the partons
except $i$ or $j$. We use the splitting deconstruction transformation
from Eq.~(\ref{Qijkdef}) to define $m$ parton variables and splitting
variables according to
\begin{equation}
\{\{p,f\}_{m}^{ij,k},Y_{ij,k}\} =
{\cal Q}_{ij,k}(\{\hat p,\hat f\}_{m+1})\;\;.
\label{Qijkdefencore}
\end{equation}
The $ij,k$ superscripts or subscripts remind us that the definition of
these variables is different for each choice of parton indices
$i,j,k$.

The subtraction term in Eq.~(\ref{realandsubtraction}) contains a
dipole subtraction function ${\cal D}_{ij,k}$. We state the
definitions in Sec.~\ref{sec:splittingfctns},
Eqs.~(\ref{newDijkencore},
\ref{VtoS},
\ref{flavorDipoleFF:S},
\ref{DipoleFF:Sqg},
\ref{DipoleFF:Vqqb},
\ref{DipoleFF:Vgg}),
but here simply note that
${\cal D}_{ij,k}$ has the structure
\begin{equation}
\begin{split}
{\cal D}_{ij,k}&(\{\hat p,\hat f\}_{m+1}) ={}
\\
&
\frac{1}{2 \hat p_i\cdot \hat p_j}\
\bra{{\cal M}(\{p,f\}_{m}^{ij,k})}
\frac{\bm{T}_{ij}\cdot \bm{T}_k}{-\bm{T}_{ij}^2}\,
{\bm{V}_{ij}}(\{\tilde p_{ij},y,z,\phi\}_{ij,k},f_i,f_j)\,
\ket{{\cal M}(\{p,f\}_{m}^{ij,k})}\;\;.
\label{newDijk}
\end{split}
\end{equation}
The function ${\cal D}_{ij,k}$ is based on the Born level amplitude
$|{\cal M}(\{p,f\}_{m}^{ij,k}) \rangle$ for the flavors and momenta with
partons $i$ and $j$ combined. There is an operator
${\bm{T}_{ij}\cdot\bm{T}_k}/{\bm{T}_{ij}^2}$ that acts on the color
degrees of freedom of $|{\cal M}\rangle$. Specifically,
$\bm{T}_{ij}^a$ is the infinitesimal SU(3) generator matrix in the $a$
direction acting on the color indices for the mother parton that results
from combining partons $i$ and $j$ (which, to be precise, has been placed
in the second-to-last parton slot by the transformation ${\cal
Q}_{ij,k}$). Similarly, $\bm{T}_{k}^a$ is the generator matrix acting on
the color indices for the spectator parton.  Then
$\bm{T}_{ij}\cdot\bm{T}_k \equiv \sum_a \bm{T}_{ij}^a \bm{T}_{k}^a$. In
the denominator, $\bm{T}_{ij}^2 \equiv \sum_a \bm{T}_{ij}^a \bm{T}_{ij}^a$
is a number, $C_{\rm F}$ or $C_{\rm A}$, depending on whether parton $ij$
is quark or antiquark or a gluon. There is an important identity involving
the color matrices. Invariance of  $\big| {\cal M}\big\rangle$ under color
rotations implies that $ (\bm{T}_{ij} + \sum_k\bm{T}_k)\big| {\cal
M}\big\rangle = 0$. Thus
\begin{equation}
\sum_{k\ne ij}\frac{\bm{T}_{ij}\cdot \bm{T}_k}{-\bm{T}_{ij}^2}
=1
\label{coloridentity}
\end{equation}
when operating on $\big| {\cal M}\big\rangle$. Next in
Eq.~(\ref{newDijk}), there is an operator ${\bm{V}}_{\!ij,k}$ that
depends on the momenta and acts on the spin degrees of freedom of $|{\cal
M}\rangle$. The dipole subtraction functions ${\cal D}_{ij,k}$ are
constructed so that their sum matches the squared matrix element
$\big\langle{\cal M}(\{\hat p,\hat f\}_{m+1})\big| {\cal M}(\{\hat p,\hat
f\}_{m+1})\big\rangle$ in the limit in which the matrix element is
singular. 

Taking some liberties with the notation, we will also use ${\cal
D}_{l,k}(\{p,f\}_{m};Y)$ to denote the same function written
in terms of the variables $\{\{ p,f\}_{m};Y\}$ given by 
Eq.~(\ref{Qijkdef}).

Next in $d\hat\sigma^{\rm A}$ there is a measurement function
$F_J(\{p\}_m^{ij,k})$ evaluated at the momenta for the $m$-parton
state.  The two measurement functions in the two terms in
Eq.~(\ref{realandsubtraction}) match in the limits in which the matrix
element is singular because of the infrared safety property
(\ref{irsafe}).

\subsection{The virtual loop contribution}
\label{virtloop}

The virtual loop contribution along with its counterterm has the form
\begin{equation}
\sigma^{\rm V +A }
=
\int_{m}\big[d\sigma^{\rm V}+\int_{1}d\sigma^{\rm A}\big]
=  \frac{1}{m!}
\sum_{\{f\}_m} 
\int\!d\varGamma(\{p\}_m)\
V(\{p,f\}_m)\
F_m(\{p\}_m)\;\;.
\label{sigmaV}
\end{equation}
The function $V$ comes from the one loop matrix element. The
matrix element has been calculated analytically in $4-2\epsilon$
dimensions. Then its $1/\epsilon^n$ pole terms and certain finite pieces
have been subtracted. The result $V$ is constructed from Born amplitude
and certain spin and flavor dependent functions. The terms subtracted
are precisely what was subtracted from the real emission contribution,
but with the opposite sign. The details of $V$ are laid out in
Sec.~\ref{sec:virtualfunctions}, Eqs.~(\ref{NLONo:sigmaVI}, 
\ref{NLONo:Ioper}, \ref{calVdef}).

\section{Partial cross sections}
\label{sec:PartialCrossSections}

In subsequent sections, we will add parton showers to our calculation.
Before we do this, however, it is useful to divide the cross section into
partial cross sections $\sigma_m$ with $m = 2,3,\dots$ that are based on
the cross sections for 2 partons $\to$ $m$ partons and have higher order
corrections added. This construction is based on the $k_T$-jet matching
scheme of Catani, Krauss, Kuhn, and Webber \cite{CKKW}.

We consider the calculation of an infrared safe $N$-jet observable, as
described in Sec.~\ref{sec:notations}. In particular, the measurement
function $F$ obeys Eq.~(\ref{mjetsonly}), $F_{m}(\{p_1, p_2, \dots,
p_{m}\}) = 0$ for $m < N$. We organize this calculation according to the
number of partons in the final state,
\begin{equation}
\begin{split}
\sigma[F] =&{} \sum_{n=N}^\infty \frac{1}{n!}
\sum_{\{f\}_n} \int\!d\varGamma(\{p\}_n)\
G_n(\{p,f\}_n)\
F_n(\{p\}_n)\;\;.
\end{split}
\end{equation}
Here $G_n$ has a lowest order contribution proportional to
$\alpha_{\rm s}^{B_{n}}$, where $B_n = n-2$,
\begin{equation}
G_n^0 = \big\langle{\cal M}(\{p,f\}_n)\big|{\cal
M}(\{p,f\}_n)\big\rangle\;\;.
\label{lowestGn}
\end{equation}
There are higher order contributions, proportional to $\alpha_{\rm
s}^{B_{n}+L}$ with $L$ virtual loops. When the cross section is expanded
in this form, some kind of regulation is needed on the integrals; the
divergences then cancel between terms with different numbers of partons
$n$.

Given an $n$-parton final state, $\{p,f\}_n$, we group the partons into
jets. We use the recursive ``$k_T$'' jet finding algorithm
\cite{kTalgorithm}, modified slightly to make use of the parton flavor
information available. At each stage of the algorithm there is a
dimensionless jet resolution function $d_{ij}$ that approximates (for
small angles and virtualities) the squared transverse momentum of one jet
(group of partons or a single parton) with label $i$ with respect to
another with label $j$ with which it might be grouped, divided by $s$:
\begin{equation}
  \label{dij}
  d_{ij} \equiv d(p_{i}, p_{j}) = 
  \frac{2}{s} \,
\min\!\left[E_i^2,E_j^2\right]\,
\left(1-\cos(\theta_{ij})\right)
\;\;.
\end{equation}
Here we use the $e^+e^-$ c.m.\ frame to define $\theta_{ij}$ as the angle
between $\vec p_i$ and $\vec p_j$ and to define $E_i = p_i^0$, $E_j =
p_j^0$. 

We begin with $n$ partons. At each stage of the algorithm, we combine two
partons, so that as the algorithm progresses, there are fewer and fewer
partons left. There are some rules. In order to combine partons with
flavors $f_b$ and $f_c$, there must be a flavor $f_a$ such that there is
a QCD vertex for $f_a \to f_b f_c$. In addition, the initial splitting in
$e^+ e^- \to {\it hadrons}$ must be $\gamma/Z \to \bar q q$. Therefore,
when working backwards from the final state, it is not allowed to combine
the last remaining $q$-$\bar q$ pair into a gluon.  Thus at each stage, we
find all pairs $\{i,j\}$ of partons that are allowed to
be combined by these flavor considerations. For each such pair, we
calculate the distance measure $d_{ij}$ given in Eq.~(\ref{dij}). The
allowed pair with the smallest $d_{ij}$ is combined by adding the
four-momenta of the daughter partons to form the four-momentum of the
mother and determining the flavor of the mother such that there is a QCD
vertex for the mother to split  to form the daughters. 

This process gives a sequence of resolution parameters $d_J(\{p,f\}_n)$ at
which two jets were joined, reducing $J$ jets to $J-1$ jets. Typically
$d_J < d_{J-1}$. However, the flavor considerations discussed
above may invalidate this ordering when we must reject the parton pair
with the smallest $d_{ij}$ and choose a pair to combine that has a larger
$d_{ij}$. This misordering does not happen in the leading approximation
in a parton shower. If $d_{J-1}$ calculated according to Eq.~(\ref{dij})
is smaller than $d_J$, we simply redefine $d_{J-1}$ to equal $d_J$. Then
the definition gives $d_n \le d_{n-1} \le \cdots \le d_3$. We also define
$d_J(\{p,f\}_n) = 0$ for $J > n$.

We can use this $k_T$ jet algorithm to divide the cross section into
partial cross sections $\sigma_m$ with specified integration ranges,
\begin{equation}
\sigma[F] = \sum_{m = 2}^\infty \sigma_m[F]\;\;,
\end{equation}
with
\begin{equation}
\begin{split}
\sigma_m[F] =&{} \sum_{n=N}^\infty \frac{1}{n!}
\sum_{\{f\}_n} \int\!d\varGamma(\{p\}_n)\
\theta(d_{m+1}(\{p,f\}_n) < d_{\rm ini} < d_m(\{p,f\}_n))
\\ & \times
G_n(\{p,f\}_n)\
F_n(\{p\}_n)\;\;.
\end{split}
\label{partialsigma}
\end{equation}
In $\sigma_m$ there are exactly $m$ jets that are resolvable at a scale
$d_{\rm ini}$ that can be chosen to suit our purposes, as discussed
briefly in Sec.~\ref{sec:Introduction}. Although there may be more than
$m$ partons, there are not more resolvable jets. There are no infrared
divergences in $\sigma_m$ arising from two of the $m$ jets becoming
collinear or one of them becoming soft because the singular region is
removed by the cut $d_{\rm ini} < d_m(\{p,f\}_n)$. There are also no
infrared divergences in $\sigma_m$ arising from the possible subjets
becoming collinear or soft because of the cancellation between real and
virtual graphs. 

Assume that $F$ is an infrared-safe $N$-jet observable that is sensitive
only to event structure at a large momentum scale $Q \sim \sqrt s$.
Then as long as the jet resolution parameter $d_{\rm ini}$ is
appropriately chosen, the main contribution to $\sigma[F]$ comes from
$\sigma_N[F]$, with $N$ jets resolvable at scale $d_{\rm ini}$. We
touched on this topic in Sec.~\ref{sec:Introduction} and discuss it
further in Sec.~\ref{sec:smalldini}.

Our aim in this paper will be to add showers to the perturbative
calculation of $\sigma_m[F]$ in such a way that the first two terms in
the perturbative expansion of $\sigma_m[F]$ are reproduced without
``double counting'' between the shower splittings and the splittings that
are part of the contributions to $\sigma_m[F]$ of order $\alpha_{\rm
s}^{B_m+1}$. First, however, we need to indicate how the dipole
subtraction scheme for NLO perturbative calculations can be modified to
work with the partial cross sections $\sigma_m[F]$,

\section{Partial cross sections with dipole subtractions}
\label{sec:PartialCrossSectionsandDipoles}

The integrals in Eq.~(\ref{partialsigma}) still need regulation because
the real-virtual cancellations happen between terms with different values
of $n$. However, as long as we limit the calculation to next-to-leading
order, it is simple to adapt the dipole subtraction scheme to make the
cancellations happen inside of each of the integrations. The idea is to
write each partial cross section $\sigma_m$ as a sum of three terms, with
corrections suppressed by two powers of $\alpha_{\rm s}$,
\begin{equation}
\sigma_m = \sigma_m^{\rm B}
+ \sigma_m^{\rm R-A}
+ \sigma_m^{\rm V+A}
+ {\cal O}(\alpha_{\rm s}^{B_m + 2})\;\;.
\label{sigmam}
\end{equation}

For the Born contribution, we simply modify Eq.~(\ref{sigmaB}) by
inserting the appropriate cut,
\begin{equation}
\begin{split}
\sigma^{\rm B}_m
={}& \frac{1}{m!}
\sum_{\{f\}_m} 
\int\!d\varGamma(\{p\}_m)\,
\theta(d_{\rm ini} < d_m(\{p,f\}_m))
\\ &\quad\quad\quad\times
\big\langle{\cal M}(\{p,f\}_m)\big|
{\cal M}(\{p,f\}_m)\big\rangle\,
F_m(\{p\}_m)\;\;.
\end{split}
\label{sigmaBm}
\end{equation}
(Note that there is no cut $d_{m+1} < d_{\rm ini}$ since there are only
$m$ partons.)

In the real emission contribution, we follow
Eq.~(\ref{realandsubtraction}), inserting the $d_{m+1} < d_{\rm ini} <
d_m$ cut in the main term and a similar cut in the subtraction term,
\begin{equation}
\begin{split}
\sigma^{\rm R - A}_m = {}&
  \frac{1}{(m+1)!}  \sum_{\{\hat f\}_{m+1}} \int\!d\varGamma(\{\hat
  p\}_{m+1})
  \\
  &\times \biggl[ \big\langle{\cal M}(\{\hat p,\hat f\}_{m+1})\big|
  {\cal M}(\{\hat p,\hat f\}_{m+1})\big\rangle\ F_{m+1}(\{\hat
  p\}_{m+1})
\\ &\qquad \times
\theta(d_{m+1}(\{\hat p,\hat f\}_{m+1}) < d_{\rm ini} < 
d_m(\{\hat p,\hat f\}_{m+1}))
  \\
  &\qquad - \sum_{\substack{{i,j}\\{\rm pairs}}} \sum_{k \ne i,j}
  {\cal D}_{ij,k} (\{\hat p,\hat f\}_{m+1})\, F_m(\{p\}_{m}^{ij,k})
\\
    &\qquad\quad\times
    \theta\big(\tilde d(\{p,f\}_{m}^{ij,k}, \{l,y,z \}_{ij,k}) 
            < d_{\rm ini}< d_{m}(\{p,f\}_{m}^{ij,k})\big)
  \biggr]\;\;.
\end{split}
\label{realandsubtractionbis}
\end{equation}
In the subtraction term, there is a sum over parton pairs $\{ij\}$ and
spectator partons $k$. For each term, there is an appropriate cut on the
momenta. First, the $m$-parton state that results from combining partons
$i$ and $j$ must be resolvable at scale $d_{\rm ini}$. Second, the
splitting of the mother parton thus obtained must be unresolvable
according to a resolution function $\tilde d$,
\begin{equation}
  \tilde d(\{p,f\}_m,l,y_l,z_l) = \frac{s_l}{s}\,
  y_{l} \min\left\{\frac{1-z_{l}}{z_{l}},
\frac{z_{l}}{1-z_{l}}\right\}\;\;.
\label{tildeddef}
\end{equation}
Here $s_l$ is a virtuality scale appropriate to parton $l$. The simplest
choice would be $s_l = s$. However, for our purposes, it will prove useful
to choose a value obtained from $\{p,f\}_m$ that we will specify in
Sec.~\ref{sec:CKKWSudakov}, Eq.~(\ref{sldef}). Aside from the factor
$s_l/s$, the function $\tilde d(\{p,f\}_m,l,y_l,z_l)$ is derived from the
resolution variable of the $k_T$ jet finding algorithm, Eq.(\ref{dij}).
Except for a $y$ and $z$ independent factor, it gives an approximate
version of the resolution variable associated with the splitting $\{p\}_m
\to \{\hat p\}_{m+1}$ generated with the splitting parameters
$y_l,z_l$ with the help of spectator parton $k_l$,
\begin{equation}
  d(\hat p_{l,1}, \hat p_{l,2}) = \frac{2p_l\cdot p_{k_l}}{s_l}\,
  \tilde d(\{p,f\}_m, l, y_l, z_l) \, [1+{\cal O}(\sqrt{y_l})]\;\;.
\label{tildedapprox}
\end{equation}  
We describe this more precisely in Sec.~\ref{sec:splittingkinematics}
at Eq.~(\ref{dijmod}).  The effect of the cuts on the dipole terms is
easiest to understand in the case that for the $m+1$-parton state that we
start with, one pair has a very small resolution parameter, while once
that pair is combined the other pairs are well separated. In this case,
the cuts provide that only the dipole term for the pair with the small
resolution parameter contributes. Note that the cancellation needed as
any two of the partons $\{\hat p\}_{m+1}$ become collinear or one of them
becomes soft is left intact.

We mention here a subtle point with respect to infrared
singularities in Eq.~(\ref{realandsubtractionbis}). Consider the
$\{ij,k\}$ subtraction term. As we integrate over $\{\hat
p,\hat f\}_{m+1}$, we can encounter a point at which two of the momenta
in $\{p\}_{m}^{ij,k}$ are collinear or one is zero. (The simplest
situation is that two of the original $\{\hat p\}_{m+1}$ other than $\hat
p_i$ and $\hat p_j$ are collinear or one is zero.) At this point, 
${\cal D}_{ij,k} \propto \langle{\cal M}(\{p,f\}_{m}^{ij,k}) |{\cal
M}(\{p,f\}_{m}^{ij,k}) \rangle$ is singular. With an $N$-jet observable
and with $m=N$, the function $F(\{p\}_{m}^{ij,k})$ vanishes at this
point, so the singularity is weakened or eliminated. This is what happens
in the standard dipole subtraction method \cite{CataniSeymour}.  However,
suppose that $m>N$. Then  $F(\{p\}_{m}^{ij,k})$ does not vanish at the
point in question, but still $d_m(\{p,f\}_{m}^{ij,k})$ vanishes, so that
the cut $d_{\rm ini} < d_m(\{p,f\}_{m}^{ij,k})$ eliminates the
singularity.

In the virtual contribution, we write
\begin{equation}
\begin{split}
\sigma^{\rm V + A}_m
= {}& \frac{1}{m!}
\sum_{\{f\}_m} 
\int\!d\varGamma(\{p\}_m)\
 \theta\big( d_{\rm ini} < d_{m}(\{p,f\}_{m})\big)\,
F_m(\{p\}_m)
\\ &\times
\biggl\{
V(\{p,f\}_m)
-
\sum_{l}\sum_{k\neq l} C_{l,k}(\{p,f\}_{m}, d_{\rm ini})
\biggr\}\;\;.
\end{split}
\label{sigmaVbis}
\end{equation}
This follows Eq.~(\ref{sigmaV}), with the addition of a cut $d_{\rm ini} <
d_m$. There are also added terms involving functions $C_{l,k}$,
\begin{equation}
\begin{split}
  C_{l,k}(\{p,f\}_{m}, d_{\rm ini}) ={}& 
  \int_0^1 \frac{dy}{y}
\int_0^1\!dz\,\int_0^{2\pi}\frac{d\phi}{2\pi}\,
\frac{1}{2}\!\sum_{\hat f_{1},\hat f_{2}}\!
\delta_{\hat f_{1} + \hat f_{2}}^{f}
  \theta\big(d_{\rm ini} <\tilde d(\{p,f\}_{m},l, y, z) \big)
\\ &\times
\frac{y(1-y)\,2 p_l\cdot p_k}{16\pi^2}\
  {\cal D}_{l,k}(\{p,f\}_{m};Y)\;\;.
\end{split}
\label{Clkdef}
\end{equation}
This arises because we modified the dipole subtractions in
Eq.~(\ref{realandsubtractionbis}) by imposing an extra cut
$\tilde d(\{p,f\}_{m}^{ij,k}, \{l,y,z \}_{ij,k})  < d_{\rm ini}$ that
was not present in Eq.~(\ref{realandsubtraction}). What we removed from
the real emission subtraction we must also remove from the virtual loop
subtraction. Doing that and changing integration variables from
$\{\hat p,\hat f\}_{m+1}$ to $(\{p,f\}_{m};Y_l)$ gives
Eq.~(\ref{sigmaVbis}). The details of the variable transformation can be
found in Sec.~\ref{sec:splittingkinematics}, Eq.~(\ref{jacobian3}). The
functions $C_{l,k}(\{p,f\}_{m}, d_{\rm ini})$ are evaluated in
Sec.~\ref{sec:virtualfunctions}, Eqs.~(\ref{Clkform}, \ref{calCresult})

Using dipole subtraction with cuts as defined above, one could construct
a computer code that would calculate at next-to-leading order the
expectation values of infrared safe $N$-jet measurement functions for $N =
2,3,4,\dots$ up to the value for which one had the required calculated
matrix elements. The program would produce partonic events with weights,
and the same events could be used for the calculation of different $N$-jet
observables with different values of $N$. The practical value of such a
program would be minimal, since it would not add anything to having
separate programs for each value of $N$. However, if one could add parton
showers to the calculation thus organized, the program would have some
added value over the separate perturbative programs. It is to this goal
that we now turn.

\section{Partial cross sections with showers}
\label{sec:PartialCrossSectionswithShowers}

In the following sections, we discuss the construction of a parton shower
algorithm that matches the dipole subtraction scheme for NLO
calculations. We incorporate the $k_T$-jet matching scheme \cite{CKKW},
so that the cross section is generated as a sum of partial cross sections
$\sigma_m$ as in Secs.~\ref{sec:PartialCrossSections} and
\ref{sec:PartialCrossSectionsandDipoles}, but this time with showers
added. We suppose that we can calculate the perturbative $m$-parton
cross section, $\sigma^{\rm B}_{m}$, at the Born level,  $\alpha_{\rm
s}^{B_m} \equiv \alpha_{\rm s}^{m - 2}$ for $m = 2,3,\dots,m_{\rm max}$.
Furthermore, we suppose that we can calculate the order $\alpha_{\rm
s}^{B_m + 1}$ corrections, $\sigma^{\rm R-A}_{m}$ and $\sigma^{\rm
V+A}_{m}$, for $m = 2,3,\dots,m_{\rm NLO}$. The values of $m_{\rm NLO} $
and $m_{\rm max}$ depend on our ability to calculate, but we can always
assume that $m_{\rm max} \ge m_{\rm NLO} + 1$.

The cross section computed with parton showers will consist of
contributions from each available $m$,
\begin{equation}
\sigma^{\rm NLO+S} = \sum_{m=2}^{m_{\rm NLO}}
\left[
\sigma^{\rm B+S}_{m} + \sigma^{\rm R+S}_{m} + \sigma^{\rm V+S}_{m} 
\right]
+ \sum_{m=m_{\rm NLO}+1}^{m_{\rm max}} \sigma^{\rm B+S}_{m}
\;\;.
\label{sigmaNLOplusS}
\end{equation}
For the contributions at NLO level, there are three terms, which
correspond to Born, real emission, and virtual loop  terms
with showers added (``$+S$''). For the remaining contributions there is
only a  Born term. We will arrange that (for a suitably behaved
observable)
\begin{equation}
\begin{split}
\sigma^{\rm B+S}_{m}  + \sigma^{\rm R+S}_{m} + \sigma^{\rm V+S}_{m}
={}&
\sigma^{\rm B}_{m} + \sigma^{\rm R-A}_{m} + \sigma^{\rm V+A}_{m} 
\\ &
+{\cal O}(\alpha_{\rm s}^{B_m + 2})
+{\cal O}(\alpha_{\rm s}^{B_m}\times 1\, {\rm GeV}/\sqrt s\,)
\end{split}
\label{showernet}
\end{equation}
and 
\begin{equation}
\begin{split}
\sigma^{\rm B+S}_{m}
={}&
\sigma^{\rm B}_{m} 
+{\cal O}(\alpha_{\rm s}^{B_m + 1})
+{\cal O}(\alpha_{\rm s}^{B_m}\times 1\, {\rm GeV}/\sqrt s\,)
\label{showernetBorn}
\;\;.
\end{split}
\end{equation}

The main body of this construction is in the following section (with
several subsections). There we construct $\sigma^{\rm B+S}_{m}$, 
$\sigma^{\rm R+S}_{m}$, and $\sigma^{\rm V+S}_{m}$ for $2 \le m \le m_{\rm
NLO}$ and show that these quantities sum to $\sigma^{\rm B}_{m} +
\sigma^{\rm R-A}_{m} + \sigma^{\rm V+A}_{m}$ to NLO accuracy. We also
obtain the leading-order result in Eq.~(\ref{showernetBorn}) as a
byproduct.

\section{The shower construction}
\label{sec:ShowerConstruction}

In this section, we discuss $\sigma^{\rm B+S}_{m}$, 
$\sigma^{\rm R+S}_{m}$, and $\sigma^{\rm V+S}_{m}$ for $2 \le m\le m_{\rm
NLO}$ and show that these quantities sum to $\sigma^{\rm B}_{m} +
\sigma^{\rm R-A}_{m} + \sigma^{\rm V+A}_{m}$ to NLO accuracy. We begin
with $\sigma^{\rm B + S}_m$ in Sec.~\ref{sec:BornwithShowers} and the
following subsections. Then $\sigma^{\rm
R+S}_{m}$ is described in Sec.~\ref{sec:NLOrealonShower}. The virtual
contribution $\sigma^{\rm V+A}_{m}$ is described in
Sec.~\ref{sec:NLOvirtualonShower}. We sum the contributions in
Sec.~\ref{sec:netresult}.

\subsection{Born term with showers}
\label{sec:BornwithShowers}

Our discussion begins in this subsection with $\sigma^{\rm B +
S}_m$. We define
\begin{equation}
\begin{split}
  \sigma^{\rm B+S}_{m} = {}& \frac{1}{m!}  \sum_{\{f\}_m}
  \int\!d\varGamma(\{p\}_m)\, 
  \theta(d_{\rm ini}< d_m(\{p,f\}_m))\,W_{m}(\{p,f\}_m)
  \\
  &\times \sum_{l=1}^m \sum_{k\ne l}
  \big\langle {\cal M}(\{p,f\}_m)\big| 
   \int\!dY_l\
  \bm{E}_{l,k}(\{p,f\}_m,Y_l)\, \big|{\cal M}(\{p,f\}_m)\big\rangle
  \\
  &\times I(\{p,f\}_m;l,k,Y_l) \;\;.
\label{sigmaBS}
\end{split}
\end{equation}
This formula is illustrated in Fig.~\ref{fig:sigmaB}.
The first line contains integrals over Born level parton momenta and a
corresponding sum over parton flavors. The second line contains sums over
choices $l$ of the parton that splits and a spectator parton $k$ along
with an integral over the splitting variables $Y_{l}$ and a matrix element
of certain operators $\bm{E}_{l,k}$ acting on the Born amplitude
$|{\cal M}\rangle$. The operators $\bm{E}_{l,k}$ together with the other
factors in the formula describe the formation of showers from the Born
level partons.

\FIGURE{
\centerline{
\includegraphics[width = 6 cm]{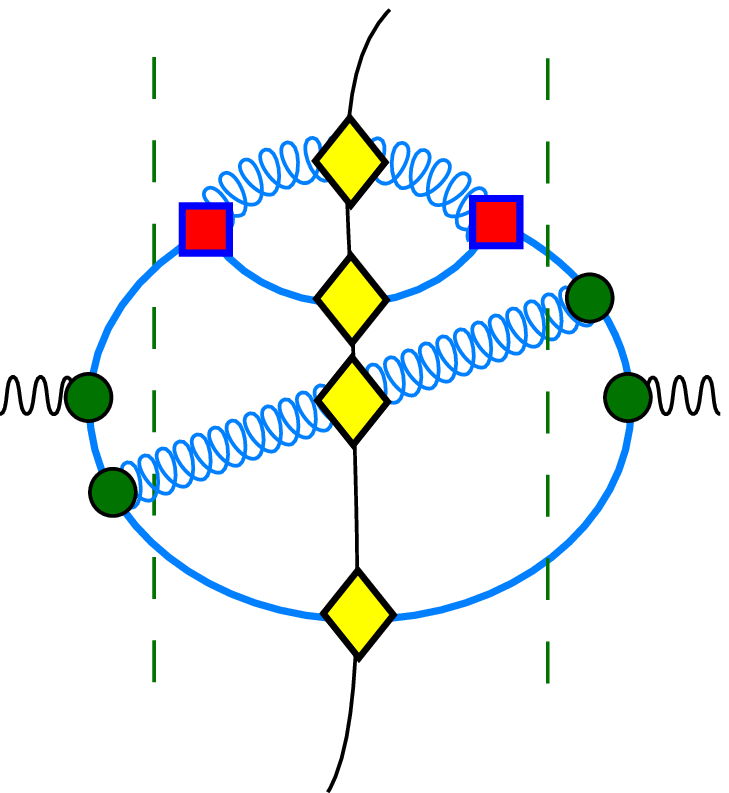}
}
\medskip
\caption{
Illustration of $\sigma_3^{\rm B + S}$, Eq.~(\ref{sigmaBS}). 
The basic Born graph lies outside the dashed lines. The three partons are
required to be resolvable at scale $d_{\rm ini}$. To the left of the
first dashed line, we illustrate one contribution to $|{\cal M}\rangle$.
To the right, we illustrate one contribution to  $\langle{\cal M}|$. We
multiply by a reweighting factor $W$, not illustrated, that contains
Sudakov factors for a constructed parton splitting history.  Between the
dashed lines there is a splitting of one of the partons $l$, using a
spectator parton $k$, as described by the splitting operator ${\bm
E}_{l,k}$. Finally, there is secondary showering represented in the
figure by diamonds and in Eq.~(\ref{sigmaBS}) by the function $I$. The
diamonds split the partons into a shower of partons and then into jets of
hadrons. }
\label{fig:sigmaB}
}

Our goal in the following subsections is first to explain the parts of
Eq.~(\ref{sigmaBS}) and then to show that, with suitable definitions
of the functions in Eq.~(\ref{sigmaBS}), the perturbative expansion of
$\sigma^{\rm B + S}_m$ gives the Born cross section $\sigma^{\rm B}_m$ at
order $\alpha_{\rm s}^{B_m}$. Expanding to one more order, we get certain
order $\alpha_{\rm s}^{B_m+1}$ terms that we need to keep track of. We
will then be able to add correction terms that allow us to match the
$\alpha_{\rm s}^{B_m+1}$ contributions in the dipole subtraction scheme.

\subsection{The Born level partons}
\label{sec:bornpartons}

The first line of Eq.~(\ref{sigmaBS}) concerns what we call the Born
level partons. These are labelled by an index $l$ that takes the
values $l \in \{1,2,\dots, m\}$. The Born level partons are produced
in the hard scattering and evolve into final state showers.

There is an integration over the phase space for the $m$ final state
partons with momenta $\{p\}_m$. There is also a sum over the flavors
$\{f\}_m$ of the final state partons. The integration over the momenta
$\{p\}_m$ is restricted by a factor
\begin{displaymath}
\theta\big(d_{\rm ini}< d_m(\{p,f\}_m)\big)\;\;.
\end{displaymath}
Here $d_m(\{p,f\}_m)$ is defined by applying the $k_T$ jet finding
algorithm to the $m$ parton momenta, as in Eq.~(\ref{dij}).
As discussed in Secs.~\ref{sec:PartialCrossSections} and
\ref{sec:PartialCrossSectionsandDipoles}, we classify a final state of $n$
partons by the values of the jet resolution parameters $d_m$. The
contribution to the observable from states with $d_{m+1} < d_{\rm ini} <
d_m$ are calculated using $\sigma^{\rm B+S}_m$ and its corrections. The
cut on $d_m$ in Eq.~(\ref{sigmaBS}) implements the first part of this
general plan.

There is also a factor $W_{m}(\{p,f\}_m)$, which is the product of
factors associated with the splitting history that matches the
found jet structure, following the method of Ref.~\cite{CKKW}. Imagine
that the found jet structure had been generated by parton shower Monte
Carlo program. Then for each vertex $V$ in the graph that represents the
splitting history, there would be a factor $\alpha_{\rm s}(\sqrt {d_V
s})$, where $d_V$ is the $k_T$ scale of the splitting at that vertex. In
the exact squared matrix element $\langle {\cal
M}(\{p,f\}_m) |{\cal M}(\{p,f\}_m)\rangle$, there is a factor 
$\alpha_{\rm s}(\mu_{\rm R})$ for each vertex. We switch from 
$\alpha_{\rm s}(\mu_{\rm R})$ to $\alpha_{\rm s}(\sqrt {d_V
s})$ by including in $W_{m}(\{p,f\}_m)$ a factor 
\begin{equation}
\frac{\alpha_{\rm s}(\sqrt {d_V s})}
{\alpha_{\rm s}(\mu_{\rm R})}
\label{CKKWfactor1}
\end{equation}
for each vertex. The parton shower Monte Carlo program would also
generate a Sudakov factor that represents the probability that the
partons did not split between each scale $d_V$ at which a splitting
occurred and the next smaller scale $d_V$ and the probability that the
partons did not split between the smallest scale $d_V$ at which a
splitting occurred and the limiting scale $d_{\rm ini}$. This Sudakov
factor is also included in $W_{m}(\{p,f\}_m)$. The details are given in
Sec.~\ref{sec:CKKWSudakov}.

All that we need to know at present about $W_{m}(\{p,f\}_m)$ is that
it has a perturbative expansion in powers of $\alpha_{\rm
s}(\mu_{\rm R})$ so that
\begin{equation}
W_{m}(\{p,f\}_m) = 1 + \frac{\alpha_{\rm s}(\mu_{\rm R})}{2\pi}\,
W_m^{(1)}(\{p,f\}_m)
+ \cdots\;\;.
\label{Wexpansion}
\end{equation}
We will use this property in our perturbative analysis.

In the second line of Eq.~(\ref{sigmaBS}) there is a matrix element of
certain operators $\bm{E}_{l,k}$ that act on the Born amplitude
$\big|{\cal M}(\{p,f\}_m)\big\rangle$ to generate the splitting of the
partons. Our next task is to describe this parton splitting.

\subsection{Description of parton splitting}
\label{sec:partonsplitting}

Each Born level parton has the opportunity to split into two daughter
partons. The one that splits with the largest (suitably defined) evolution
parameter is designated as parton $l$. The splitting of the remaining
partons is left to be described by the function $I$ in Eq.~(\ref{sigmaBS}). This
splitting of parton $l$ with the help of spectator parton $k$ is
described by an operator $\bm{E}_{l,k}$.

There are a number of parameters that describe the splitting of parton
$l$,
\begin{equation}
Y_l = \{y_l, z_l,\phi_l, \hat f_{l,1}, \hat f_{l,2}\}\;\;.
\end{equation}
as in Eq.~(\ref{Yldef}).  We integrate over these variables in
Eq.~(\ref{sigmaBS}), using $\int\! dY_l$ as defined in
Eq.~(\ref{dYldef}).

The momenta and flavors of the $m+1$ daughter partons are given in
terms of the momenta and flavors of the $m$ mother partons and the
splitting variables $Y_l$ by the transformation (\ref{Rldef}),
\begin{equation}
\{\hat p,\hat f\}_{m+1}
=
{\cal R}_{l,k}(\{p,f\}_m,Y_l)\;\;.
\label{Rldefencore}
\end{equation}

\subsection{The splitting functions}
\label{sec:splittingfunctions}

Now, we turn to the splitting function $\bm{E}_{l,k}$ in
Eq.~(\ref{sigmaBS}), which is an operator on the color and spin space of
parton $l$ in the vector $|{\cal M}\rangle$. This operator has the
following form,
\begin{equation}
\label{Elfinal}
\begin{split}
  \bm{E}_{l,k} = {}& 
 \int_{0}^{\infty}\! dr\,
   \delta(r - R_{l}(\{p,f\}_{m}, y_{l}, z_{l}))\,
   \theta(\tilde d(\{p,f\}_m, l, y_l,z_l) < d_{\rm ini})
  \\ &\times\frac{\bm{T}_{l}\cdot
    \bm{T}_{k}}{-\bm{T}_{l}^2}\, 
  \frac{\alpha_{\rm s}(k_T(\{p,f\}_{m},l,y,z))}{2\pi}\,
  \bm{S}_{l}(p_l,f_l,Y_l)
  \\
  &\times\exp\!\biggl( -\int_{r}^{\infty}\!\! dr' \int_{0}^1
  \frac{dy'}{y'}\int_0^1\!\! dz'\, 
   \sum_{l'}
  \delta(r' - R_{l'}(\{p,f\}_{m}, y',z'))
  \\
  &\hskip 2.0 cm \times 
 \theta(\tilde d(\{p,f\}_m, l', y',z') < d_{\rm ini})\,
  \frac{\alpha_{\rm s}(k_T(\{p,f\}_{m},l', y',z'))}{2\pi}
\\
  &\hskip 2.0 cm \times 
  \big\langle\bm{S}(y',z',f_{l'})\big\rangle \biggr)\;\;.
\end{split}
\end{equation}
The parton splitting is organized according to an evolution parameter
$r$, which is defined in Eq.~(\ref{Elfinal}) to be a certain function
$R_{l}$ of the hard parton momenta and flavors $\{p,f\}_{m}$ and the
splitting parameters $y_l,z_l$. The function $R_{l}$ measures the hardness
of the splitting. There is some flexibility in choosing this function as
long as $y \to 0$ implies $R_{l} \to 0$.  Some of the choices available
may be understood using the kinematic variables described in
Sec.~\ref{sec:splittingkinematics}. If the secondary shower encoded in
the Monte Carlo interface function to be described in
Sec.~\ref{sec:MCinterface} is based on the P{\small YTHIA} algorithm it
might be sensible to set $R_{l}$ to a scaled virtuality, 
\begin{equation}
  R_{l}(\{p,f\}_{m}, y,z) = s_l\, y\;\;,
\label{virtualityevolution}
\end{equation}
where $s_l$ is the virtuality scale associated with parton $l$ as
discussed at Eq.~(\ref{tildeddef}). In the H{\small ERWIG} case one might
choose a scaled squared transverse momentum,
\begin{equation}
  R_{l}(\{p,f\}_{m}, y,z) =s_l\, y z(1-z)\;\;.
\label{ktevollution}
\end{equation}
We have indicated the scale of $\alpha_{\rm s}$ as a function $k_T$,
defined by
\begin{equation}
k_T(\{p,f\}_{m},l, y,z) = [s_l\, y z(1-z)]^{1/2}\;\;.
\label{ktdef}
\end{equation}
Other choices for the scale in $\alpha_{\rm s}$ are possible. The
matching to fixed order perturbation theory will work as long as we can
write $\alpha_{\rm s}(k_T(\{p,f\}_{m},l, y,z)) = \alpha_{\rm s}(\mu_R) +
{\cal O}(\alpha_{\rm s}^2)$. Although our notation allows the possibility
of a variety of choices for the functions $R_l$ and $k_T$, for purposed
of minimizing the dependence on the arbitrary parameter $d_{\rm ini}$ our 
favored choices are given by Eqs.~(\ref{ktevollution}) and
(\ref{ktdef}) with $s_l$ a certain function of $\{p,f\}_m$ that is defined
later in Eq.~(\ref{sldef}).

In Eq.~(\ref{Elfinal}), we also make use of a function $\tilde d$ defined
in Eq.~(\ref{tildeddef}) that gives an approximate version of the
resolution variable associated with the splitting generated with the
splitting parameters $y_l,z_l$. With the use of this function, we
limit the splitting in $\bm{E}_{l}$ to be unresolvable at a scale $d$ that
is approximately $d_{\rm ini}\times 2p_l\cdot p_{k_l}/s_l$ (see 
Eq.~(\ref{tildedapprox})).

In each $\bm{E}_{l,k}$ operator, there is an operator on the parton color
space, ${\bm{T}_l\cdot \bm{T}_{k}}/[{-\bm{T}_l^2}]$, that is familiar
from the dipole splitting formulas. Next, there is an operator on the
parton spin space, $\bm{S}_{l}$, with the interpretation that
$(\alpha_{\rm s}/(2\pi))\,\bm{S}_l$ is the probability for the parton to
split at a given evolution parameter $r$ if it has not split at a higher
evolution parameter.  The splitting function $\bm{S}_{l}$ depends on the
splitting parameters  $Y_l = \{y_l,z_l,\phi_l,\hat f_{l,1},\hat f_{l,2}\}$
for parton $l$ as well as on the momentum $p_l$, which is needed to fully
specify the meaning of $\phi_l$. We will specify this function later in
Sec.~\ref{sec:splittingfctns}, Eq.~(\ref{flavorDipoleFF:S}).

The next factor, the Sudakov exponential, gives the probability that
{\it none} of the partons has split at a higher evolution scale. Thus we
work in a scheme similar to that of Sj\"ostrand and Skands
\cite{SjostrandSkands} and of Nason \cite{Nasonshowers}, picking out the
hardest splitting. In the exponent there is a sum over partons
$l'$ and an integration over virtualities $y'$ and the momentum fractions
$z'$ of the ``virtual'' splittings. The corresponding evolution parameter
$r'$ is required to be bigger than $r$.

The remaining factor in the Sudakov exponent is the average over angle
and flavors of $\bm{S}$ for parton $l'$,
\begin{equation}
\big\langle \bm{S}(y_{l'},z_{l'},f_{l'})\big\rangle
\equiv
\int \frac{d\phi_l}{2\pi}\
\frac{1}{2}\!\sum_{\hat f_{{l'},1},\hat f_{{l'},2}}\!
\delta_{\hat f_{{l'},1} + \hat f_{{l'},2}}^{f_{l'}}
\bm{S}_{l'}(p_{l'},y_{l'},z_{l'},\phi_{l'},\hat f_{{l'},1},\hat
f_{{l'},2})\;\;.
\label{Slaveragedef}
\end{equation}
With our definitions, this is a numerical function times a unit operator
on the partonic spin space. Explicit expressions are given later in
Sec.~\ref{sec:splittingfctns}, Eq.~(\ref{avSquark}), (\ref{avSgluon}).

\subsection{Monte Carlo interface function}
\label{sec:MCinterface}

The last factor in Eq.~(\ref{sigmaBS}) is
\begin{equation}
I(\{p,f\}_m;l,k,Y_l)\;\;.
\end{equation}
We imagine that after the parton splitting represented by the
splitting function $\bm{E}_{l,k}$, which we may call the {\it primary}
splitting, there is further parton showering, which we may call {\it
secondary showering}. This showering is to be carried out by a shower
Monte Carlo style computer program. The factor $I$ represents the average
value of the observable corresponding to the daughter hadrons after
secondary showering when the shower starts with initial conditions
specified by the variables $(\{p,f\}_m;l,k,Y_l)$. Here the initial
conditions include not only the partonic state 
\begin{equation}
\{\hat p,\hat f\}_{m+1} = {\cal R}_{l,k}(\{p,f\}_m,Y_l)
\end{equation}
generated from the partonic
state $\{p,f\}_m$ by splitting parton $l$ according to the splitting
variables $Y_l$, but also the history of the first step of showering as
specified by $Y_l$. 

We can think of $I$ as being an integral,
\begin{eqnarray}
I
&=&
\sum_N\frac{1}{N!}
\sum_{\{\tilde f\}_N} 
\prod_{i=1}^N
\left( \int\!d\tilde p_i\ \delta(\tilde p_i^2 - m^{\!2}(\tilde f_i))
\right)
\nonumber \\
&&\times
P(
\{\tilde p,\tilde f\}_N|
\{p,f\}_m;l,k,Y_l
)\
F_N(\{\tilde p\}_N )\;\;.
\label{Ifull}
\end{eqnarray}
There is a sum over the number $N$ of final state hadrons that are
generated by the shower, a symmetry factor $1/N!$, a sum over the
flavors $\{\tilde f\}_N$ of these hadrons and an integration over their
momenta $\{\tilde p\}_N$.  In the next line there is a factor
$P$ that represents the probability density to produce the final state
hadrons given the starting conditions represented by
$(\{p,f\}_m;l,k,Y_l)$.  The final factor in Eq.~(\ref{Ifull}) is the
measurement function evaluated with the produced final state hadrons. 
If one were to substitute $F = 1$, one would get the normalization
condition for the conditional probability $P$, namely $I = 1$.
Equivalently, if we substitute $F = \lambda$, then 
\begin{equation}
I(\{p,f\}_m;l,k,Y_l)\big|_{F=\lambda} = \lambda\;\;.
\label{Inorm}
\end{equation}
Here $\lambda$ could depend on the initial conditions for the secondary
shower as represented by the variables $(\{p,f\}_m;l,k,Y_l)$.

A simple model for $I$ is obtained by omitting all secondary showering.
Then $I$ becomes
\begin{equation}
I^{(0)}(\{p,f\}_m;l,k,Y_l) =
F_{m+1}({\cal R}_{l,k}(\{p,f\}_m,Y_l))\;\;.
\label{I0}
\end{equation}

We will have more to say about the construction of the Monte Carlo
showering program represented by function $I$ in
Sec.~\ref{sec:moreMCinterface}. One important feature is that the primary
splitting, the splitting of parton $l$, should be the hardest splitting in
the shower (according to the hardness measure $R$ in
Eq.~(\ref{Elfinal})).   However, we leave the choice of this program
largely open.  Here we simply note two properties that $I$ should have
when the measurement $F$ is an infrared safe $N$-jet measurement
function.

First, secondary showering should provide perturbative and power
suppressed corrections to the simple function $I^{(0)}$ (assuming,
always, an infrared safe observable):
\begin{eqnarray}
I(\{p,f\}_m;l,k,Y_l) &=&
F_{m+1}(
{\cal R}_{l,k}(\{p,f\}_m,Y_l))\,
\nonumber\\
&&\times [1 +{\cal O}(\alpha_{\rm s}) + {\cal O}(1\ {\rm GeV}/\sqrt s)]\;\;.
\label{Iproperty1}
\end{eqnarray}
Here the order $\alpha_{\rm s}$ correction corresponds to splitting with a
substantial virtuality, while the power suppressed correction corresponds
to hadronization. 

Second, when $y_l$ vanishes, $I$ should reduce to $I^{(0)}$
with only power corrections,
\begin{equation}
I(\{p,f\}_m;l,k,0,z_l,\phi_l,\hat f_{l,1},\hat f_{l,2}) =
F_m(\{p\}_m)
\times[1 + {\cal O}\left({1\ {\rm GeV}}/{\sqrt s}\right)]\;\;.
\label{Iproperty0}
\end{equation}
The requirement here is that having the initial splitting
virtuality $y_l$ equal to zero should set the maximum hardness for all
of the secondary splittings to zero and thus turn the secondary showering
off except for hadronization. The hadronization model should then turn the
partons into jets of a limited mass, leading to only power suppressed
contributions to the measurement function. Note that $F_{m+1}(\{\hat
p,\hat f\}_{m+1})$ reduces to $F_m(\{p\}_m)$ here because of the infrared
safety property of the measurement function.

In Secs.~\ref{sec:NLOrealonShower} and \ref{sec:NLOvirtualonShower}, we
will also need a function $\tilde I(\{p,f\}_{n})$ that
represents the average value of the observable corresponding to the
daughter hadrons after secondary showering when the shower starts with a
partonic state $\{p,f\}_n$ with no other information (other than the
$d_{\rm ini}$ cut) given as to previous shower history. The function
$\tilde I(\{p,f\}_n)$ obeys the normalization condition
\begin{equation}
\tilde I(\{p,f\}_n)\big|_{F=\lambda} = \lambda\;\;,
\label{tildeInorm}
\end{equation}
where $\lambda$ could depend on the initial conditions for the secondary
shower as represented by the variables $(\{p,f\}_n)$.

As in the case of the version of $I$ above, we assume that $\tilde I$ is
constructed so that the effects on the measurement of the showering are
suppressed either by a power of $\alpha_{\rm s}$ or a power of $1\ {\rm
GeV}/\sqrt s$,
\begin{equation}
\tilde I(\{p,f\}_{n}) =
F_{n}(\{p\}_{n})
\times[ 1 + {\cal O}(\alpha_{\rm s}) + {\cal O}(1\ {\rm GeV}/\sqrt s)]\;\;.
\label{tildeIproperty}
\end{equation}

\subsection{Perturbative expansion}
\label{sec:perturbativeexpansion}

We now seek the perturbative expansion of $\sigma^{\rm B+S}_{m}$ in
Eq.~(\ref{sigmaBS}). Insert into Eq.~(\ref{sigmaBS}) a factor $1 = T_l +
\Delta_l$, where $T_l$ sets $y_l$ to zero in $I$ and $\Delta_l \equiv1 -
T_l$. Then
\begin{equation}
\begin{split}
  \sigma^{\rm B+S}_{m} = {}& \frac{1}{m!}  \sum_{\{f\}_m}
  \int\!d\varGamma(\{p\}_m)\, 
   \theta\big(d_{\rm ini}< d_m(\{p,f\}_m)\big)\,
   W_{m}(\{p,f\}_m)
  \\&\times \sum_{l=1}^m \sum_{k\ne l}
  \big\langle {\cal M}(\{p,f\}_m)\big| 
   \int\!dY_l\
  \bm{E}_{l,k}(\{p,f\}_m,Y_l)\, \big|{\cal M}(\{p,f\}_m)\big\rangle
  \\
  &\times (T_l + \Delta_l) I(\{p,f\}_m;l,k,Y_l)
  \;\;.
\label{sigmaBS1}
\end{split}
\end{equation}

Our strategy will be to separate the $T_l$ term from the
$\Delta_l$ term. We call the the $T_l$ term $\sigma^{\rm B+S}_{m,T}$ and
the $\Delta_l$ term $\sigma^{\rm B+S}_{m,\Delta}$. Thus we write
\begin{equation}
\sigma^{\rm B+S}_{m}
=
\sigma^{\rm B+S}_{m,T}
+\sigma^{\rm B+S}_{m,\Delta}
\;\;.
\label{DeltaExpansion}
\end{equation}

Then we will expand each term in powers of $\alpha_{\rm s}$, up to
next-to-leading order. We will find that, to this order, there are two
terms in the expansion of $\sigma^{{\rm B}+{\rm S}}_{m,T}$
\begin{equation}
\sigma^{\rm B+S}_{m,T} = 
\sigma^{\rm B+S}_{m,\{T,0\}}
+ \sigma^{\rm B+S}_{m,\{T,1\}}
+ {\cal O}(\alpha_{\rm s}^{B_m + 2})\;\;.
\label{sigmaBSm0expansion}
\end{equation}
Here $\sigma^{\rm B + S}_{m,\{T,0\}}$ is proportional to
$\alpha_{\rm s}^{B_m}$ and is, in fact, the Born contribution,
$\sigma^{\rm B}_{m}$, to $\sigma_m$. Then $\sigma^{\rm B + S}_{m,\{T,1\}}$
consists of certain $\alpha_{\rm s}^{B_m+1}$ corrections. We will then
find that the expansion of $\sigma^{\rm B + S}_{m,\Delta}$ begins
at order $\alpha_{\rm s}^{B_m + 1}$, so that we need only the first term
in this expansion in order to evaluate $\sigma^{\rm B + S}_m$ to
order $\alpha_{\rm s}^{B_m + 1}$.

\subsubsection{The $T_l$ contribution}
\label{sec:BornSTterm}

We begin by analyzing the term $\sigma^{\rm B + S}_{m,T}$. We use
\begin{equation}
  \label{Rexpand0}
  \begin{split}
     T_l\, &I(\{p,f\}_m,l,k,Y_l)) \equiv
    I(\{p,f\}_m;l,k,0,z_l,\phi_l,\hat f_{l,1},\hat f_{l,2})\;\;.
  \end{split}
\end{equation}
Using the property (\ref{Iproperty0}) of $I$ we have
\begin{equation}
  \begin{split}
    I(\{p,f\}_m;l,k,0,z_l,\phi_l,\hat f_{l,1},\hat f_{l,2}) =
F\big(\{p\}_m\big) 
   \times[1+
  {\cal O}\left({1\ {\rm GeV}}/{\sqrt s}\right)]\;\;.
  \end{split}
\label{R0}
\end{equation}
Then
\begin{equation}
\label{Rexpand1}
\begin{split}
  \sigma^{\rm B+S}_{m,T} = {}& \frac{1}{m!}  \sum_{\{f\}_m}
  \int\!d\varGamma(\{p\}_m) 
   \theta\big(d_{\rm ini}< d_m(\{p,f\}_m)\big)\,
    W_{m}(\{p,f\}_m)\, F_m\big(\{p\}_m\big)
  \\&\times \sum_{l=1}^m \sum_{k\ne l}
 \big\langle {\cal M}(\{p,f\}_m)\big| 
   \int\!dY_l\
  \bm{E}_{l,k}(\{p,f\}_m,Y_l)\, \big|{\cal M}(\{p,f\}_m)\big\rangle
  \\
  &\times[ 1 + {\cal O}(1\,\text{GeV}/\sqrt{s})] \;\;.
\end{split}
\end{equation}

Let us look at the sum of the $\bm {E}_{l,k}$ operators integrated over
the corresponding splitting variables. We have
\begin{equation}
  \label{Elexpansion}
  \begin{split}
    \sum_{l,k} \int\!dY_l\, \bm {E}_{l,k}
    ={}& 
   \sum_{l=1}^m
    \sum_{k\ne l} \frac{\bm{T}_{l}\cdot \bm{T}_{k}}{-\bm{T}_{l}^2}
    \int_{0}^{\infty}\! dr \int_0^1\! \frac{dy_l}{y_l} \int_0^1\!
    dz_l\,  \delta(r - R_{l}(\{p,f\}_{m}, y_{l},z_{l}))
    \\
    &\quad\times 
    \theta( \tilde d(\{p,f\}_m,l,y_l,z_l) < d_{\rm ini})\
   \frac{\alpha_{\rm s}(k_T(\{p,f\}_{m},l,y,z))}{2\pi}
   \\
    &\quad\times  
   \frac{1}{2}\!\sum_{\hat
      f_{l,1},\hat f_{l,2}}\!  \delta_{\hat f_{l,1} + \hat
      f_{l,2}}^{f_l} \int_0^{2\pi}
    \frac{d\phi_l}{2\pi}\,\bm{S}_{l}(p_l,f_l,Y_l)\,
    \\
    &\quad\times\exp\!\biggl( - \int_{r}^{\infty}\!\! dr' \int_0^1
    \frac{dy'}{y'}\int_0^1\!\! dz' \sum_{l'}\,
    	\delta(r' - R_{l}(\{p,f\}_m,y',z'))
    \\
    &\qquad\qquad\qquad \times 
    \theta( \tilde d(\{p,f\}_m,l,y',z') < d_{\rm ini})\
    \frac{\alpha_{\rm s}(k_T(\{p,f\}_{m},l', y',z'))}{2\pi}
   \\
    &\qquad\qquad\qquad \times 
   \big\langle
    \bm{S}(y',z',f_{l'})\big\rangle \biggr)\;\;,
  \end{split}
\end{equation}
where $\langle\bm{S}\rangle$ was defined in Eq.~(\ref{Slaveragedef}). 
According to Eq.~(\ref{coloridentity}), the sum over $k$ of the color
factors is
\begin{equation}
  \sum_{k}
  \frac{\bm{T}_{l}\cdot\bm{T}_{k}}
  {-\bm{T}_{l}^2} = 1
\end{equation}
when operating on the state $\big|{\cal M}(\{p,f\}_m)\big\rangle$. Having
eliminated the color factor, we have an integral of a derivative,
\begin{equation}
  \label{Elterm1}
\begin{split}
  \int_0^{\infty}\!\! dr\ \frac{d}{dr} & \exp\!\biggl( -
  \int_{r}^{\infty}\!\! dr' \int_0^1 \frac{dy'}{y'}\int_0^1\!\!
  dz'\, \delta(r' - R_{l'}(\{p,f\}_m,y',z'))
  \\
  &\hskip 1.5 cm \times 
    \theta( \tilde d(\{p,f\}_m,l',y',z') < d_{\rm ini})\
    \frac{\alpha_{\rm s}(k_T(\{p,f\}_{m},l', y',z'))}{2\pi}
   \\
   &\hskip 1.5 cm \times 
   \big\langle
    \bm{S}(y',z',f_{l'})\big\rangle \biggr)
  \\
  &={} 1\;\;,
  \end{split}
\end{equation}
since the exponent is 0 at $r = \infty$ and $-\infty$ at $r = 0$
(since the integration region for $r=0$ includes the $y \to 0$
singularity). Thus (when operating on the state
$\big|{\cal M}(\{p,f\}_m)\big\rangle$)
\begin{equation}
  \label{Elintegral}
    \sum_{l,k}\int\!dY_l\, \bm {E}_{l,k}
    =1\;\;.
\end{equation}

We consider next the factor $W_{m}(\{p,f\}_m)$ in Eq.~(\ref{Rexpand1}),
which we write as 1 plus an order $\alpha_{\rm s}$ contribution
$\alpha_{\rm s}\,W_m^{(1)}(\{p,f\}_m)$ according to
Eq.~(\ref{Wexpansion}),
\begin{equation}
W_{m}(\{p,f\}_m) = 1 + \frac{\alpha_{\rm s}(\mu_{\rm R})}{2\pi}\,
W_m^{(1)}(\{p,f\}_m)
+ \cdots\;\;.
\end{equation}

Taking the lowest order term in $\sigma^{\rm B+S}_{m,T}$ gives the lowest
order term in the perturbative expansion of $\sigma^{\rm B+S}_m$. This
term is
\begin{equation}
\label{sigmaBS00}
\begin{split}
  \sigma^{\rm B+S}_{m, \{T,0\}} = {}& \frac{1}{m!}  \sum_{\{f\}_m}
  \int\!d\varGamma(\{p\}_m)\ \theta\big(d_{\rm ini}< d_m(\{p,f\}_m)\big)
  \\
  &\times \big\langle {\cal M}(\{p,f\}_m) \big|{\cal
    M}(\{p,f\}_m)\big\rangle\, F_m\big(\{p\}_m\big)
  \\
  &\times \big[1 + {\cal O}(1\,\text{GeV}/\sqrt{s})\big] \;\;.
\end{split}
\end{equation}
Comparing to Eq.~(\ref{sigmaBm}), we see that
\begin{equation}
\sigma^{\rm B+S}_{m, \{T,0\}} = \sigma^{\rm B}_m
\times
\big[1 + {\cal O}(1\,\text{GeV}/\sqrt{s})\big]\;\;.
\label{sigmaBS00is}
\end{equation}

The only order $\alpha_{\rm s}^1$ term in $\sigma^{\rm B+S}_{m, T}$ comes
from the expansion of the reweighting function $W$. This term is
\begin{equation}
  \label{sigmaBS01}
  \begin{split}
    \sigma^{\rm B+S}_{m,\{T,1\}} = {}& \frac{1}{m!}\sum_{\{f\}_m}
    \int\!d\varGamma(\{p\}_m)\, 
     \theta\big(d_{\rm ini}< d_m(\{p,f\}_m)\big)\,
    F_m(\{p\}_m)
    \\
    &\qquad\qquad\times \frac{\alpha_{\rm s}(\mu_{\rm R})}{2\pi}\,
    W_{m}^{(1)}(\{p,f\}_m)\, \big\langle {\cal M}(\{p,f\}_m)
    \big|{\cal M}(\{p,f\}_m)\big\rangle
  \\
  &\qquad\qquad\times \big[1 + {\cal O}(1\,\text{GeV}/\sqrt{s})\big] \;\;,
  \end{split}
\end{equation}
where $W_m^{(1)}$ is the first order contribution to $W$,
Eq.~(\ref{Wexpansion}).

\subsubsection{The $\Delta_l$ contribution}
\label{sec:BornSDeltaterm}

Now we turn to the contribution $\sigma^{\rm B + S}_{m,\Delta}$ in
Eq.~(\ref{DeltaExpansion}).  From the  definition in
Eq.~(\ref{DeltaExpansion}) we have
\begin{equation}
  \label{sigmaBSJ0}
  \begin{split}
    \sigma^{\rm B + S}_{m,\Delta} = {}& \frac{1}{m!} \sum_{\{f\}_m}
    \int\!d\varGamma(\{p\}_m)\ 
    \theta\big(d_{\rm ini}< d_m(\{p,f\}_m)\big)\,
    W_{m}(\{p,f\}_m)
    \\
    &\qquad\times  \sum_{l=1}^m\sum_{k\ne l}
    \big\langle {\cal M}(\{p,f\}_m)\big|
   \int\!dY_l\
  \bm{E}_{l,k}(\{p,f\}_m,Y_l)\, \big|{\cal M}(\{p,f\}_m)\big\rangle
    \\
    &\qquad\times \Delta_l 
    I(\{p,f\}_m;l,k,Y_l) \;\;.
  \end{split}
\end{equation}
Here
\begin{equation}
\label{DeltaJ}
  \begin{split}
    \Delta_l\,
    I(\{p,f\}_m;l,k,Y_l)  ={}& I(\{p,f\}_m; l,k,Y_l) 
- I(\{p,f\}_m;l,k,0,z_l,\phi_l,\hat f_{l,1},\hat f_{l,2})\;\;.
  \end{split}
\end{equation}
For $I$ with $y_l = 0$ we can use Eq.~(\ref{R0}). For $I$
with nonzero $y_l$ we use Eq.~(\ref{Iproperty1}). This gives
\begin{equation}
  \begin{split}
\label{DeltaJfinal}
 \Delta_l\,
    I(\{p,f\}_m;l,k,Y_l) 
= {}& \Big\{ F_{m+1}({\cal R}_{l,k}(\{p,f\}_m,Y_l)) - F_m\big(\{p\}_m\big)
\Big\}\\
&\times [1+ {\cal O}(\alpha_{\rm s}) 
+ {\cal O}\left({1\ {\rm GeV}}/{\sqrt s}\right)]\;\;.
  \end{split}
\end{equation}

We use this result in  Eq.~(\ref{sigmaBSJ0}), giving
\begin{equation}
  \label{sigmaBSJ1}
  \begin{split}
    \sigma^{\rm B + S}_{m,\Delta} = {}& \frac{1}{m!} \sum_{\{f\}_m}
    \int\!d\varGamma(\{p\}_m)\ 
     \theta\big(d_{\rm ini}< d_m(\{p,f\}_m)\big)\,
     W_{m}(\{p,f\}_m)
    \\
    &\quad
   \times \sum_{l=1}^m\sum_{k\ne l}
    \big\langle {\cal M}(\{p,f\}_m)\big|
   \int\!dY_l\
   \bm{E}_{l,k}(\{p,f\}_m,Y_l)\, \big|{\cal M}(\{p,f\}_m)\big\rangle
    \\
    &\quad\times \Big\{ F_{m+1}({\cal R}_{l,k}(\{p,f\}_m,Y_l)) -
    F_m\big(\{p\}_m\big) \Big\}
    \\
    &\quad\times [1+ {\cal O}(\alpha_{\rm s}) + {\cal O}\left({1\ {\rm
          GeV}}/{\sqrt s}\right)]\;\;.
  \end{split}
\end{equation}

As we will see, the leading contribution to $\sigma^{\rm B
+ S}_{m,\Delta}$ is of order $\alpha_{\rm s}^{B_m+1}$. Therefore, we can
expand all of the factors and keep only the leading order terms.  Using
Eq.~(\ref{Wexpansion}) we can replace $W$ by 1. We then use
Eq.~(\ref{Elfinal}) for $\bm{E}_{l,k}$, replacing the Sudakov exponential
by 1, since we want only the first perturbative contribution (and since,
because of the subtraction at $y_l = 0$, the integrand is not divergent
at $y_l = 0$).  We can also replace the running coupling
$\alpha_{\rm s}(k_T)$ in $\bm{E}_{l,k}$ by $\alpha_{\rm s}(\mu_R)$ at
leading perturbative order. Then
\begin{equation}
  \label{sigmaBSJ}
  \begin{split}
    \sigma^{\rm B + S}_{m,\Delta} = {}& \frac{1}{m!} \sum_{\{f\}_m}
    \int\!d\varGamma(\{p\}_m)\ \theta\big(d_{\rm ini}< d_m(\{p,f\}_m)\big)
    \\
    &\qquad\times   \sum_{l=1}^m\sum_{k\ne l}\int\!dY_l\, 
    \theta( \tilde d(\{p,f\}_m,l,y_l,z_l) < d_{\rm ini})
    \\
    &\qquad\times 
   \frac{\alpha_{\rm s}(\mu_{\rm R})}{2\pi}\ \big\langle 
   {\cal M}(\{p,f\}_m)\big|
    \frac{\bm{T}_l\cdot \bm{T}_{k}}{-\bm{T}_l^2} \bm{S}_l(p_l,f_l,Y_l)
    \big|{\cal M}(\{p,f\}_m)\big\rangle
    \\
    &\qquad\times \Big\{ F_{m+1}({\cal R}_{l,k}(\{p,f\}_m,Y_l)) -
    F_m\big(\{p\}_m\big) \Big\}
    \\
    &\qquad\times [1+ {\cal O}(\alpha_{\rm s}) + {\cal O}\left({1\ {\rm
          GeV}}/{\sqrt s}\right)]  \;\;.
  \end{split}
\end{equation}

In this equation there is an integration $\int\! dY_l$, which we can
write out in full using Eq.~(\ref{dYldef}). Then we can change integration
variables to momenta $\{\hat p\}_{m+1}$ after the splitting described by
$Y_l$. The jacobian is given in Sec.~\ref{sec:splittingkinematics},
Eq.~(\ref{jacobian3}). After the change of variables, we have
\begin{equation}
  \label{sigmaBSJsum2}
  \begin{split}
    \sigma^{\rm B + S}_{m,\Delta} = {}& \frac{1}{(m+1)!}
    \sum_{\{f\}_{m+1}} \int\!d\varGamma(\{\hat p\}_{m+1})
    \\
    &\times \sum_{\substack{{i,j}\\{\rm pairs}}}\sum_{k\ne i,j}
     \theta(\tilde d(\{p,f\}_{m}^{ij,k}, \{l,y,z\}_{ij,k}) <
     d_{\rm ini} < d_{m}(\{p,f\}_{m}^{ij,k}) )
    \\
    &\times
    \frac{\alpha_{\rm s}(\mu_{\rm R})}{2\pi}\
     \frac{16\pi^2}{2 \hat p_i \cdot \hat p_j\, (1-y_{ij,k})}\ 
    \\
    &\times 
    \big\langle {\cal M}(\{p,f\}_m^{ij,k})\big| \frac{\bm{T}_{ij}\cdot
      \bm{T}_{k}}{-\bm{T}_{ij}^2} 
       \bm{S}_{ij}(\{p_{ij},f_{ij},Y_{ij}\}_{ij,k}) 
 \big|{\cal M}(\{p,f\}_m^{ij,k})\big\rangle
    \\
    &\times \Big\{ F_{m+1}(\{\hat p\}_{m+1}) -
    F_m\big(\{p\}_m^{ij,k}\big) \Big\}
    \\
    &\times [1+ {\cal O}(\alpha_{\rm s}) + {\cal O}\left({1\ {\rm
          GeV}}/{\sqrt s}\right)] \;\;.
  \end{split}
\end{equation}
Using Eq.~(\ref{newDijk}),
\begin{equation}
  \label{dipoleterm}
\begin{split}
  &{\cal D}_{ij,k}\big(\{\hat p,\hat f\}_{m+1}\big) =
\\ &\quad
        \frac{1}{2\hat p_i\!\cdot\!\hat p_j}\,
  \big\langle {\cal M}\big(\{p,f\}_m^{ij,k}\big)\big|
        \frac{\bm{T}_{ij}\cdot\bm{T}_{k}}{-\bm{T}_{ij}^{2}}
    {\bm{V}_{ij}}(\{\tilde p_{ij},y,z,\phi\}_{ij,k},f_i,f_j)
    \big|{\cal M}\big(\{p,f\}_m^{ij,k}\big)\big\rangle\;\;,
\end{split}
\end{equation}
we see that if we identify
\begin{equation}
\frac{\alpha_{\rm s}}{2\pi}\, \frac{16\pi^2}{1-y}\
\bm{S}_{l}(p,f,Y )
=
\bm{V}_{l}(\{p,y,z,\phi\},f_1,f_2)\;\;,
\label{identifyS}
\end{equation}
we will have
\begin{equation}
  \label{sigmaBSJsum3}
  \begin{split}
    \sigma^{\rm B+S}_{m,\Delta} = {}& \frac{1}{(m+1)!}
    \sum_{\{f\}_{m+1}} \int\!d\varGamma(\{\hat p\}_{m+1}) 
    \sum_{\substack{{i,j}\\{\rm pairs}}}\sum_{k\ne i,j} 
     {\cal D}_{ij,k}\big(\{\hat p,\hat f\}_{m+1}\big)
    \\
    &\times
    \theta\big(\tilde d(\{p,f\}_{m}^{ij,k}, \{l,y,z \}_{ij,k}) 
            < d_{\rm ini}< d_{m}(\{p,f\}_{m}^{ij,k})\big)
    \\
    &\times \Big\{ F_{m+1}(\{\hat p\}_{m+1}) -
    F_m\big(\{p\}_m^{ij,k}\big) \Big\}
    \\
    &\times [1+ {\cal O}(\alpha_{\rm s}) + {\cal O}\left({1\ {\rm
          GeV}}/{\sqrt s}\right)]  \;\;.
  \end{split}
\end{equation}

We will use $\sigma^{\rm B+S}_{m,\Delta}$ in this form to combine with
$\sigma^{\rm R+S}_{m}$, which is discussed in the following section.

\subsection{NLO real emission corrections with shower}
\label{sec:NLOrealonShower}

We turn to the discussion of the NLO corrections. Let us start with
the real contribution. Define
\begin{equation}
  \label{sigmaRS}
  \begin{split}
    \sigma^{\rm R+S}_{m} = {}& \frac{1}{(m+1)!}\sum_{\{\hat f\}_{m+1}}
    \int\!d\varGamma(\{\hat p\}_{m+1})\, 
     \tilde I\big(\{\hat p,\hat f\}_{m+1}\big)\,
      W_{m}^{\rm R+S}(\{p,f\}_{m+1})
    \\
    &\times\bigg\{ \big\langle{\cal M}(\{\hat p,\hat f\}_{m+1})\big|
    {\cal M}(\{\hat p,\hat f\}_{m+1})\big\rangle\, 
    \\
    &\qquad\quad\times 
    \theta\big(d_{m+1}(\{\hat p, \hat f\}_{m+1}) < d_{\rm ini}
    	< d_{m}(\{\hat p, \hat f\}_{m+1})\big)
    \\
    &\qquad -\sum_{\substack{{i,j}\\{\rm pairs}}} \sum_{k\neq i,j}
    {\cal D}_{ij,k} \big(\{\hat p,\hat f\}_{m+1}\big)\, 
    \\
    &\qquad\quad\times
    \theta\big(\tilde d(\{p,f\}_{m}^{ij,k}, \{l,y,z \}_{ij,k}) 
            < d_{\rm ini}< d_{m}(\{p,f\}_{m}^{ij,k})\big)
    \bigg\}\;\;.
  \end{split}
\end{equation}
This formula is illustrated in Fig.~\ref{fig:sigmaR}. The first term is
the $m+1$-parton matrix element squared and the second term is the sum of
the dipole contributions to eliminate the infrared singularities. There
is a Monte Carlo interface function $\tilde I$ with the property
Eq.~(\ref{tildeIproperty}). There is a reweighting factor 
$W_{m}^{\rm R+S}(\{p,f\}_{m+1})$ that is similar to the reweighting
factor $W$ in $\sigma^{\rm B+S}_{m}$. We discuss this factor in
Sec.~\ref{sec:CKKWSudakov}, Eq.~(\ref{WRplusS}).  All that we need to
know here is that $W_{m}^{\rm R+S}(\{p,f\}_{m+1})$ has a perturbative
expansion that begins with 1: $W_{m}^{\rm R+S}(\{p,f\}_{m+1}) = 1 + {\cal
O}(\alpha_{\rm s})$. There is a cut on $d_m$ and $d_{m+1}$. The
$m+1$-parton state should be not resolvable at scale $d_{\rm ini}$ but,
having put the two closest partons together, the resulting $m$-parton
state should be resolvable.  In the dipole terms we have cuts on $d_{m}$
and on
$\tilde d_{m+1}$ for the splitting in question, Eq.~(\ref{tildeddef}).
The $m$-parton states defined by the dipole momentum set
$\{p\}_{m}^{ij,k}$ should be resolvable at the scale $d_{\rm ini}$.
However, every dipole splitting gives only non-resolvable radiations
according to the $\tilde d$ measure.

In Eq.~(\ref{sigmaRS}) there is a potential infrared singularity when two
of the partons, $\hat p_i$ and $\hat p_j$, become collinear or one
vanishes. This singularity is rendered harmless by the corresponding
$\{ij,k\}$ subtraction terms. Each of the $\{ij,k\}$ subtraction terms can
have other singularities when two of the momenta $\{p\}_{m}^{ij,k}$ are
collinear or one is zero. However at such an extra singularity,
$d_m(\{p,f\}_{m}^{ij,k})$ vanishes, so that the cut $d_{\rm ini} <
d_m(\{p,f\}_{m}^{ij,k})$ eliminates the singularity.

\FIGURE{
\centerline{
\includegraphics[width = 12 cm]{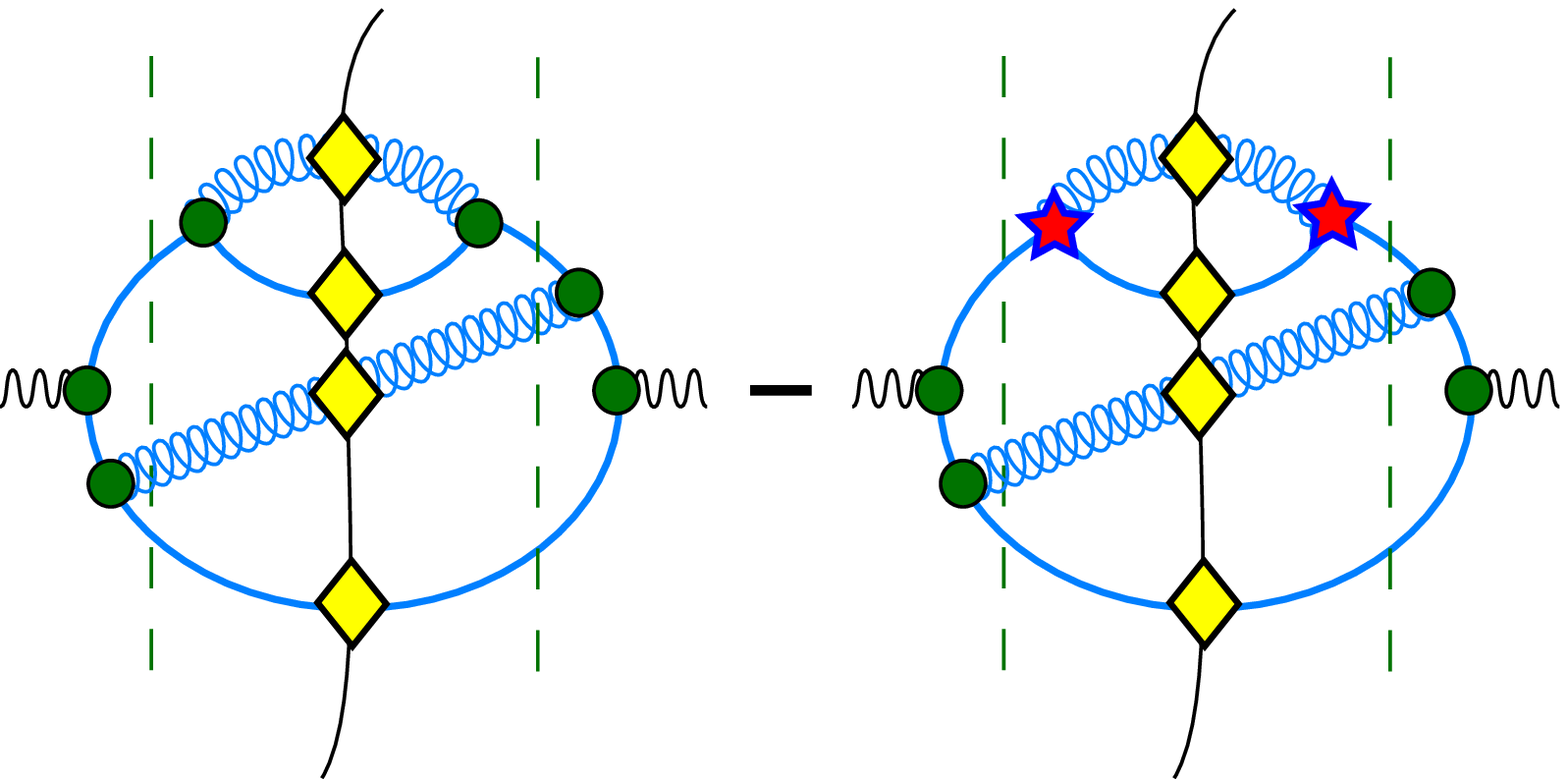}
}
\medskip
\caption{
Illustration of $\sigma_3^{\rm R + S}$, Eq.~(\ref{sigmaRS}). As
in Fig.~\ref{fig:sigmaB}, the three partons that cross the dash lines are
resolvable at scale $d_{\rm ini}$. In the first graph, there is an
additional perturbative splitting that is {\it not} resolvable at scale
$d_{\rm ini}$. There is a subtraction graph in which the perturbative
splitting is replaced by a splitting according to the dipole splitting
formula, which is the order $\alpha_{\rm s}$ contribution to the splitting
represented by the box in  Fig.~\ref{fig:sigmaB}. The subtraction removes
the soft/collinear divergence from the splitting. Finally, there is
secondary showering represented in the figure by diamonds and in
Eq.~(\ref{sigmaRS}) by the function $\tilde I$. There is a reweighting
factor $W$ that is not represented by a graphical symbol. This
illustration is an attempt to depict the main idea behind
Eq.~(\ref{sigmaRS}). Actually, though, the full matrix element  $|{\cal
M}\rangle$ to produce four partons appears in the first diagram and the
full matrix element  $|{\cal M}\rangle$ to produce three partons appears
in the second diagram along with a sum over the dipole subtraction terms. 
 }
\label{fig:sigmaR}
}

The perturbative expansion of this is simple. Using
Eq.~(\ref{tildeIproperty}), we can replace $\tilde I$ by the jet
observable function $F$ and, using $W_{m}^{\rm R+S}(\{p,f\}_{m+1}) = 1 +
{\cal O}(\alpha_{\rm s})$, we can replace $W$ by 1. This gives
\begin{equation}
  \label{sigmaRS1}
  \begin{split}
    \sigma^{\rm R+S}_{m} = {}& \frac{1}{(m+1)!}\sum_{\{\hat f\}_{m+1}}
    \int\!d\varGamma(\{\hat p\}_{m+1})\, F_{m+1}\big(\{\hat p\}_{m+1}\big)
    \\
    &\times\bigg\{ \big\langle{\cal M}(\{\hat p,\hat f\}_{m+1})\big|
    {\cal M}(\{\hat p,\hat f\}_{m+1})\big\rangle
    \\
    &\qquad\quad\times 
    \theta\big(d_{m+1}(\{\hat p, \hat f\}_{m+1}) < d_{\rm ini}
    	< d_{m}(\{\hat p, \hat f\}_{m+1})\big)
    \\
    &\qquad -\sum_{\substack{{i,j}\\{\rm pairs}}} \sum_{k\neq i,j}
    {\cal D}_{ij,k} \big(\{\hat p,\hat f\}_{m+1}\big)
    \\
    &\qquad\quad\times
    \theta\big(\tilde d(\{p,f\}_{m}^{ij,k}, \{l,y,z \}_{ij,k}) 
            < d_{\rm ini}< d_{m}(\{p,f\}_{m}^{ij,k})\big)
    \bigg\}
   \\
    &\times \big[1+ {\cal O}(\as) + {\cal O}(1\ {\rm
  GeV}/\sqrt{s})\big]\;\;.
  \end{split}
\end{equation}

Now if we add $\sigma^{\rm R+S}_{m}$ to $\sigma^{\rm R+S}_{m,\Delta}$
from the previous section, Eq.~(\ref{sigmaBSJsum3}), we
get
\begin{equation}
  \label{sigmaRsum}
  \begin{split}
   \sigma^{\rm R+S}_{m}+&  \sigma^{\rm B + S}_{m,\Delta}
\\
={}&
  \frac{1}{(m+1)!}  \sum_{\{\hat f\}_{m+1}} \int\!d\varGamma(\{\hat
  p\}_{m+1})
  \\
  &\times \biggl[\
  \big\langle{\cal M}(\{\hat p,\hat f\}_{m+1})\big|
  {\cal M}(\{\hat p,\hat f\}_{m+1})\big\rangle\ F_{m+1}(\{\hat
  p\}_{m+1})
\\ &\qquad \times
\theta(d_{m+1}(\{\hat p,\hat f\}_{m+1}) < d_{\rm ini} < d_m(\{\hat
p,\hat f\}_{m+1}))
  \\
  &\qquad - \sum_{\substack{{i,j}\\{\rm pairs}}} \sum_{k \ne i,j}
  {\cal D}_{ij,k} (\{\hat p,\hat f\}_{m+1})\, F_m(\{p\}_{m}^{ij,k})
\\
    &\qquad\times
    \theta\big(\tilde d(\{p,f\}_{m}^{ij,k}, \{l,y,z\}_{ij,k}) 
            < d_{\rm ini}< d_{m}(\{p,f\}_{m}^{ij,k})\big)
  \biggr]
\\ &\times
\Big[1+ {\cal O}(\alpha_{\rm s}) + {\cal O}\left({1\ {\rm
          GeV}}/{\sqrt s}\right)\Big]\;\;.
  \end{split}
\end{equation}
Comparing with Eq.~(\ref{realandsubtractionbis}), we see that
\begin{equation}
  \label{sigmaRsumis}
   \sigma^{\rm R+S}_{m}+ \sigma^{\rm B + S}_{m,\Delta}
=
  \sigma^{\rm R - A}_m\times
\Big[1+ {\cal O}(\alpha_{\rm s}) + {\cal O}\left({1\ {\rm
          GeV}}/{\sqrt s}\right)\Big]\;\;.
\end{equation}

\subsection{NLO virtual corrections with shower}
\label{sec:NLOvirtualonShower}

In this subsection, we turn to the virtual loop corrections. We
define\footnote{The expression in Eq.~(\ref{sigmaVS}) can be simplified
using Eq.~(\ref{Clkform}). }
\begin{equation}
\begin{split}
\sigma^{\rm V + S}_m
= {}& 
\frac{1}{m!}
\sum_{\{f\}_m} 
\int\!d\varGamma(\{p\}_m)\
 \theta\big( d_{\rm ini} < d_{m}(\{p,f\}_{m})\big)\,
\tilde I(\{p,f\}_m)\,
W_{m}^{\rm V+S}(\{p,f\}_{m})
\\ &\times
\biggl\{
V(\{p,f\}_m)
-
\sum_{l}\sum_{k\neq l} C_{l,k}(\{p\}_{m}, d_{\rm ini})
\\ &\qquad
-
\frac{\alpha_{\rm s}(\mu_{\rm R})}{2\pi}\,
    W_{m}^{(1)}(\{p,f\}_m)\, \big\langle {\cal M}(\{p,f\}_m)
    \big|{\cal M}(\{p,f\}_m)\big\rangle
\biggr\}\;\;.
\end{split}
\label{sigmaVS}
\end{equation}
This formula is illustrated in Fig.~\ref{fig:sigmaV}. We integrate over
the phase space for $m$ partons with a cut on $d_{m}(\{p,f\}_{m})$. There
is a Monte Carlo interface function $\tilde I$ with the property
(\ref{tildeIproperty}). There is a reweighting factor $W_{m}^{\rm
V+S}(\{p,f\}_{m})$, which we discuss in
Sec.~\ref{sec:CKKWSudakov}, Eq.~(\ref{WRplusS}).  This function has a
perturbative expansion that begins with 1: $W_{m}^{\rm V+S}(\{p,f\}_{m})
= 1 + {\cal O}(\alpha_{\rm s})$. Then there is a factor with three terms.
The first two contain the functions $V$ and
$C_{l,k}$ from the perturbative virtual loop contribution,
Eq.~(\ref{sigmaVbis}). The third contains the first order contribution
$W_m^{(1)}$ to the Sudakov factor and is similar in structure to
$\sigma^{\rm B+S}_{m,\{T,1\}}$, Eq.~(\ref{sigmaBS01}). 

\FIGURE{
\centerline{
\includegraphics[width = 6 cm]{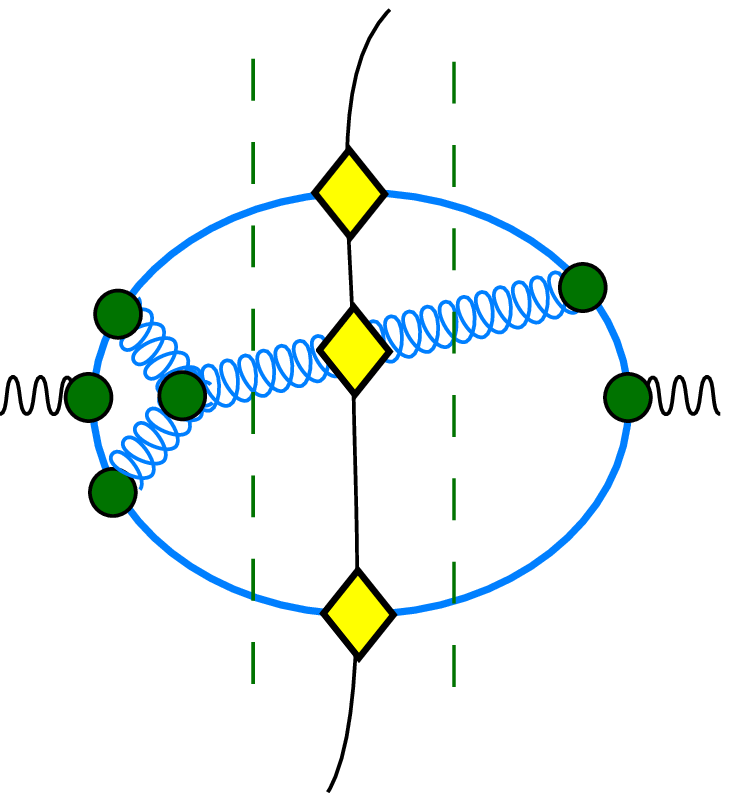}
}
\medskip
\caption{
Illustration of $\sigma_3^{\rm V + S}$, Eq.~(\ref{sigmaVS}). As in
Fig.~\ref{fig:sigmaB}, the three partons that cross the dash lines are
resolvable at scale $d_{\rm ini}$.  To the left of the first dashed line,
we illustrate one contribution to the one loop amplitude $|{\cal
M}^{(1)}\rangle$. The infrared poles of $|{\cal M}^{(1)}\rangle$ are
subtracted according to the dipole subtraction scheme. To the right, we
illustrate one contribution to $\langle{\cal M}|$. All of this is
included in Eq.~(\ref{sigmaVS}) in the function $V$, which is described
in Sec.~\ref{sec:virtualfunctions}.  Finally, there is secondary
showering represented in the figure by diamonds and in
Eq.~(\ref{sigmaVS}) by the function $\tilde I$. There is a reweighting
factor $W$ that is not represented by a graphical symbol and there are two
additional terms in Eq.~(\ref{sigmaVS}) that are not illustrated in the
figure.
 }
\label{fig:sigmaV}
}

To find the perturbative expansion of this, we replace $\tilde I$
by the jet observable function $F$, using the
property (\ref{tildeIproperty}) of $\tilde I$ and, using $W_{m}^{\rm
V+S}(\{p,f\}_{m}) = 1 + {\cal O}(\alpha_{\rm s})$, we replace $W$
by 1. This gives
\begin{equation}
\begin{split}
\sigma^{\rm V + S}_m
= {}& \frac{1}{m!}
\sum_{\{f\}_m} 
\int\!d\varGamma(\{p\}_m)\
 \theta\big( d_{\rm ini} < d_{m}(\{p,f\}_{m})\big)\,
F_m(\{p\}_m)
\\ &\times
\biggl\{
V(\{p,f\}_m)
-
\sum_{l}\sum_{k\neq l} C_{l,k}(\{p\}_{m}, d_{\rm ini})
\\ &\qquad
-
\frac{\alpha_{\rm s}(\mu_{\rm R})}{2\pi}\,
    W_{m}^{(1)}(\{p,f\}_m)\, \big\langle {\cal M}(\{p,f\}_m)
    \big|{\cal M}(\{p,f\}_m)\big\rangle
\biggr\}
\\ &\times
\Big[1+ {\cal O}(\alpha_{\rm s}) + {\cal O}\left({1\ {\rm
          GeV}}/{\sqrt s}\right)\Big]\;\;.
\end{split}
\label{sigmaVS2}
\end{equation}
We now add $\sigma^{\rm B+S}_{m,\{T,1\}}$ from Eq.~(\ref{sigmaBS01}),
\begin{equation}
\begin{split}
\sigma^{\rm V + S}_m
+\sigma^{\rm B+S}_{m,\{T,1\}}
= {}& 
\frac{1}{m!}
\sum_{\{f\}_m} 
\int\!d\varGamma(\{p\}_m)\
 \theta\big( d_{\rm ini} < d_{m}(\{p,f\}_{m})\big)\,
F_m(\{p\}_m)
\\ &\times
\biggl\{
V(\{p,f\}_m)
-
\sum_{l}\sum_{k\neq l} C_{l,k}(\{p\}_{m}, d_{\rm ini})
\biggr\}
\\ &\times
\Big[1+ {\cal O}(\alpha_{\rm s}) + {\cal O}\left({1\ {\rm
          GeV}}/{\sqrt s}\right)\Big]\;\;.
\end{split}
\label{sigmaVS3}
\end{equation}
Comparing with Eq.~(\ref{sigmaVbis}), we see that
\begin{equation}
\sigma^{\rm V + S}_m
+\sigma^{\rm B+S}_{m,\{T,1\}}
= {}
\sigma^{\rm V + A}_m\,
\Big[1+ {\cal O}(\alpha_{\rm s}) + {\cal O}\left({1\ {\rm
          GeV}}/{\sqrt s}\right)\Big]\;\;.
\label{sigmaVSis}
\end{equation}

\subsection{Net result}
\label{sec:netresult}

Combining the contributions 
$\sigma^{\rm B+S}_{m,\{T,0\}}$
from Eq.~(\ref{sigmaBS00is}),
$\sigma^{\rm R+S}_{m} + \sigma^{\rm B + S}_{m,\Delta}$ 
from Eq.~(\ref{sigmaRsumis}), and
$\sigma^{\rm V + S}_m +\sigma^{\rm B+S}_{m,\{T,1\}}$
from Eq.~(\ref{sigmaVSis}) we see that the calculation of $\sigma_m$ with
showers, using both the Born term and the correction terms, reproduces
$\sigma_m$ to next-to-leading order accuracy,
\begin{equation}
\begin{split}
\sigma^{\rm B+S}_{m}  + \sigma^{\rm R+S}_{m} + \sigma^{\rm V+S}_{m}
={}&
\sigma^{\rm B}_{m} + \sigma^{\rm R-A}_{m} + \sigma^{\rm V+A}_{m} 
\\ &
+{\cal O}(\alpha_{\rm s}^{B_m + 2})
+{\cal O}(\alpha_{\rm s}^{B_m}\times 1\, {\rm GeV}/\sqrt s\,)\;\;.
\end{split}
\label{showernetagain}
\end{equation}
This is the result claimed in Eq.~(\ref{showernet}).

We also note that the calculation of $\sigma_m$ with showers using
only the Born term reproduces $\sigma_m$ to leading order accuracy,
\begin{equation}
\begin{split}
\sigma^{\rm B+S}_{m}
={}&
\sigma^{\rm B}_{m} 
+{\cal O}(\alpha_{\rm s}^{B_m + 1})
+{\cal O}(\alpha_{\rm s}^{B_m}\times 1\, {\rm GeV}/\sqrt s\,)
\;\;.
\end{split}
\label{Bornshowernetagain}
\end{equation}

\section{Alternative LO partial cross sections with showers}
\label{sec:LOalternative}

In Eq.~(\ref{sigmaNLOplusS}), we imagined that the perturbative partial
cross sections  $\sigma_m$ are known at order $\alpha_{\rm s}^{B_m + 1}$
but that beyond a certain value $m_{\rm NLO}$ of $m$ only order
$\alpha_{\rm s}^{B_m}$ results are known. Furthermore, we suppose that
exact order $\alpha_{\rm s}^{B_m}$ results are known only up to certain
value $m_{\rm max}$ of $m$. We needed a construction of the corresponding
contributions $\sigma^{\rm B+S}_{m}$ including showers. Our choice was to
simply define $\sigma^{\rm B+S}_{m}$  for $m > m_{\rm NLO}$ according to
Eq.~(\ref{sigmaBS}). This reproduces $\sigma_m$ to leading order.

In this section, we wish to point out an alternative. For $m > m_{\rm
NLO}$ one could replace $\sigma^{\rm B+S}_{m}$ by $\tilde \sigma^{\rm
B+S}_{m}$ defined by
\begin{equation}
\begin{split}
  \tilde \sigma^{\rm B+S}_{m} = {}& \frac{1}{m!}  \sum_{\{f\}_m}
  \int\!d\varGamma(\{p\}_m)\, 
  \theta\big(d_{\rm ini}< d_m(\{p,f\}_m)\big)\,W_{m}(\{p,f\}_m)
  \\
  &\times 
  \big\langle {\cal M}(\{p,f\}_m)\big| 
      {\cal M}(\{p,f\}_m)\big\rangle
\ \tilde I(\{p,f\}_m) \;\;.
\label{sigmaBSLO}
\end{split}
\end{equation}
Here we use the $m$-parton Born cross section modified by the reweighting
factor $W$ with a cut $d_{\rm ini}< d_m(\{p,f\}_m$. We omit the dipole
style splitting that we used in Eq.~(\ref{sigmaBS}). Instead, we pass the
partonic state $\{p,f\}_m$ to the shower Monte Carlo program, represented
here by the Monte Carlo interface function $\tilde I$ with
the property Eq.~(\ref{tildeIproperty}), as used already in
Eqs.~(\ref{sigmaRS}) and (\ref{sigmaVS}). The shower Monte Carlo program
generates parton showers from the scale $d_{\rm ini}$ downwards. 

This is to say that for $m > m_{\rm NLO}$ we simply use the scheme of
Ref.~\cite{CKKW}. This is perhaps a little simpler than the use of a
splitting based on the dipole subtraction scheme for the first splitting.

There is a largest value, $m_{\rm max}$, of $m$ such that we have
available exact leading order matrix elements. Thus the sum in
Eq.~(\ref{sigmaNLOplusS}) stops with $\sigma^{\rm B+S}_{m}$ or  $\tilde
\sigma^{\rm B+S}_{m}$ for $m = m_{\rm max}$. This leaves us with a zero
cross section for producing more than $m_{\rm max}$ jets that are
resolvable at scale $d_{\rm ini}$. Even though we lack the matrix elements
for calculating exactly at leading order the cross section for producing
more than $m_{\rm max}$ jets, we can still have an approximate cross
section for producing many resolvable jets. Assuming that we are using
$\tilde \sigma^{\rm B+S}_{m}$, all that we need to do is make a simple
modification in the formula (\ref{sigmaBSLO}) for $\tilde \sigma^{\rm
B+S}_{m}$ for $m = m_{\rm max}$: in $W_{m}(\{p,f\}_m)$ and in $\tilde
I(\{p,f\}_m)$ we should replace $d_{\rm ini}$ by $d_m(\{p,f\}_m)$. That
is, there are splittings that produce $m_{\rm max}$ partons according to
the exact matrix element, with the corresponding Sudakov factors and
$\alpha_{\rm s}$ factors, represented in the modified $W_{m}(\{p,f\}_m)$.
We remove from $W_{m}(\{p,f\}_m)$ the factor representing the probability
that there was no splitting between $d_m(\{p,f\}_m)$ and $d_{\rm ini}$
because we now allow such splittings as part of the modified $\tilde
I(\{p,f\}_m)$, which describes an indefinite number of splittings at
scales below the scale $d_m(\{p,f\}_m)$ of this last splitting according
to the exact matrix element.

\section{The kinematics of parton splitting}
\label{sec:splittingkinematics}

In this section, we review the dipole splitting construction of Catani
and Seymour \cite{CataniSeymour}, using the notation adopted for this
paper. This covers one parton splitting into two partons (with the
participation of a spectator parton). With a trivial extension that we
have already described in Sec.~\ref{sec:splittingconstruction}, this
construction also covers one of $m$ partons splitting to create an
$m+1$-parton state. 

The construction and deconstruction of dipole splitting outlined in
Sec.~\ref{sec:splittingconstruction} is based on a transformation   
\begin{equation}
\{\hat p_i,\hat p_j,\hat p_k\} \leftrightarrow
\{\tilde p_{ij}, \tilde p_k, y, z, \phi\}\;\;.
\end{equation}
The transformation from left to right combines partons $i$ and $j$. With
the inclusion of the rather trivial transformation to combine the
flavors, this transformation was called ${\cal Q}_{ij,k}$ in
Eq.~(\ref{Qijkdef}). The transformation from right to left splits parton
$ij$ and, with the inclusion of the flavor splitting, was called ${\cal
R}_{ij}$ (or ${\cal R}_l$) in Eq.~(\ref{Rldef}).

Combining parton momenta works as follows. Starting with three
massless momenta  $\{\hat p_i,\hat p_j,\hat p_k\}$, Catani and Seymour
define $\tilde p_{ij}$ and $\tilde p_k$ by
\begin{equation}
\begin{split}
\tilde p_{ij} ={}&
\hat p_i + \hat p_j 
- \frac{\hat p_i\!\cdot\! \hat p_j}
       {(\hat p_i + \hat p_j)\!\cdot\! \hat p_k}\
   \hat p_k
\;,
\\
\tilde p_{k} ={}&
\hat p_k
+ \frac{\hat p_i\!\cdot\! \hat p_j}
       {(\hat p_i + \hat p_j)\!\cdot\! \hat p_k}\
   \hat p_k
\;\;.
\end{split}
\end{equation}
Then it is evident that 
\begin{equation}
\tilde p_{ij} + \tilde p_k= \hat p_i + \hat p_j + \hat p_k
\end{equation}
and that
\begin{equation}
\tilde p_{ij}^2 = \tilde p_k^2 = 0\;\;.
\end{equation}
The splitting parameter $y$ is defined by
\begin{equation}
y = \frac{\hat p_i\!\cdot\! \hat p_j}
  {\hat p_i\!\cdot\!\hat p_j + (\hat p_i + \hat p_j)\!\cdot\!\hat p_k}\;\;,
\end{equation}
so that
\begin{equation}
\begin{split}
\tilde p_{ij} ={}&
\hat p_i + \hat p_j 
- \frac{y} {1-y}\ \hat p_k
\;,
\\
\tilde p_{k} ={}&
  \frac{1} {1-y}\ \hat p_k
\;\;.
\label{alttildepdef}
\end{split}
\end{equation}
The splitting parameter $z$ is defined by
\begin{equation}
z=
\frac{\hat p_i\!\cdot\! \hat p_k}
  {(\hat p_i + \hat p_j)\!\cdot\!\hat p_k}\;\;.
\end{equation}
Then $(1-z)$ is given by the same expression as for $z$ but with $i
\leftrightarrow j$. Alternative formulas for $y$ and $z$ are
\begin{equation}
\label{yandzdef}
y = \frac{\hat p_i\!\cdot\! \hat p_j}{\tilde p_{ij}\!\cdot\!\tilde p_k}
\;\;,\qquad\qquad
z= \frac{\hat p_i\!\cdot\! \tilde p_k}
  {\tilde p_{ij}\!\cdot\!\tilde p_k}\;\;.
\end{equation}

To define the azimuthal angle $\phi$ we first define $k_{\perp}$ to be the
part of $(\hat p_i - \hat p_j)/2$ that is orthogonal to both $\tilde
p_{ij}$ and $\tilde p_k$,
\begin{equation}
k_\perp = {\textstyle \frac{1}{2}}\,(\hat p_i - \hat p_j) 
- (z-{\textstyle \frac12})\,\tilde p_{ij}
+ (z-{\textstyle \frac12})\,y\,\tilde p_k
\;\;.
\label{kperpdef}
\end{equation}
One can show that $k_\perp$ obeys $k_\perp\!\cdot\!\tilde p_{ij} = 0$ and
$k_\perp\!\cdot\!\tilde p_{k} = 0$ using the easily proved relations
$(\hat p_i - \hat p_j)\!\cdot\! \tilde p_k = (2z-1)\,
\tilde p_{ij}\!\cdot\! \tilde p_k$ and $(\hat p_i -
\hat p_j)\!\cdot\! \tilde p_{ij} = (2z-1)\, y\,\tilde p_{ij}\!\cdot\!
\tilde p_k$. A convenient alternative formula for $k_\perp$ is
\begin{equation}
k_{\perp} =  z p_i - (1-z) p_j
  - (2 z - 1)\, \tilde p_{ij} 
\;\;.
\end{equation}
The squared length of $k_\perp$ is
\begin{equation}
 k_{\perp}^2 = 
-2y\,z (1 - z)\,
  \tilde p_{ij}\!\cdot\!\tilde p_k\;\;.
\end{equation}
The unit vector
\begin{equation}
\kappa_\perp = \frac{k_\perp}{\sqrt{-k_\perp^2}}
\end{equation}
defines the plane of the splitting in a reference frame in which the
vector parts of $\tilde p_{ij}$ and $\tilde p_{k}$ lie along the positive
and negative z-axis, respectively. The azimuthal angle of $\kappa_\perp$
with respect to some convenient reference direction is $\phi$ . Stated
perhaps more precisely, we say that one specifies $\phi$ as a
shorthand for saying that one specifies the unit vector
$\kappa_\perp$. Integrating over $\phi$ means integrating over
$\kappa_\perp$ subject to the conditions that it is a unit vector
orthogonal to $\tilde p_{ij}$ and $\tilde p_k$.

The inverse transformation, giving $\{\hat p_i,\hat p_j,\hat p_k\}$ in
terms of $\{\tilde p_{ij}, \tilde p_k, y, z, \phi\}$, is easily obtained
by combining Eq.~(\ref{kperpdef}) and Eq.~(\ref{alttildepdef}),
\begin{equation}
\begin{split}
\label{hatpijkdef}
\hat p_i ={}& z \tilde p_{ij} 
  + y (1-z) \tilde{p}_k   
  +[2y\,z (1 - z)\,\tilde p_{ij}\!\cdot\!\tilde p_k]^{1/2}\,
   \kappa_{\perp}\;\;,
\\
\hat p_j ={}& (1-z) \tilde p_{ij} 
  + y z \tilde{p}_k   
  - [2y\,z (1 - z)\,\tilde p_{ij}\!\cdot\!\tilde p_k]^{1/2}\,
   \kappa_{\perp}\;\;,
\\
\hat p_k ={}& (1-y)\,\tilde p_k\;\;.
\end{split}
\end{equation}

It is reasonably straightforward to work out the jacobian for this
transformation. Defining
\begin{equation}
d\varGamma^{(n)}(\{p\}_{n},Q)
=
\prod_{l = 1}^n\biggl( (2\pi)^{-3} d^4p_l\ \delta_+(p_l^2)\biggr)\,
(2 \pi)^4
\delta^4\!\left(\sum_{l = 1}^n p_l - Q\right)\;\;,
\label{Gammadefmod}
\end{equation}
one finds \cite{CataniSeymour}
\begin{equation}
\begin{split}
\label{jacobian1}
d\varGamma^{(3)}(\hat p_i,\hat p_j,\hat p_k;Q)
={}&
d\varGamma^{(2)}(\tilde p_{ij},\tilde p_k;Q)
\
     dy\,dz\,\frac{d\phi}{2\pi} \
\frac{2\tilde{p}_{ij}\!\cdot\!\tilde{p}_k}{16\pi^2}\,
     (1-y)
\\
    &\times\,\Theta(z(1-z) > 0)\,
     \Theta(y(1-y) > 0)\;\;,
\end{split}
\end{equation}
or, equivalently,
\begin{equation}
\begin{split}
\label{jacobian2}
d\varGamma^{(3)}(\hat p_i,\hat p_j,\hat p_k;Q)\
\frac{16\pi^2}{2\hat{p}_{i}\!\cdot\!\hat{p}_j}
={}&
d\varGamma^{(2)}(\tilde p_{ij},\tilde p_k;Q)
\
     dy\,dz\,\frac{d\phi}{2\pi} \
    \frac{1-y}{y}
\\
    &\times\,\Theta(z(1-z) > 0)\,
     \Theta(y(1-y) > 0)\;\;.
\end{split}
\end{equation}

This transformation between two parton momenta plus splitting variables
and three parton momenta is trivially extended to a map between $m$
parton momenta plus splitting variables and $m+1$ parton momenta, as
described in Sec.~\ref{sec:splittingconstruction}. If the parton that
splits is labelled $l$, the jacobian is
\begin{equation}
\label{jacobian3}
d\varGamma(\{p\}_m)\
dy_l\,\frac{1-y_{l}}{y_l}
dz_l\
\frac{d\phi_l}{2\pi}
= d\varGamma(\{\hat p\}_{m+1})\
\frac{16\pi^2}{2 \hat p_{l,1} \cdot \hat p_{l,2}}\;\;.
\end{equation}

In our applications, this result will appear with symmetry factors and
sums over the indices of splitting partons. Then it looks like
\begin{equation}
\begin{split}
{}&\frac{1}{m!} \sum_{\{f\}_m}
    \int\!d\varGamma(\{p\}_m)\  \sum_{l=1}^m\sum_{k\ne l}
    \\
    &\qquad\times \int_0^1 \frac{dy_l}{y_l}
\int_0^1\!dz_l\,\int_0^{2\pi}\frac{d\phi_l}{2\pi}\,
\frac{1}{2}\!\sum_{\hat f_{l,1},\hat f_{l,2}}\!
\delta_{\hat f_{l,1} + \hat f_{l,2}}^{f_l}
(1-y_l)\,
h(\{\hat p, \hat f\}_{m+1})
\\
& \quad\quad = \frac{1}{(m+1)!}
    \sum_{\{f\}_{m+1}} \int\!d\varGamma(\{\hat p\}_{m+1}) 
    \sum_{\substack{{i,j}\\{\rm pairs}}}\sum_{k\ne i,j}\,
\frac{16\pi^2}{2\hat p_i\cdot \hat p_j}\,
h(\{\hat p, \hat f\}_{m+1})
\;\;,
\label{jacobian4}
\end{split}
\end{equation}
where $h(\{\hat p, \hat f\}_{m+1})$ represents any well behaved function.
To prove this, note that the left hand side has symmetry factors $1/m!$
and $1/2$ and a sum over $m$ values of the index $l$ and $(m-1)$ values
of the index $k$. On the right hand side we have a symmetry factor 
$1/(m+1)!$ and a sum over $(m+1)m/2$ values of the index pair $i,j$ and
$(m-1)$ values of the index $k$. If we put the sums over parton indices
outside the integrals, each term is the same, so we can just take one
term and multiply by the number of terms. We note that the net counting
factor $m(m-1)/[2m!]$ on the left hand side equals the net counting
factor $(m+1)m(m-1)/[2(m+1)!]$ on the right hand side. Thus
Eq.~(\ref{jacobian3}) implies Eq.~(\ref{jacobian4}).

We now examine the relation between parton splitting and the jet
resolution functions $d_{ij}$ defined in Eq.~(\ref{dij}) and
$\tilde d$ defined in Eq.~(\ref{tildeddef}). Taking massless parton
momenta in Eq.~(\ref{dij}) and using Eq.~(\ref{yandzdef}), we have
\begin{equation}
  \label{dijmod}
  d_{ij} \equiv d(\hat p_{i}, \hat p_{j}) 
= 
\frac{2 \hat p_i\cdot \hat p_j}{s} \, 
\min\!\left\{
  \frac{\hat p_{i}\!\cdot\!n}{\hat p_{j}\!\cdot\!n},\,
  \frac{\hat p_{j}\!\cdot\!n}{\hat p_{i}\!\cdot\!n}
  \right\}
=
  \frac{2\tilde p_{ij}\!\cdot\!\tilde p_{k}}{s}\
  y\
  \min\!\left\{
  \frac{\hat p_{i}\!\cdot\!n}{\hat p_{j}\!\cdot\!n},\,
  \frac{\hat p_{j}\!\cdot\!n}{\hat p_{i}\!\cdot\!n}
  \right\}\;\;,
\end{equation}
where $n$ is a unit vector that defines the time axis in the $e^+ e^-$
c.m.\ frame. For $\hat p_{i}\!\cdot\!n$ and $\hat p_{j}\!\cdot\!n$,
we can use Eq.~(\ref{hatpijkdef}). Of particular interest is the case $0<
y \ll z < 1$, in which $\hat p_i$ is either collinear to $\tilde p_{ij}$
or both soft and collinear. Under this condition we have
\begin{equation}
  \label{dijmod1}
  d_{ij}  \sim 
  \frac{2\tilde p_{ij}\!\cdot\!\tilde p_{k}}{s}\
  y\
  \min\!\left\{
  \frac{z}{1-z},\,
  \frac{1-z}{z}
  \right\}
\equiv \frac{2\tilde p_{ij}\!\cdot\!\tilde p_{k}}{s_{ij}}\,\tilde d
\;\;.
\end{equation}
Here $\tilde d$ is the function defined Eq.~(\ref{tildeddef}) (with $y_l
\to y$, $z_l \to z$ and with the scale $s_l$ for parton $l$ renamed to
$s_{ij}$). We used this function to limit parton splitting in
Eq.~(\ref{realandsubtractionbis}) and Eq.~(\ref{Elfinal}):
$\tilde d < d_{\rm ini}$. We see that, under these conditions,  limiting
$\tilde d$ limits $d_{ij}$, although there is a factor ${2\tilde
p_{ij}\!\cdot\!\tilde p_{k}}/{s_{ij}}$ that relates the size of the two
resolution measures.

Also if interest is the case $0 <  z \ll 1$, $0 <  y \ll 1$, $y/z \sim 1$,
in which $\hat p_i$ is soft but not  collinear to $\tilde p_{ij}$ or
$\tilde p_{k}$. Under this condition, we have
\begin{equation}
  \label{dijmod2}
  d_{ij}  \sim 
  \frac{2\tilde p_{ij}\!\cdot\!\tilde p_{k}}{s}\
  y\,z\
  \left\{
  1 + \frac{y}{z}\ \frac{\tilde p_k\!\cdot\!n}{\tilde p_{ij}\!\cdot\! n}
+ \sqrt{\frac{y}{z}}\
\frac
{
\left[ 2 \tilde
p_{ij}\!\cdot\!\tilde p_k\right]^{1/2}\,\kappa_\perp\!\cdot\! n }
{
\tilde p_{ij}\!\cdot\! n
}
  \right\}\;\;.
\end{equation}
Thus in the soft limit, $d_{ij}$ has the same scaling behavior as
$\tilde d$, {\it i.e.} a factor $yz$, but the ratio of these functions is
not generally 1 and depends on the ratio of $y$ to $z$.

We should also be concerned about the region $0 < z \ll y <
1$. This region is not enhanced in the integration over $y$ and $z$
because the relevant splitting functions in
Eqs.~(\ref{realandsubtractionbis}) and (\ref{Elfinal}) are not singular
for $z \to 0$ at fixed $y$. Nevertheless, since $\tilde d  \ll
d_{ij}$ in this region, we should understand what the restriction
$\tilde d  \ll d_{\rm ini}$ means. Consider, then, the resolution
parameter for combining parton $i$ with the spectator parton $k$. Using
the definition (\ref{dij}) together with Eq.~(\ref{yandzdef}), we have
\begin{equation}
  \label{dikmod}
  d_{ik} \equiv d(\hat p_{i}, \hat p_{k}) = 
  \frac{2\tilde p_{ij}\!\cdot\!\tilde p_{k}}{s}\
  z\,(1-y)
  \min\!\left\{
  \frac{\hat p_{i}\!\cdot\!n}{\hat p_{k}\!\cdot\!n},\,
  \frac{\hat p_{k}\!\cdot\!n}{\hat p_{i}\!\cdot\!n}
  \right\}\;\;.
\end{equation}
Under the condition  $0 < z \ll y < 1$, this is 
\begin{equation}
  \label{dikmod2}
  d_{ik}  \sim 
  \frac{2\tilde p_{ij}\!\cdot\!\tilde p_{k}}{s}\
  y\,z 
\min\!\left\{
  1,\,
  \frac{(1-y)^2}{y^2}
  \right\}
< \frac{2\tilde p_{ij}\!\cdot\!\tilde p_{k}}{s_{ij}}\,
 \tilde d\;\;.
\end{equation}
Thus when $ z \ll y$ the condition $\tilde d < d_{\rm ini}$ restricts
parton $i$ to be close to the spectator parton $k$.

\section{Splitting functions}
\label{sec:splittingfctns}

In this section we review the functions ${\cal D}_{ij,k}$ used in the
dipole subtraction algorithm \cite{CataniSeymour}. We then specify the
functions $\bm{S}_{l}(p, f ,Y)$ and $\big\langle
\bm{S}(y,z,f)\big\rangle$ that we use to describe parton splitting
and that are derived from the ${\cal D}_{ij,k}$ functions.

We begin with the functions ${\cal D}_{ij,k}$ that form the basis for the
dipole subtraction algorithm. The basic idea is that the squared matrix
element $\la{\cal M}(\{\hat p\}_{m+1})|{\cal M}(\{\hat p\}_{m+1})\ra$ for
$m+1$ partons has a potential singularity that could lead to a divergent
integral when the dot product of any pair of the
momenta, say $p_i \cdot p_j$, goes to zero. The matrix element takes a
rather simple form in this limit. In fact, the form would be extremely
simple were it not for the fact that one has $p_i \cdot p_j\to 0$ not
only when $p_i$ becomes collinear with $p_j$ but also when $p_i \to 0$ 
with $p_j$ fixed and $p_i$ not necessarily collinear with $p_j$ (or
when $p_j$ becomes soft with fixed $p_i$). This configuration can lead
to a divergent integral if the soft parton $i$ (or $j$) is a gluon.
When
the soft gluon $i$ couples to parton $j$ and another parton $k$, the
structure of the limiting function depends on what $k$ is. For this
reason, Catani and Seymour write the matrix element in the limit $p_i
\cdot p_j\to 0$ as a sum of terms labelled by the index $k$ of a
``spectator'' parton.  That is
\begin{equation}
  \label{DipoleFF:Fac}
  \big\la{\cal M}\big(\{\hat p,\hat f\}_{m+1}\big)\big|
  {\cal M}\big(\{\hat p, \hat f\}_{m+1}\big)\big\ra =
  \sum_{k\neq i,j} {\cal D}_{ij,k}(\{\hat p,\hat f\}_{m+1}) + \cdots
\;,
\end{equation}
where the dots stand for terms that are nonsingular in the limit
$p_i\!\cdot\!p_j\to 0$. The dipole functions ${\cal D}_{ij,k}$ have a
simple structure of the form that we reviewed in Eq.~(\ref{newDijk}),
\begin{equation}
\begin{split}
{\cal D}_{ij,k}&(\{\hat p,\hat f\}_{m+1}) ={}
\\
&
\frac{1}{2 \hat p_i\cdot \hat p_j}\
\bra{{\cal M}(\{p,f\}_{m}^{ij,k})}
\frac{\bm{T}_{ij}\cdot \bm{T}_k}{-\bm{T}_{ij}^2}\,
{\bm{V}_{ij}}(\{\tilde p_{ij},y,z,\phi\}_{ij,k},f_i,f_j)\,
\ket{{\cal M}(\{p,f\}_{m}^{ij,k})}
\;\;.
\label{newDijkencore}
\end{split}
\end{equation}
First, there is a singular factor $1/(2 \hat p_i\cdot \hat p_j)$. Then
there is the Born amplitude $|{\cal M}(\{p,f\}_{m}^{ij,k})\rangle$ and
its complex conjugate for $m$ parton momenta and flavors formed by
combining partons $i$ and $j$ with the help of the spectator $k$ according
to the formulas of the previous section. These amplitudes are vectors in
the color and spin space of the partons. More precisely, for each of the
$m$ partons there is a spin space spanned by two basis vectors
$|s\rangle$ and a color space spanned by three basis vectors $|c\rangle$
in the case of a quark or antiquark or eight basis vectors
$|c\rangle$ in the case of a gluon. The amplitude $|{\cal
M}(\{p,f\}_{m})\rangle$ lies in the direct product of the $m$ spin spaces
and $m$ color spaces. In Eq.~(\ref{newDijkencore}) there are color and
spin operators that act on given single parton factors in the direct
product space, with the parton factor affected labelled by a subscript
$ij$ (for the parton obtained by combining partons $i$ and $j$) or $k$
(for the spectator parton).

For the color, there is an operator, the SU(3) generator
$\bm{T}^a_{ij}$, that acts on the color space for parton $ij$ and there is
another SU(3) generator  $\bm{T}^a_{k}$ that acts on the color space for
the spectator parton $k$. The dot product indicates a sum over $a$ from 1
to 8. In the denominator there is a factor $\bm{T}_{ij}\cdot
\bm{T}_{ij}$, which is simply a number equal to $C_{\rm A}$ if parton $ij$
is a gluon and  $C_{\rm F}$ if it is a quark.

The remaining factor, ${\bm{V}}_{ij}$, is a function of variables defined
by considering that one combines partons $i$ and $j$ with the help of the
spectator $k$, as in Eqs.~(\ref{Qijkdef}), (\ref{Yldef}) and
(\ref{Qijkdefencore}). Specifically, ${\bm{V}}_{ij}$ is an operator on
the spin space of the mother parton before the splitting and depends on
the momentum $\tilde p_{ij}$ of this parton. It is also a function of 
the splitting variables $\{y,z,\phi\}$ and the daughter flavors
$\{f_i,f_j\}$. 

The shower algorithm of this paper makes use of splitting functions
$\bm{S}_{l}(p ,Y)$. We found it useful to take these functions to be
proportional to the functions ${\bm{V}}_{ij}$, using
Eq.~(\ref{identifyS}),
\begin{equation}
\frac{\alpha_{\rm s}}{2\pi}\, \frac{16\pi^2}{1-y}\
\bm{S}_{l}(p,f ,Y )
=
\bm{V}_{l}(\{p,y,z,\phi\},f_1,f_2)\;\;,
\label{VtoS}
\end{equation}
where $Y = \{y,z,\phi,f_1,f_2\}$. (Note that the subscript $l$ does not
enter into the functional dependence of these functions but merely tells
on what parton's spin space the operator acts.) Since we use mostly the
functions $\bm{S}_{l}(p,f,Y )$, we present here the standard definition of
${\bm{V}}_{ij}$ simply translated into the new notation.

The definition begins by separating the possibilities for
flavors,
\begin{equation}
\begin{split}
\label{flavorDipoleFF:S}
\bm{S}_{l}(p,f ,Y ) ={}&\delta_{gf_1}\delta_{gf_2}
   \bm{S}_{gg}(p,z,y,\kappa_\perp)
+ \sum_{r=u,\bar u,d,\bar d,\dots}\!\!\biggl[
      \delta_{rf_1}\delta_{\bar{r}f_2}
      \bm{S}_{q\bar{q}}(p,z,y,\kappa_\perp)
\\  &
      +\delta_{rf_1}\delta_{gf_2}\bm{S}_{qg}(z,y)
      + \delta_{gf_1}\delta_{rf_2} \bm{S}_{qg}(1-z,y)\biggr]
\;\;.
\end{split}
\end{equation}
Then for the splitting of a quark or antiquark into the same flavor quark
or antiquark plus a gluon the splitting function is  
\begin{equation}
  \label{DipoleFF:Sqg}
  \la s|\bm{S}_{qg}(z, y)|s'\ra  = 
    C_{\rm F}\, (1-y)\left[\frac2{1-z(1-y)} 
    -(1+z)\right]\delta_{ss'}\;\;,
\end{equation}
where the $s$ and $s'$ are the spin indices of the emitter quark or
antiquark.  For the splitting of a gluon into a quark and an antiquark, we
again denote the spin indices of the emitter gluon by $s$ and $s'$ and
define
\begin{equation}
  \label{DipoleFF:Vqqb}
  \la s|\bm{S}_{q\bar{q}}(p,z,y,\kappa_\perp)|s'\ra 
  =    T_{\rm R}\,(1-y)\
  \epsilon^*_\mu(p,s)
  \left[- g^{\mu\nu}  
   - 4z(1-z)\,{\kappa_{\perp}^\mu \kappa_{\perp}^\nu}
  \right]
\epsilon_\nu(p,s')\;\;.
\end{equation} 
Finally for the splitting of a gluon into two gluons the splitting
function is
\begin{equation}
  \label{DipoleFF:Vgg}
  \begin{split}
\la s|\bm{S}_{gg}(p,z,y,\kappa_\perp)|s'\ra
   =& 
\\
 2 C_{\rm A}\,(1-y)\
 \epsilon^*_\mu(p,s)
    \bigg[ & -g^{\mu\nu}\left(\frac1{1-z(1-y)}+\frac1{1-(1-z)(1-y)}-2
    \right)
\\&
+ 2\,z(1-z)\,{\kappa_{\perp}^\mu \kappa_{\perp}^\nu}
    \bigg]
\epsilon_\nu(p,s')\;\;.
  \end{split}
\end{equation} 
In the gluon splitting functions, $\epsilon(p,s)$ is the polarization
vector for the emitter gluon. With a change of the gauge
used in defining $\epsilon(p,s)$, one has $\epsilon(p,s) \to \epsilon(p,s)
+ \lambda\, p$. However, the matrix elements are unchanged because
$\epsilon(p,s) \cdot p = 0$ and $\kappa_\perp \cdot p = 0$. 

It is also of interest to know whether these matrix elements depend on
the spectator momentum, $p_k$. No spectator momentum appears explicitly
in Eqs.~(\ref{DipoleFF:Vqqb}) and (\ref{DipoleFF:Vgg}), but recall that
$\kappa_\perp$ is a unit vector orthogonal to $p$ and to $p_{k}$. If we
change $p_{k}$ to $p_{k}'$, we can define a new vector $\kappa_\perp'$ to
specify the azimuthal angle of the splitting by
\begin{equation}
\kappa_\perp' = \kappa_\perp - 
\frac{p_{k}' \cdot \kappa_\perp}
{p_{k}' \cdot   p}\
p \;\;.
\end{equation}
This new vector is still a unit vector, still orthogonal to $p$, but now
is orthogonal to $p_{k}'$ instead of $p_{k}$. The change does not affect
the matrix element because  $\epsilon(p,s) \cdot p = 0$. This justifies
the notation that $\bm{V}_{ij}$ depends on $p$ and on the splitting
variables (including $\phi$ or, equivalently, $\kappa_\perp$) but not on
the spectator momentum $p_{k}$.

The Sudakov exponent in the showering formula contains the average over
angle and flavors of $\bm{S}_{l}$, Eq.~(\ref{Slaveragedef}),
\begin{equation}
\big\langle \bm{S}(y,z,f)\big\rangle
\equiv
\int \frac{d\phi}{2\pi}\
\frac{1}{2}\!\sum_{ f_1,  f_2}\!
\delta_{ f_1 +   f_2}^{f}
\bm{S}_{l}(p,Y)\;\;.
\label{Slaveragebis}
\end{equation}
Here the angular average is in 2 transverse dimensions. There are extra terms
in $\langle \bm{S}\rangle$ if one works in $2- 2\epsilon$ dimensions, but
we do not need these terms in this paper.
For the angular average, we can use
\begin{equation}
\int \frac{d\phi}{2\pi}\
\epsilon^*_\mu(p,s)\
\kappa_{\perp}^\mu \kappa_{\perp}^\nu\
\epsilon_\nu(p,s')
=  \frac12\, \delta_{s s'}
\;\;.
\end{equation}
Then we see that $\big\langle \bm{S}_{l}(y,z,f)\big\rangle$ is a number
times the unit operator on the partonic spin space. A simple calculation
gives for a quark or antiquark emitter
\begin{equation}
\big\langle \bm{S}(y,z,f)\big\rangle
=
 C_{\rm F}\, (1-y)\left[\frac2{1-z(1-y)} 
    -(1+z)\right]\;,
\hskip 0.5 cm
f \in \{{\rm u},\bar {\rm u},{\rm d},\dots\} \;.
\label{avSquark}
\end{equation}
For a gluon emitter, including splittings into both gluon and
quark-antiquark pairs, one finds
\begin{equation}
\begin{split}
\big\langle \bm{S}(y,z,g)\big\rangle={}&
  C_{\rm A}\,(1-y)\
    \bigg[ \frac1{1-z(1-y)}+\frac1{1-(1-z)(1-y)}-2
+ z(1-z)
    \bigg]
\\&+
N_{\rm f}\,T_{\rm R}\,(1-y)\
  \left[1 
   - 2z(1-z)
  \right]
\;\;.
\label{avSgluon}
\end{split}
\end{equation}

The $y \to 0$ limits of these functions match the Altarelli-Parisi
splitting functions,
\begin{equation}
\begin{split}
\big\langle \bm{S}(y,z,f)\big\rangle \to{}&
 C_{\rm F}\, \frac{1+z^2}{1-z} 
= P_{{\rm q}/{\rm q}}(z)\,,
\hskip 0.5 cm
f \in \{{\rm u},\bar {\rm u},{\rm d},\dots\}\;,
\\
\big\langle \bm{S}(y,z,g)\big\rangle \to {}&
  C_{\rm A}\,
       \frac{[1-z(1-z)]^2 }{z(1-z)} 
+
N_{\rm f}\,T_{\rm R}\,
  \left[1 
   - 2z(1-z)
  \right]
\\ & =  \frac12\,P_{{\rm g}/{\rm g}}(z) 
+ N_{\rm f} P_{{\rm q}/{\rm g}}(z)
\;\;.
\end{split}
\end{equation}
Note that some of the Altarelli-Parisi functions are singular as $z\to 0$
or $z \to 1$. However, for $y>0$ the splitting functions $\bm{S}$ are
not singular as $z \to 0,1$.

\section{Virtual contributions}
\label{sec:virtualfunctions}

In Eq.~(\ref{sigmaV}), we specified the virtual loop
contribution in the dipole subtraction scheme in terms of a function
$V(\{p,f\}_m)$,
\begin{equation}
\sigma^{\rm V+A }
=
\int_{m}\big[d\sigma^{\rm V}+\int_{1}d\sigma^{\rm A}\big]
= 
\frac{1}{m!}
\sum_{\{f\}_m} 
\int\!d\varGamma(\{p\}_m)\
V(\{p,f\}_m)\
F_m(\{p\}_m)\;\;.
\end{equation}
The definition \cite{CataniSeymour} is
\begin{equation}
\begin{split}
  \label{NLONo:sigmaVI}
    V(\{p,f\}_m) ={}&\Big[
    2\,{\rm Re}\,\big\langle{\cal M}(\{p\}_m)\big|
    {\cal M}^{(1)}(\{p\}_m;\epsilon) \big\rangle
\\ & +
{\rm Re}\,\big\langle{\cal M}(\{p\}_m)\big|
\bm{I}(\{p,f\}_m; \epsilon)
\big|{\cal M}(\{p\}_m)\big\rangle
    \Big]_{\epsilon =0}\;\;.
\end{split}
\end{equation}
Here $\big|{\cal M}^{(1)}\left(\{p\}_m;\epsilon\right)\big\rangle$ denotes
the virtual one-loop correction to the $m$-parton matrix element in
$4-2\epsilon$ dimensions. This matrix element is singular as $\epsilon
\to 0$. The singularity is cancelled by the integral of $d\sigma^{\rm
A}$, performed in $4-2\epsilon$ dimensions. The result of this
integration is the Born amplitude $|{\cal M}(\{p\}_m) \rangle$ times an
operator $\bm{I}(\epsilon)$  in the color space of the final-state partons,
\begin{equation}
  \label{NLONo:Ioper}
  \bm{I}(\{p,f\}_m; \epsilon) =  \frac\as{2\pi}\frac1{\Gamma(1-\epsilon)}
  \sum_{i=1}^m
    {\cal V}(f_i,\epsilon) 
  \sum_{\substack{j=1\\ j\neq i}}^m
  \frac{\bm{T}_i\cdot \bm{T}_j}{-\bm{T}_i^2}
  \left(\frac{4\pi\mu^2}{2p_i\!\cdot\!p_j}\right)^\epsilon\;\;.
\end{equation}
The function ${\cal V}(f,\epsilon)$ has a simple expansion about
$\epsilon = 0$,
\begin{equation}
\label{calVdef}
  {\cal V}(f,\epsilon)= C_f\left(\frac1{\epsilon^2} -\frac{\pi^2}3\right)
  +\gamma_f\left(\frac{1}{\epsilon} + 1\right) 
  + K_f +{\cal O}(\epsilon)\;\;,
\end{equation}
where $C_f$, $\gamma_f$, and $K_f$ are given in 
Eq.~(\ref{calVcoefficients}).

When we adapt the dipole subtraction scheme to the calculation of partial
cross sections based on a cut $d_{\rm ini}$ on the ``distance'' between
partons, we need a correction to the subtraction term for virtual
graphs. The correction involves a function $C_{l,k}$ defined in 
Eq.~(\ref{Clkdef}),
\begin{equation}
\begin{split}
  C_{l,k}(\{p,f\}_{m}, d_{\rm ini}) ={}& 
  \int_0^1 \frac{dy}{y}
\int_0^1\!dz\,\int_0^{2\pi}\frac{d\phi}{2\pi}\
\frac12\!\sum_{\hat f_{1},\hat f_{2}}\!
\delta_{\hat f_{1} + \hat f_{2}}^{f_l}
  \theta\big(d_{\rm ini} < \tilde d(\{p,f\}_{m},l, y, z) \big)
\\ &\times
\frac{y(1-y)\,2 p_l\cdot p_k}{16\pi^2}\
  {\cal D}_{l,k}(\{p,f\}_{m};Y)\;\;.
\end{split}
\label{Clkdefecore}
\end{equation}
Here the distance function $\tilde d(\{p,f\}_{m},l, y, z)$ is defined in
Eq.~(\ref{tildeddef}). To find the explicit form of $C_{l,k}(\{p,f\}_{m},
d_{\rm ini})$, we use Eqs.~(\ref{newDijkencore}), (\ref{VtoS}) and
(\ref{Slaveragebis}) to relate the sum over daughter flavors and average
over azimuthal angle $\phi$ of ${\cal D}_{l,k}$ to $\big\langle
\bm{S}(y,z,f)\big\rangle$. This gives 
\begin{equation}
\label{Clkform1}
  C_{l,k}(\{p,f\}_{m}, d_{\rm ini}) =
 \frac{\as}{2\pi}\,
 {\cal C}\!\left(\frac{s\,d_{\rm ini}}{s_l},
f_{l}\right) 
 \big\langle {\cal M}(\{p,f\}_m)\big|
    \frac{\bm{T}_l\cdot \bm{T}_{k}}{-\bm{T}_{l}^2}
    \big|{\cal M}(\{p,f\}_m)\big\rangle\;\;,
\end{equation}
where
\begin{equation}
\begin{split}
\label{calCintegral}
\frac{\as}{2\pi}\,
{\cal C}\!\left(x, f\right)={}&
\int_{0}^{1} \frac{dy}{y}\int_{0}^{1/2}\!dz\,
\theta(z/(1-z) \geq x/y) 
\left[
\big\langle \bm{S}(y,z,f)\big\rangle
+ 
\big\langle \bm{S}(y,1-z,f)\big\rangle
\right]\;\;.
\end{split}
\end{equation}
When we sum over the spectator label $k$, Eq.~(\ref{Clkform1}) simplifies
to
\begin{equation}
\label{Clkform}
  \sum_{k\ne l} C_{l,k}(\{p,f\}_{m}, d_{\rm ini}) =
\frac{\as}{2\pi}\,
 {\cal C}\!\left(\frac{s\,d_{\rm ini}}{s_l},
f_{l}\right) 
 \big\langle {\cal M}(\{p,f\}_m)\big|
     {\cal M}(\{p,f\}_m)\big\rangle\;\;.
\end{equation}
Using the explicit form of $\big\langle
\bm{S}(y,z,f)\big\rangle$ given in Eqs.~(\ref{avSquark}) and
(\ref{avSgluon}), we can perform the integrations over $y$ and $z$ to
obtain
\begin{equation}
\begin{split}
\label{calCresult}
 {\cal C}\!\left(x, f\right)={}&
	\frac{1}{2}\,C_f\log^{2} (x) + \gamma_f\log(4x)
	- 2C_f\log(x) \log(1 + x) - \frac{\pi^2}{6}\,C_f
	\\
	&\qquad
	+\gamma_f 2x\log(2x) -2\gamma_{f} (1+x) \log(1+x)
	+ \gamma_f(1 - x) - 2C_f\,\mathrm{Li_{2}}(-x)
	\\
	&\qquad
	+\delta_{gf}(C_{\rm A}-2T_{\rm R}N_{\rm
F})\left(\frac13x\log\frac{2x}{1+x} 
			+\frac{1+2x-3x^2}{12(1+x)}\right)
\;\;.
\end{split}
\end{equation}

\section{Sudakov factors}
\label{sec:CKKWSudakov}

The functions $\sigma^{\rm B + S}_m$, $\sigma^{\rm R + S}_m$, and
$\sigma^{\rm V + S}_m$ contain reweighting functions $W$. These functions
contain rather trivial ratios of $\alpha_{\rm s}$ at different scales and
not so trivial Sudakov exponentials. In addition, the splitting operators
${\bm E}_{l,k}$ in $\sigma^{\rm B + S}_m$ contain Sudakov exponentials.
In this section, we discuss these factors and, in particular, give
definitions for the functions $W$ that we use. The presentation also
touches on the Sudakov exponentials in the the Monte Carlo interface
functions $I$ and $\tilde I$ in $\sigma^{\rm B + S}_m$, $\sigma^{\rm R +
S}_m$, and $\sigma^{\rm V + S}_m$.

\subsection{The standard NLL Sudakov factor and its interpretation}
\label{sec:NLLSudakov}

The formula (\ref{sigmaBS}) for $\sigma^{\rm B + S}_m$ contains a
reweighting factor $W_{m}(\{p,f\}_m)$ consisting of a product of ratios
of $\alpha_{\rm s}$ at different scales, Eq.~(\ref{CKKWfactor1}), and a
Sudakov exponential.  Following Ref.~\cite{CKKW}, we base this factor on a
construction of a splitting history corresponding to the final state
$\{p,f\}_m$ using the ``$k_T$'' jet algorithm (slightly extended to
include flavor information) as described in
Sec.~\ref{sec:PartialCrossSections}. The algorithm combines the final
state partons so as to produce a QCD tree graph for $e^+ e^-
\to \{p,f\}_m$. Let the lines in this tree graph be labeled by an index
$L$. We can label the vertices according to the order in which the partons
were combined. We start with $m$ partons and combine two of them at a
vertex with label $V = m$ to produce an $m-1$ parton state. Then we
combine two of these $m-1$ partons at a vertex with label $V = m-1$ to
produce an $m-2$ parton state. Finally, the last 2 partons are combined
at vertex 2 to form a $\gamma$ or $Z$. At each stage between vertices $V$
and $V+1$ there are $n = V$ partons. The line labels $L$ for these
partons form a set that we can call $I(V)$ (for the intermediate state
after vertex $V$).

Each vertex $V$ has an associated distance measure $d(V)$, with $d(V) \ge
d(V+1)$. The values of the $d(V)$ will appear in $W_{m}(\{p,f\}_m)$.
Typically, $d(V)$ is the $k_T$ measure (\ref{dij}) for the splitting at
vertex $V$. In some cases, we have redefined $d(V)$ to equal $d(V+1)$. 
 
With these definitions, for each line $L$ in the splitting graph there is
initial resolution parameter $d(V_{\rm i}(L))$ equal to $d(V)$ for the
vertex $V_{\rm i}(L)$ at which the line originates and a final resolution
parameter $d(V_{\rm f}(L))$ equal to $d(V)$ for the vertex $d(V_{\rm
f}(L))$ at which the line splits.  For a line that enters the final state,
we take $d(V_{\rm f}(L)) \equiv d_{\rm ini}$. Thus $1 \ge d_{\rm i}(L) \ge
d_{\rm f}(L) \ge d_{\rm ini}$. We denote the flavor of line $L$ by
$f(L)$. 

With this notation, we can state the definition of the reweighting factor
$W$ as given in \cite{CKKW}. Briefly, the idea is the following.
Consider the Born cross section $\sigma_m^{\rm B}$ with observation
function $F_m(\{p\}_m) = 1$. This is the Born level cross section for
making $m$ jets, using resolution parameter $d_{\rm ini}$. By
inserting a factor $W_{m}(\{p,f\}_m)$ into the integral for
$\sigma_m^{\rm B}$, we obtain the approximation to the $m$-jet cross
section that sums logarithms of  $d_{\rm ini}$ at the leading log and
next-to-leading log level. The required reweighting factor is a product of
$\alpha_{\rm s}$ factors, one for each vertex, and a product of Sudakov
factors, one for each line,
\begin{equation}
W_{m}^{\rm NLL}(\{p,f\}_m) = \prod_{V>2}
\frac{\as\big(\sqrt{d(V)s}\big)}
 	{\as\big(\mu_{\rm R}\big)}\
\prod_L\frac{\Delta_{f(L)}(d(V_{\rm i}(L)),d_{\rm ini})}
            {\Delta_{f(L)}(d(V_{\rm f}(L)),d_{\rm ini})}
\;\;,
\label{CKKWfactor0}
\end{equation}
where
\begin{equation}
\Delta_f(d_{\rm max},d_{\rm min}) = \exp\left[
-\int^{d_{\rm max}}_{d_{\rm min}}\!\frac{d\lambda}{\lambda}\
\frac{\alpha_{\rm s}(\sqrt{\lambda s})}{2\pi}\,
\left\{
C_{f}\,\log\!\left(\frac{d_{\rm max}}{\lambda}\right) - \gamma_{f} 
\right\}
\right]\;\;.
\label{Deltadef}
\end{equation}
The coefficients $C_f$ and $\gamma_f$ are given in 
Eq.~(\ref{calVcoefficients}). Thus one multiplies the perturbative matrix
element by Sudakov factors that would have been associated
with the lines if the final state had been generated by a suitable parton
shower algorithm according to the specified QCD tree graph with the
restriction that no splittings were to be generated that were
unresolvable at resolution parameter $d_{\rm ini}$. For each QCD vertex
we also use the reweighting factor to replace $\alpha_{\rm s}(\mu_{\rm
R})$ used in the perturbative matrix element by the running coupling used
in a standard parton shower. Our default reweighting factor
$W_{m}(\{p,f\}_m)$ will be a slightly modified version of $W_{m}^{\rm
NLL}(\{p,f\}_m)$ that is equivalent if we neglect terms suppressed by a
color factor $1/N_{\rm c}^2$. We define the modified version below after
examining the structure of $W_{m}^{\rm NLL}(\{p,f\}_m)$.

The function $W_{m}^{\rm NLL}(\{p,f\}_m)$ as given by
Eq.~(\ref{CKKWfactor0}) is a product of factors, one for each propagator
in the jet-structure graph. There is another way to write it that focuses
on the evolution from one value of $d(V)$ to the next (See
Fig.~\ref{fig:kTshower}).  

\FIGURE{
\centerline{
\includegraphics[width = 3 cm]{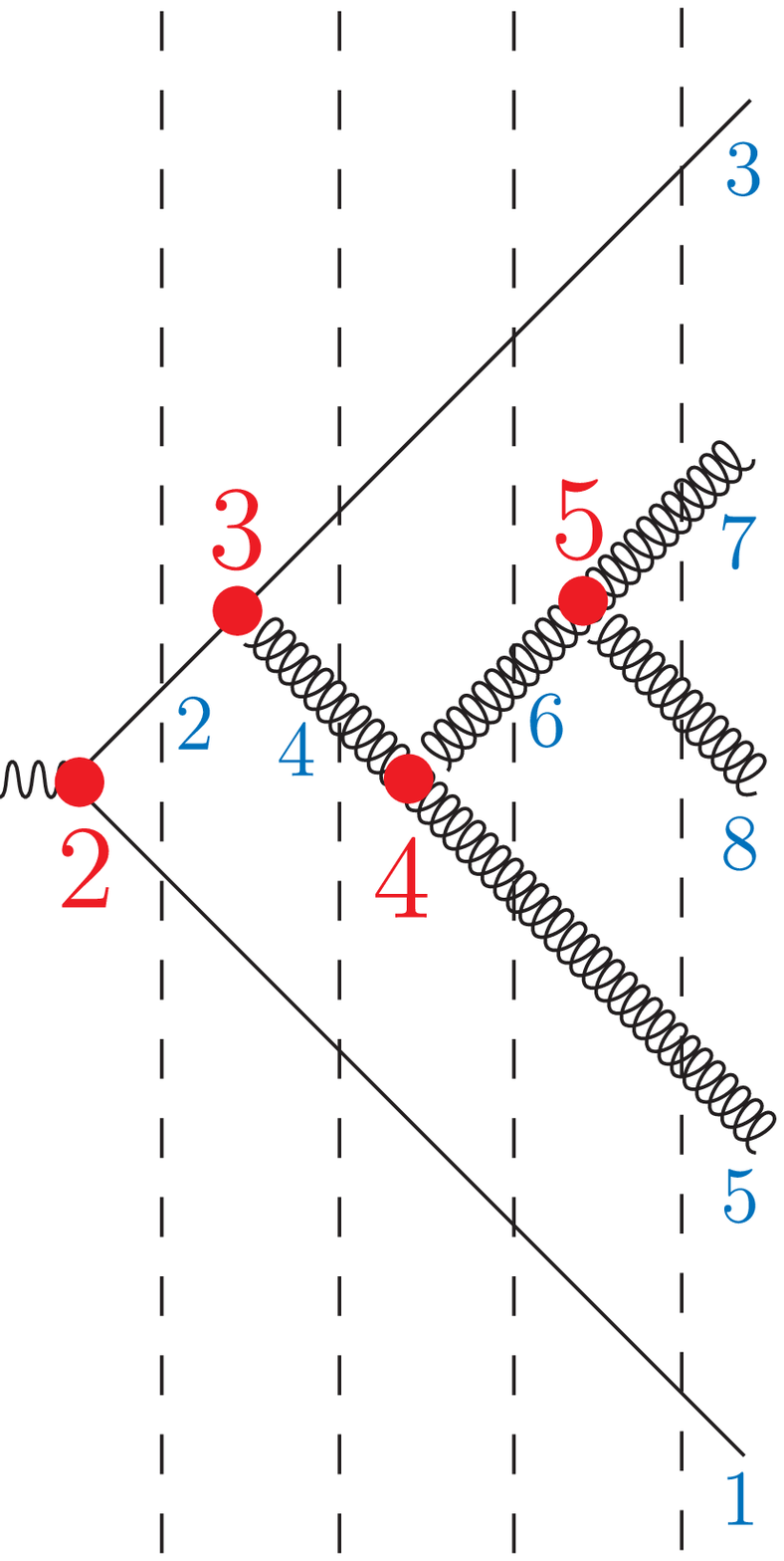}
}
\medskip
\caption{
Illustration of the tree graph for a shower constructed from a five
parton final state. The lines in the graph are numbered $L = 1,\dots 8$.
The vertices are numbered $V = 2,\dots,5$. The resolution parameters
$d(V)$ for the vertices decrease: $d(2) \ge d(3) \ge d(4) \ge
d(5) > d_{\rm ini}$. There is a starting vertex $V_{\rm s}(L)$ for each
parton line $L$. Here $V_{\rm s}(1) = V_{\rm s}(2) = V_{\rm s}(3) = 2$,  
$V_{\rm s}(4) = V_{\rm s}(5) = 3$, $V_{\rm s}(6) = V_{\rm s}(7) = 4$,
and $V_{\rm s}(8) = 5$. We will write the Sudakov factor as a product of
factors that represent the probabilities that the partons in each of the
indicated intermediate state evolve from scale $d(V)$ to scale $d(V+1)$
without splitting. For the last final state, the Sudakov factor is the
probability for the five partons to evolve from scale $d(5)$ to $d_{\rm
ini}$ without splitting.}
\label{fig:kTshower}
}

To reorganize $W_{m}^{\rm NLL}(\{p,f\}_m)$, we define a special vertex
$V_{\rm s}(L)$ corresponding to each line $L$ in the graph. We can think
of $V_{\rm s}(L)$ as the vertex where the history of line $L$ started
({\it cf.} Sec. 3.2 of \cite{CKKW}). The identification of the $V_{\rm
s}(L)$ is recursive. Consider the starting vertex $2:\gamma/Z \to
q(L_1) \bar q(L_2)$, where our notation indicates that $2$ is the label
of the vertex, $L_1$ is the label of the quark line and $L_2$ is the
label of the antiquark line.  We define $V_{\rm s}(L_1)
= V_{\rm s}(L_2) = 2$. Then for any vertex $V_i: q(L_0) \to q(L_1)
g(L_2)$,  we define $V_{\rm s}(L)$ for the two daughter lines by $V_{\rm
s}(L_{1}) = V(L_0)$ while $V_{\rm s}(L_2) = V_i$. Similarly for any
vertex $V_i: \bar q(L_0) \to \bar q(L_1) g(L_2)$,  we define  $V_{\rm
s}(L_{1}) = V_{\rm s}(L_0)$ while $V_{\rm s}(L_2) = V_i$. For a vertex
$V_i: g(L_0) \to g(L_1) g(L_2)$, we first provide a definite labelling
for the daughters by, say, letting $L_1$ be the label of the daughter
gluon with splitting fraction $z_1 > 1/2$. Then we define $V_{\rm s}(L_1)
= V(L_0)$ while $V_{\rm s}(L_2) = V_i$. Finally, for any vertex $V_i:
g(L_0) \to q(L_1)\bar q(L_2)$, we define $V_{\rm s}(L_1) = V_{\rm s}(L_2)
= V_{\rm s}(L_0)$.

In the case that $L$ labels a quark or antiquark line, we define another
special vertex $V_{\rm o}(L)$ that is the vertex where the quark line
originated in the jet-structure graph: either the $\gamma/Z \to  q \bar q$
vertex $V=2$ or else a later $g \to  q \bar q$ vertex $V$. (The subscript
is ``o'' for ``originated.'')

The reader can check that with this definition
\begin{equation}
W_{m}^{\rm NLL}(\{p,f\}_m)
=
\prod_{V>2}
\frac{\as\big(\sqrt{d(V)s}\big)}
 	{\as\big(\mu_{\rm R}\big)}\
\prod_{V=2}^{m}
\prod_{L \in I(V)} 
\exp\{- {\cal S}^{\rm NLL}(V,L)\}
\label{CKKWfactoralt}
\end{equation}
with
\begin{equation}
\label{Sdef}
\begin{split}
{\cal S}^{\rm NLL}(V,L) ={}&
\int^{d(V)}_{d(V+1)}\!\frac{d\lambda}{\lambda}\
\frac{\alpha_{\rm s}(\sqrt{\lambda s})}{2\pi}\,
\Biggl\{
C_{f(L)}\,\log\!\left(\frac{d(V_{\rm s}(L))}{\lambda}\right)
 - \gamma_{f(L)}  
\\
& \hskip 1 cm
+ \theta\bigl(f(L) \in \{u,\bar u, d, \dots\}\bigr)
\left[\frac{C_{\rm A}}{2} - C_{\rm F}\right]\,
\log\!\left(\frac{d(V_{\rm s}(L))}{d( V_{\rm o}(L))}\right)
\Biggr\}\;\;.
\end{split}
\end{equation}
Here we interpret the lower limit $d(V+1)$ to be $d_{\rm ini}$ in the
case $V=m$. This represents a simple algebraic reshuffling of the
original expression that appears in Eq.~(\ref{CKKWfactor0}). What is
more significant than the algebra is the interpretation. The total
Sudakov factor is a product of factors, one for each splitting vertex
$V$. Each of these factors represents the propagation of the system
between scales $d(V)$ and $d(V+1)$. The factor for propagation from
$d(V)$ and $d(V+1)$ is a product of factors, one for each parton line
$L$ that occurs in the intermediate state after splitting vertex $V$.
Each of these factors, $\exp\{- {\cal S}(V,L)\}$, represents the
probability that parton $L$ in the intermediate state following vertex
$V$ did not split at a resolution parameter between $d(V)$ and $d(V+1)$.
In the case of the last vertex, $V = m$, $\exp\{- {\cal S}(V,L)\}$
represents the probability that parton $L$ in the intermediate state
following vertex $m$ did not split at a resolution parameter between
$d(m)$ and $d_{\rm ini}$.

In the Sudakov factor, we integrate over the resolution parameter
$\lambda = k_T^2/s$ of a virtual splitting. As we will see, we can
understand the logarithms in the integrand as the result of integrating
over the angle of a virtual splitting:
\begin{equation}
\begin{split}
\label{angleranges}
\log\!\left(\frac{d(V_{\rm s}(L))}{\lambda}\right) = {}&
\int_{\lambda\, s/ E_{\rm s}^2}^{d(V_{\rm s}(L))\,s/ E_{\rm s}^2} 
\frac{d\theta^2}{\theta^2}\,,
\\
\log\!\left(\frac{d(V_{\rm s}(L))}{d( V_{\rm o}(L))}\right) ={}&
\int_{d(V_{\rm o}(L))\,s/ E_{\rm s}^2}^{d(V_{\rm s}(L))\,s/ E_{\rm s}^2} 
\frac{d\theta^2}{\theta^2}\,,
\\
\log\!\left(\frac{d(V_{\rm s}(L))}{\lambda}\right) 
-\log\!\left(\frac{d(V_{\rm s}(L))}{d( V_{\rm o}(L))}\right)
= {}&
\int_{\lambda\, s/ E_{\rm s}^2}^{d(V_{\rm o}(L))\,s/ E_{\rm s}^2} 
\frac{d\theta^2}{\theta^2}\,.
\end{split}
\end{equation}
Here $E_{\rm s}$ is the energy of the least energetic of the two partons
that were produced at vertex $V_{\rm s}(L)$.

The angle limits that appear in these formulas can be understood on the
basis of ``angular ordering'' \cite{ESWbook}. There
are three types of splitting vertices to consider,
$g \to g g$, $q \to q g$, and $g \to q \bar q$. Then the result in
Eq.~(\ref{Sdef}) can be understood as the result of stringing together
the three cases in all possible combinations.

The case of a $g \to g g$ vertex, $V$, is easiest. Typically, one of the
daughter gluons is ``hard'' and moves in almost the direction of the
mother gluon. The other gluon is softer and moves at an angle
$\theta_V = k_{T,V}/E_V$ with respect to the mother direction, where $E_V$
is the energy of the softer daughter gluon. That is,
\begin{equation}
\theta_V^2 = d(V) s/E_V^2 .
\end{equation}
Suppose that we now emit a soft gluon with energy $E$ from the two
daughter partons.  The soft gluon has transverse momentum $k_T$ with
respect to the daughter parton from vertex $V$. The resolution parameter
for the soft gluon parameter is $\lambda = k_T^2/s$ and we integrate over
$\lambda$ in Eq.~(\ref{Sdef}). The emission angle $\theta$ of this soft
gluon is given by $\theta^2 = \lambda\,s/E^2$. The smallest that
$\theta^2$ could be is  $\theta_{\rm min}^2 = \lambda\,s/E_{\rm max}^2$,
where we can take $E_{\rm max}$ to be the energy of whichever parton is
the source of the emission. Now suppose that the emission angle is less
than the angle $\theta_V$ of the $g \to g g$ splitting. Then we get
incoherent emission, with color factor $C_{\rm A}$, from each of the
daughters. If we emit a soft gluon at a larger angle from the two
daughters, the gluon does not resolve the daughters and instead sees the
color of the mother. Thus large angle emission comes with a color factor
$C_{\rm A}$. Thus we get soft gluon emission with a color factor $C_{\rm
A}$ for either small or large angles with respect to the hard daughter
direction, and we get {\it additional} radiation with a color factor
$C_{\rm A}$ for small angles with respect to the direction of the soft
daughter gluon.

For emission from the soft daughter, with a factor $C_{\rm A}$, we have
an angular range $d(V) s/E_V^2 > \theta^2 > \lambda\,s/E_{V}^2$, where we
have set $E_{\rm max} = E_V$. In this case, vertex $V$ is the starting
vertex for the soft daughter, so that the angular range is $d(s) s/E_s^2
> \theta^2 > \lambda\,s/E_{s}^2$, as in the first line of 
Eq.~(\ref{angleranges}). 

For emission from the hard daughter plus coherent emission from both
daughters, we have, to start with, a maximum angle equal to the angle of
the vertex at which the mother gluon was a daughter gluon in a splitting.
However, if the mother was the hard daughter parton in a $g \to g g$
splitting, we can also have coherent emission involving the mother's
sister. Continuing in this way, we get emission with color factor
$C_{\rm A}$ up to a maximum squared angle $\theta_s^2$ where vertex $s$
is the vertex at which the mother of the hard daughter started as a
(relatively) soft daughter of energy $E_s$ (either in a $g \to g g$
splitting or in a $q \to q g$ splitting). That is  $\theta_s^2 = d(s)
s/E_s^2$. The minimum angle is given by 
$\theta_{\rm min}^2 = \lambda\,s/E_{\rm max}^2$ where we can approximate
the hard daughter energy $E_{\rm max}$ by $E_s$. Thus the angular range
for emissions from the hard daughter is $d(s) s/E_s^2 > \theta^2 >
\lambda\,s/E_{s}^2$, as in the first line of  Eq.~(\ref{angleranges}).

What about a $q \to q g$ vertex? Typically, the daughter
quark is ``hard'' and moves in almost the direction of the mother quark.
The daughter gluon is softer and moves at an angle $k_T/E$ with respect to
the mother direction. Suppose that we emit a soft gluon from the daughter
partons with an emission angle less than the angle of the $q \to q g$
splitting. Then we get incoherent emission, with color factors $C_{\rm F}$
or $C_{\rm A}$ respectively, from the quark and gluon daughters.
If we emit a soft gluon at a larger angle from the two daughters, the
gluon does not resolve the daughters and instead sees the color of the
mother. Thus large angle emission comes with a color factor $C_{\rm F}$.
That is, we get soft gluon emission with a color factor $C_{\rm F}$ for
either small {\it or} large angles, and we get additional radiation with
a color factor $C_{\rm A}$ for small angles. 

These two cases account for the factor $\log\!\left({d(V_{\rm
s}(L))}/{\lambda}\right)$ that appears in the first term
in Eq.~(\ref{Sdef}). To understand the remaining term in Eq.~(\ref{Sdef}),
we need to consider a $g \to q \bar q$ vertex.

In the case of a $g \to q \bar q$ vertex, both of the daughter partons
are typically quite hard. We have incoherent soft gluon emission with
color $C_{\rm F}$ from the two daughter partons as long as the emission
angle is less than the splitting angle at the $g \to q \bar q$ vertex. For
larger angles, we have soft gluon emission with color factor $C_{\rm A}$
because the soft gluon sees the net color of the $q\bar q$ pair, which is
the color of the mother gluon. It is convenient to ascribe half of the
large angle emission probability to each of the quark and the antiquark
daughters. That gives an emission probability $C_{\rm A}/2$ for each. We
can understand the Sudakov factor in Eq.~(\ref{Sdef}) in the case that 
$f(L) \in \{u,\bar u, d, \dots\}$ as describing an emission probability
$C_{\rm F}$ for emissions from the quark line with angles smaller than
the angle of the $q\bar q$ vertex at which the quark line was created and
with a factor $C_{\rm A}$ for larger angles, up to the angle of the
vertex $V_{\rm s}(L)$ at which the gluon that split to make the $q\bar q$
pair had its start.

\subsection{Modified $W$ for the Born cross section}
\label{sec:Wmod}

Now that we understand the physical picture behind
Eqs.~(\ref{CKKWfactoralt}) and (\ref{Sdef}), we can propose a small
modification. As was pointed out in \cite{CKKW} using rather
different language, the term in Eq.~(\ref{Sdef}) proportional to $[C_{\rm
A}/2 - C_{\rm F}]$ may seem unnecessarily complicated and could perhaps be
eliminated.  Note that this contribution arises only when there is a
$g\to q\bar q$ vertex, whereas $g\to q\bar q$ splitting is rather
unlikely, particularly because the  $g \to g g$ comes with two logarithms
while the $g \to q \bar q$ splitting comes with only one logarithm.
Furthermore, the color factor is small compared to $C_{\rm F}$: $[C_{\rm
A}/2 - C_{\rm F}]/C_{\rm F} = 1/(N_{\rm c}^2 - 1) = 1/8$. Shower Monte
Carlo programs typically drop contributions that are of relative order
$1/N_{\rm c}^2$, so it seems appropriate to do so here. For these
reasons, our default choice is to drop this term and instead define
$W_{m}(\{p,f\}_m)$ using 
\begin{equation}
W_{m}(\{p,f\}_m)
=
\prod_{V>2}
\frac{\as\big(\sqrt{d(V)s}\big)}
 	{\as\big(\mu_{\rm R}\big)}\
\prod_{V=2}^{m}
\prod_{L \in I(V)} 
\exp\{- {\cal S}(V,L)\} \,,
\label{CKKWfactor}
\end{equation}
with (using $d(m+1) \equiv d_{\rm ini}$)
\begin{equation}
\label{Sdefault}
\begin{split}
{\cal S}(V,L) ={}&
\int^{d(V)}_{d(V+1)}\!\frac{d\lambda}{\lambda}\
\frac{\alpha_{\rm s}(\sqrt{\lambda s})}{2\pi}\,
\Biggl\{
C_{f(L)}\,\log\!\left(\frac{d(V_{\rm s}(L))}{\lambda}\right)
 - \gamma_{f(L)}  
\Biggr\}\;\;.
\end{split}
\end{equation}

Since we include the factor $W$ in the perturbative matrix element in 
$\sigma^{\rm B + S}_m$ and we wish to maintain the NLO accuracy of the
calculation, we need to keep track of the first term in the perturbative
expansion of $W$,  Eq.~(\ref{Wexpansion}),
\begin{equation}
W_{m}(\{p,f\}_m) = 1 + \frac{\alpha_{\rm s}(\mu_{\rm R})}{2\pi}\,
W_m^{(1)}(\{p,f\}_m)
+ \cdots  \;.
\label{Wexpansionencore}
\end{equation}
Using Eqs.~(\ref{CKKWfactor}) and (\ref{Sdefault}) and
using Eq.~(\ref{runningalphas}) to express $\as\big(\sqrt{d_Vs}\big)$ in
terms of $\as\big(\mu_{\rm R}\big)$, we have
\begin{equation}
\begin{split}
\label{W1exact}
W_m^{(1)}(\{p,f\}_m) ={}&
-\sum_{V=2}^{m}
\sum_{L \in I(V)} 
\int^{d(V)}_{d(V+1)}\!\frac{d\lambda}{\lambda}\
\left\{
C_{f(L)}\,\log\!\left(\frac{d(V_{\rm s}(L))}{\lambda}\right)
 - \gamma_{f(L)}  
\right\}
\\&-
\sum_{V=3}^{m}\beta_0
\log\left(\frac{d(V) s}{\mu_{\rm R}^2}\right)
\;\;,
\end{split}
\end{equation}
where, again, $d(m+1)\equiv d_{\rm ini}$.

\subsection{Sudakov factors in dipole splitting}
\label{sec:dipoleSudakov}

Let us now look at the Sudakov exponential that occurs in the function
${\bm E}_{l,k}$, Eq.~(\ref{Elfinal}). We set the evolution variable to be
$k_T^2/s$, as in Eq.~(\ref{ktevollution}). Here $k_T^2 = s_l yz(1-z)$ and
$k_T$ is also the argument of the running coupling according to
Eq.~(\ref{ktdef}). Then the Sudakov exponential in Eq.~(\ref{Elfinal})
is independent of $l$ and $k$  and is
\begin{equation}
\begin{split}
  \Delta(r) = {}& \prod_{l'}
   	\exp\!\biggl( -\int_{r}^{\infty}\!\! dr' \int_{0}^1
  \frac{dy'}{y'}\int_0^1\!\! dz'\, 
  \delta(r' - s_{l'} y'z'(1-z'))
\\&\qquad\qquad\times
\theta\!\left(\frac{r'}{\max\{z^{\prime 2},(1-z')^2\}}< 
         d_{\rm ini}s\right)
\\&\qquad\qquad\times
  \frac{\alpha_s(\sqrt{r'})}{2\pi}\,
  \big\langle\bm{S}(y',z',f_{l'})\big\rangle \biggr)\;\;.
\end{split}
\end{equation}
Here the product runs over all final state partons $l'$. After using
Eqs.~(\ref{avSquark}) and (\ref{avSgluon}) for the splitting
functions, we can perform the $z'$and $y'$ integrals in the exponent to obtain
\begin{equation}
\begin{split}
\Delta(r) ={}& \prod_{l'}\exp\biggl(
-\int_{r/s}^{d_{\rm ini}} \frac{d\lambda}{\lambda}\,
\frac{\as(\sqrt{\lambda s})}{2\pi}\,
\biggl[C_{f(l')}\log\left(\frac{s_{l'}/s}{\lambda}\right) -\gamma_{f(l')}
\\&\qquad\quad 
+{\cal O}\left(h\left(\frac{d_{\rm ini}}{\lambda}\right)\right)
+{\cal O}\left({
\frac{\lambda}{s_{l'}/s}\log\left(\frac{s_{l'}/s}{\lambda}\right)}\right)
\biggr]
\biggr)\;\;.
\end{split}
\end{equation}
Here $s_{l'}/s$ lies between $d_{\rm ini}$ and 1. This form for the exponent applies
when $\lambda \ll s_{l'}/s $, with a correction that vanishes when $\lambda /
(s_{l'}/s) \to 0$. There is a second correction term that is bounded by a
constant times
\begin{equation}
h(d_{\rm ini}/\lambda) = 
\theta(\lambda > d_{\rm ini}/4)\,
\log(1/(2[1 - \sqrt{\lambda/d_{\rm ini}}]))\;\;.
\end{equation}
This correction comes from the difference between having a simple cut
$r'/s < d_{\rm ini}$ and having our more complicated cut 
$r'/s < d_{\rm ini}\times \max\{z^{\prime 2},(1-z')^2\}$. The two
correction terms make finite contributions to the Sudakov exponent when
$r/s \ll s_{l'}/s$ with  $d_{\rm ini}$ either small compared to $s_{l'}/s$ or of
order $s_{l'}/s$ (but in any case bigger than $r/s$).

We see that, to the accuracy of dropping non-logarithmic terms in
the exponent, we have
\begin{equation}
\label{leadingSudakov1}
\begin{split}
\Delta(r) \approx {}&
\prod_{l}
\exp\left[
-\int^{d_{\rm ini}}_{r}\!\frac{d\lambda}{\lambda}\
\frac{\alpha_{\rm s}(\sqrt{\lambda s})}{2\pi}\,
\Biggl\{
C_{f(l)}\,\log\!\left(\frac{s_l/s}{\lambda}\right)
 - \gamma_{f(l)}  
\Biggr\}
\right]\;\;.
\end{split}
\end{equation}
Compare this to the factor in Eq.~(\ref{CKKWfactor}), corresponding to
the probability that the final state particles not split between the
scale $d(m)$ and the limiting scale $d_{\rm ini}$. According to 
Eq.~(\ref{Sdefault}), this is
\begin{equation}
\label{leadingSudakov2}
\begin{split}
\prod_{l} 
e^{- {\cal S}(m,l)} = {}&
\prod_{l}
\exp\left[
-\int^{d(m)}_{d_{\rm ini}}\!\frac{d\lambda}{\lambda}\
\frac{\alpha_{\rm s}(\sqrt{\lambda s})}{2\pi}\,
\Biggl\{
C_{f(l)}\,\log\!\left(\frac{d(V_{\rm s}(l))}{\lambda}\right)
 - \gamma_{f(l)}  
\Biggr\}
\right]\;\;.
\end{split}
\end{equation}

Let us choose
\begin{equation}
s_l = s\,d_{V(l)}\;\;.
\label{sldef}
\end{equation}
Then
\begin{equation}
\label{leadingSudakov3}
\begin{split}
&\Delta(r)\prod_{l} 
e^{- {\cal S}(m,l)} = 
\\&\quad\quad
\prod_{l}
\exp\left[
-\int^{d(m)}_{r}\!\frac{d\lambda}{\lambda}\
\frac{\alpha_{\rm s}(\sqrt{\lambda s})}{2\pi}\,
\Biggl\{
C_{f(l)}\,\log\!\left(\frac{d(V_{\rm s}(l))}{\lambda}\right)
 - \gamma_{f(l)}  
\Biggr\}
\right]\;\;,
\end{split}
\end{equation}
corresponding to the final state particles not splitting between the
scale $d(m)$ and the scale $r$, at which one of them does split. There
is no dependence on $d_{\rm ini}$ in this product.

We thus see the purpose of including the factor $W_{m}(\{p,f\}_m)$ in the
formula for $\sigma_m^{\rm B+S}$: without it, there would be a dependence
on $d_{\rm ini}$ from the Sudakov factors for showering below the
resolution scale $d_{\rm ini}$. More precisely, this is the purpose of
including the factor in $W_{m}(\{p,f\}_m)$ relating to propagation from
scale $d(m)$ to $d_{\rm ini}$. The other Sudakov factors in
$W_{m}(\{p,f\}_m)$ are equally important. The complete expression for 
$\sigma_m^{\rm B+S}$ has Sudakov factors from three sources. Evolution
from $d_2$ to $d_m$ is included in $W_{m}(\{p,f\}_m)$. Evolution from
$d_m$ to $r = d_{m+1}$ is included partly in $W_{m}(\{p,f\}_m)$ and partly
in the Sudakov factor $\Delta(r)$ that is part of ${\bm E}_{l,k}$, as we
have just seen. Then there is a succession of shower splitting scales
that we have represented as being part of $I(\{p,f\}_m;l,k,Y_l)$. This
showering will involve Sudakov factors for evolution from $m+1$ to
$m+2$, from $m+2$ to $m+3$, and so forth. If we hold all of the
parton momenta fixed and change $d_{\rm ini}$ from just less than $d_m$ to
just greater than $d_m$, then the same event history counts not as part
of $\sigma_m^{\rm B+S}$ but rather as part of $\sigma_{m+1}^{\rm B+S}$.
The integrand representing the probability for this event history in 
$\sigma_m^{\rm B+S}$ will be somewhat different from the integrand for
the same event history in $\sigma_{m+1}^{\rm B+S}$, but the two integrands
will be approximately the same. The main points are that the Sudakov
factors from all three sources are approximately the same and that the
$m+1$-parton squared matrix element is approximately the $m$-parton
squared matrix element times the dipole splitting functions.

\subsection{Factors $W$ for real and virtual corrections}
\label{sec:WforRandV}

We have defined the factor $W_{m}(\{p,f\}_m)$ in $\sigma_m^{\rm B+S}$ and
examined its structure and how it meshes with the Sudakov factors in the
splitting function ${\bm E}_{l,k}$. We now turn to the function
$W_{m}^{\rm V+S}(\{p,f\}_m)$ that appears in $\sigma_m^{\rm V+S}$,
Eq.~(\ref{sigmaVS}), and the function $W_{m+1}^{\rm R+S}(\{p,f\}_{m+1})$
that appears in $\sigma_m^{\rm R+S}$, Eq.~(\ref{sigmaRS}).

For $W_{m}^{\rm V+S}(\{p,f\}_m)$, we have an $m$-parton final state that
is constructed to be resolvable at scale $d_{\rm ini}$. It thus seems
sensible to use the same reweighting function that we used for the Born
contribution,
\begin{equation}
W_{m}^{\rm V+S}(\{p,f\}_m) = W_{m}(\{p,f\}_m)
\label{WVplusS}
\;\;.
\end{equation}
It is worth noting that this choice is not as obvious as it was in the
for $\sigma_m^{\rm B+S}$. Recall that $\sigma_m^{\rm V+S}$ is based on
graphs with a virtual loop (together with a subtraction that keeps the
virtual partons from being very soft or collinear with the external
particles). If the virtual partons always had momenta of order $\sqrt s$
then the choice (\ref{WVplusS}) would be physically well motivated.
However, the momentum scales in the virtual loop can be intermediate
between $\sqrt s$ and the scales of the parton splittings in the
synthetic shower history constructed from the final state $\{p,f\}_m$.
For this reason, it could well be that the ``best'' choice for $W_{m}^{\rm
V+S}(\{p,f\}_m)$ is something more subtle than that given in 
Eq.~(\ref{WVplusS}).

For $W_{m}^{\rm R+S}(\{p,f\}_{m+1})$, we have an $m+1$-parton final
state, so that we have resolution scales $d(1) \ge \cdots \ge
d(m) \ge d(m+1)$ in the synthetic shower graph generated from the final
state $\{p,f\}_{m+1}$. In the main term of $\sigma_m^{\rm R+S}$, we have
$d(m) > d_{\rm ini} > d(m+1)$. There are then subtraction terms
that remove the leading singularity when $d(m) \gg d(m+1)$. For this
reason, $d(m+1)$ is typically not much smaller than $d_{\rm ini}$, nor is
$d(m)$ much larger. We therefore define $W_{m}^{\rm R+S}(\{p,f\}_{m+1})$
to include a reweighting factor for $\alpha_s$ at each strong
interaction vertex and a Sudakov factor giving the probability that the
partons did not radiate between each pair of vertices:
\begin{equation}
W_{m}^{\rm R+S}(\{p,f\}_{m+1})
=
\prod_{V=3}^{m+1}
\frac{\as\big(\sqrt{d(V)s}\big)}
 	{\as\big(\mu_{\rm R}\big)}\
\prod_{V=2}^{m}
\prod_{L \in I(V)} 
\exp\{- {\cal S}(V,L)\}
\label{WRplusS}
\end{equation}
with ${\cal S}(V,L)$ as defined in Eq.~(\ref{Sdefault}).  The last
Sudakov factor takes us from vertex $m$, after which there are $m$
partons, to vertex $m+1$ at which one of these partons splits with a
scale $d(m+1)$ to make an $m+1$-parton state. Subsequent evolution, as
given in the Monte Carlo interface function $\tilde I^{\rm
R+S}(\{p,f\}_{m+1})$ in Eq.~(\ref{sigmaRS}), then starts at splitting
scale $d(m+1)$. Again, it is worth noting that this choice is not as
obvious as it was in the for $\sigma_m^{\rm B+S}$ because of the
subtraction term in $\sigma_m^{\rm R+S}$. It could well be that the
``best'' choice for $W_{m}^{\rm R+S}(\{p,f\}_m)$ is something more subtle
than that given in  Eq.~(\ref{WRplusS}).

An alternative is to take 
\begin{equation}
W_{m}^{\rm V+S}(\{p,f\}_m) = W_{m}^{\rm R+S}(\{p,f\}_{m+1}) = 1
\label{Wto1}
\;\;.
\end{equation}
This choice removes some sensible physics built into $W_{m}^{\rm
V+S}$ and $W_{m}^{\rm R+S}$. However, it does have
a technically useful feature. Suppose that we take a measurement function
$F=1$, so that $\sigma_m^{\rm B+S} + \sigma_m^{\rm V+S} + \sigma_m^{\rm
R+S}$ is the contribution to the total cross section from final states
with precisely $m$ jets. The Born contribution, $\sigma_m^{\rm B+S}$,
contains terms of all orders in $\alpha_{\rm s}$ starting at order
$\alpha_{\rm s}^{B_m}$. It does not have the complete contribution at
order $\alpha_{\rm s}^{B_m+1}$ and beyond, but it does have the correct
leading and next-to-leading logarithms of $d_{\rm ini}$
\cite{CKKW} (at leading order in $1/N_{\rm c}^2$, since we have modified
$W_{m}$ a little). The addition of $\sigma_m^{\rm V+S}$ and $\sigma_m^{\rm
R+S}$ fills in the missing pieces at order $\alpha_{\rm s}^{B_m+1}$. It
also can add contributions at higher orders. However, if we
choose $W_{m}^{\rm V+S}(\{p,f\}_m) = W_{m}^{\rm R+S}(\{p,f\}_{m+1}) = 1$,
then  $\sigma_m^{\rm V+S}$ and $\sigma_m^{\rm R+S}$ are exactly proportional
to $\alpha_{\rm s}^{B_m+1}$, with no contributions at higher orders. They
provide just what is needed at order $\alpha_{\rm s}^{B_m+1}$ and no more. To see
this, note that in the expression (\ref{sigmaRS}) for
$\sigma_m^{\rm R+S}$, the factor $\tilde I$ is 1 when $F=1$ because of
the property (\ref{tildeInorm}) of $\tilde I$. When we also set
$W_{m}^{\rm R+S}(\{p,f\}_m)$ to 1, we get an expression proportional to
$\alpha_{\rm s}^{B_m+1}$ with no further dependence on $\alpha_{\rm
s}$. Similarly, the choice $W_{m}^{\rm V+S}(\{p,f\}_m) = 1$ makes
$\sigma_m^{\rm V+S}$, Eq.~(\ref{sigmaVS}), exactly proportional to
$\alpha_{\rm s}^{B_m+1}$.

\section{What size is the resolution parameter?}
\label{sec:smalldini}

The algorithm described in this paper depends on a jet resolution
parameter $d_{\rm ini}$, which plays a role similar to that of the
factorization scale in calculations of cross sections for hard processes
with one or two hadrons in the initial state. We anticipate that $d_{\rm
ini}$ would be a parameter in a computer code that implements this
algorithm, so that a user could choose its value. In this section, we
describe some considerations that would go into making the choice.

Suppose that we calculate $\sigma[F]$ for an infrared safe $N$-jet
observable that measures large momentum scale features of events and
hence does not involve large logarithms. An example for
$N=3$ is $F = T_4$, the fourth moment of the thrust distribution,
Eq.~(\ref{T4def}). More generally, we can consider the $n$th moment of
the trust distribution for $n \ge 2$,
\begin{equation}
\sigma[T_n] = \int_0^1\! dt\,(1-t)^n\, \frac{d\sigma}{dt}\;\;.
\label{Tndef}
\end{equation}
Given that we want to calculate $\sigma[F]$ for an observable in this
class, we seek to choose the resolution scale $d_{\rm ini}$ so the dominant
contribution to $\sigma[F]$ comes from $\sigma_m[F]$, with $m=N$.  Let us
examine why one would like $\sigma_N[F]$ to be dominant and how dominant
it should be. 

We first consider the comparison of $\sigma_N[F]$ to $\sigma_{N+1}[F]$,
supposing that $\sigma_{m}[F]$ for $m \le N-1$ is negligible. We note that
the cross section $\sigma[F] = \sum_m \sigma_m[F]$ has a
perturbative expansion that starts at order $\alpha_{\rm s}^{B_{N}}$ and
has higher order contributions. For a next-to-leading order calculation,
we need the next contribution, of order $\alpha_{\rm s}^{B_{N}+1}$. The
partial cross section $\sigma_N[F]$ also has a perturbative expansion
that starts at order $\alpha_{\rm s}^{B_{N}}$. As we have seen, the
algorithm  presented in this paper reproduces the order $\alpha_{\rm
s}^{B_{N}}$ and $\alpha_{\rm s}^{B_{N}+1}$ terms of this expansion. The
perturbative expansion for $\sigma_{N+1}[F]$ begins at order $\alpha_{\rm
s}^{B_{N}+1}$. Thus part of the order $\alpha_{\rm s}^{B_N+1}$ correction
to $\sigma[F]$ is in $\sigma_N^{\rm R+S}[F]$ and $\sigma_N^{\rm V+S}[F]$
and part is in $\sigma_{N+1}^{\rm B+S}[F]$. (Part is in $\sigma_m[F]$ 
for $m\le N-1$, but we suppose for the moment that this part is
negligible.)

Consistently with this counting of powers of $\alpha_{\rm s}$, we expect
$\sigma_{N+1}[F] \sim \alpha_{\rm s}\,\sigma_{N}[F]$. However, suppose
that we were to choose such a small value for $d_{\rm ini}$ that
$\alpha_{\rm s}\log^2(1/d_{\rm ini})$ is of order 1. Then the cross
section $\sigma_{N}[F]$ that we hoped was dominant would be suppressed and
the cross section would be shifted to $\sigma_{m}[F]$ with $m = N+1, N+2,
\dots$. This would not much matter at leading order since the $k_T$-jet
matching scheme arranges that the sum of the $\sigma_{m}^{\rm B+S}[F]$ is
very accurately independent of $d_{\rm ini}$. However, it is possible
that the perturbative correction terms of order $\alpha_{\rm
s}^{B_{N}+1}$ can be lost among terms like $\alpha_{\rm s}^{B_N+1}\times
\alpha_{\rm s}\log^2(1/d_{\rm ini})$. Such a term is beyond the
next-to-leading perturbative order for $\sigma_{N}[F]$ and beyond the
next-to-leading logarithm approximation also. Terms like this are always
present in $\sigma_{N+1}^{\rm B+S}[F]$, $\sigma_N^{\rm R+S}[F]$ and
$\sigma_N^{\rm V+S}[F]$, but they are numerically important only when
$d_{\rm ini}$ is very small. To ensure that these contributions are not
numerically important, we can ask that the relation between
$\sigma_{N}[F]$ and $\sigma_{N+1}[F]$ follow the perturbative expectation,
\begin{equation}
\sigma_{N+1}[F] \lesssim \alpha_{\rm s}\,\sigma_{N}[F]
\;\;.
\label{dinimin}
\end{equation}
With a computer implementation of the algorithm described here at hand,
one can also check that $\sigma_N^{\rm R+S}[F]$ and $\sigma_N^{\rm
V+S}[F]$ are of the right perturbative size (that is, $\alpha_{\rm
s}\,\sigma_{N}[F]$) and that the complete $\sigma[F]$ is accurately
reproducing the purely perturbative result for the same quantity, which
can be obtained by simply turning off the shower part of the program.

Consider next the partial cross section $\sigma_{N-1}[F]$, with ${N -
1}$ resolvable jets. The perturbative expansion for $\sigma_{N-1}[F]$ has
the form (\ref{pertexpansion})
\begin{equation}
\sigma_{N-1}[F]  =  
C_{N-1,0}[F]\, \alpha_{\rm s}^{B_N - 1}(Q) 
+ C_{N-1,1}[F]\, \alpha_{\rm s}^{B_N}(Q)
+ \cdots\;\;.
\label{pertexpansionA}
\end{equation}
This compares to the expansion for $\sigma_N[F]$,
\begin{equation}
\sigma_{N}[F]  =  
C_{N,0}[F]\, \alpha_{\rm s}^{B_N}(Q) 
+ C_{N,1}[F]\, \alpha_{\rm s}^{B_N+1}(Q)
+ \cdots\;\;.
\label{pertexpansionB}
\end{equation}
The first perturbative coefficient for $\sigma_{N-1}[F]$ vanishes since,
by assumption, $F(f) = 0$ for a state with $N-1$ partons. Fortunately, we
have the next term,\footnote{Since the required NLO calculations are
available for $N$ jets, we can assume that they are available for $N-1$
jets.} proportional to $\alpha_{\rm s}^{B_{N}}$. We do not have the
complete coefficient proportional to $\alpha_{\rm s}^{B_{N}+1}$. Thus it
seems that we do not have the $\alpha_{\rm s}^{B_{N}+1}$ accuracy that we
want. However, $F(f) = 0$ for a state with $N-1$ very narrow jets.
Furthermore $\sigma_{N-1}[F]$ is the contribution from final states with
$N-1$ jets that cannot be further resolved at scale $d_{\rm ini}$. If we
take $d_{\rm ini}$ to be small, the $N-1$ jets are guaranteed to be
narrow. We conclude that all of the coefficients $C_{N-1,j}[F]$ will
be small if we choose $d_{\rm ini}$ to be small.  Since we are missing one
factor of  $\alpha_{\rm s}$ in the expansion of $\sigma_{N-1}[F]$, we can
simply demand that $d_{\rm ini}$ be chosen so that the coefficients 
$C_{N-1,j}[F]$ are smaller than the corresponding coefficients
$C_{N,j-1}[F]$ in $\sigma_{N}[F]$ by a numerical factor that is
approximately equal to $\alpha_{\rm s}$ or else smaller. That is, we ask
that 
\begin{equation}
\sigma_{N-1}[F] \lesssim \alpha_{\rm s}\,\sigma_{N}[F]
\;\;.
\label{dinimax}
\end{equation}

Can one really satisfy Eqs.~(\ref{dinimin}) and (\ref{dinimax})? Consider
the three-jet observable $T_4$, Eq.~(\ref{Tndef}). Using Pythia with
$\sqrt s = M_Z$ to estimate $\sigma_{m}[T_4]$ we find that
$\sigma_{3}[T_4]$ is largest near $d_{\rm ini} = 0.04$. At this value of
$d_{\rm ini}$, we have $\sigma_{4}[T_4] = 0.08\,\sigma_{3}[T_4]$ and
$\sigma_{2}[T_4] = 0.12\,\sigma_{3}[T_4]$, thus satisfying
Eqs.~(\ref{dinimin}) and (\ref{dinimax}).

The case of the second moment of the thrust distribution, $F = T_2$, is
instructive. Again taking $\sqrt s = M_Z$, we find that
$\sigma_{3}[T_2]$ is largest near $d_{\rm ini} = 0.02$. At this value of
$d_{\rm ini}$, we have $\sigma_{4}[T_2] = 0.16\,\sigma_{3}[T_2]$ and
$\sigma_{2}[T_2] = 0.35\,\sigma_{3}[T_2]$. In this case the two jet
contribution is not as suppressed as one might like. We do not interpret
this as indicating a difficulty with the $k_T$-jet matching scheme.
Rather we take it as an indication that, after accounting for showering
and hadronization, $\sigma[T_2]$ gets a 30\% contribution from the two jet
region (defined with jet resolution 0.02) even though $T_2$ is nominally
an infrared-safe three-jet quantity. For this reason, a pure NLO fixed
order calculation is not as reliable a calculation as one might like and
one should not be concerned if the algorithm in this paper based on
$k_T$-jet matching does not exactly match the pure NLO calculation. To
the extent that they turn out to differ, we expect that the $k_T$-jet
matching calculation with some perturbative accuracy sacrificed in favor
of showers and hadronization is the better calculation.

These examples suggest that a choice of $d_{\rm ini}$ in the
range $\alpha_{\rm s}^2 < d_{\rm ini} < \alpha_{\rm s}$ might be
appropriate in practical cases. We can also ask whether a choice of
$d_{\rm ini}$ in the range $\alpha_{\rm s}^2 < d_{\rm ini} < \alpha_{\rm
s}$ would work in the asymptotic limit $\alpha_{\rm s} \to 0$. 

Consider first the condition in Eq.~(\ref{dinimax}). It is pretty
straightforward to see that for an infrared safe $N$-jet observable $F$,
$\sigma_{N-1}[F]$ will vanish like a power of $d_{\rm ini}$  (times
logarithms) for $d_{\rm ini} \to 0$. For example, for the $n$th moment of
the thrust distribution, Eq.~(\ref{Tndef}), one finds a limiting form
$\sigma_{2}[T_n] \propto d_{\rm ini}^{\,n/2}$. Thus if we take  $d_{\rm
ini} \sim \alpha_{\rm s}$ or smaller, Eq.~(\ref{dinimax}) will be
satisfied asymptotically as long as $n \ge 2$. Following the philosophy
stated above for $\sigma_{2}[T_2]$ when $\alpha_{\rm s} = \alpha_{\rm
s}(M_Z)$, we might take the observables $T_n$ for $n< 2$ to be
sufficiently contaminated by two jet physics that we would not need to
satisfy Eq.~(\ref{dinimax}) for those observables.

Consider next the condition in Eq.~(\ref{dinimin}). Since the
perturbative expansion of $\sigma_{N+1}[F]$ begins with one more power
of $\alpha_{\rm s}$ than the perturbative expansion of $\sigma_{N}[F]$, 
Eq.~(\ref{dinimin}) appears to follow. There is a potential problem from
the Sudakov exponentials, which could suppress $\sigma_{N}[F]$ relative
to  $\sigma_{N+1}[F]$. The exponents contain factors $\alpha_{\rm
s}\,\log^2(1/d_{\rm ini})$. With $d_{\rm ini} \sim \alpha_{\rm s}$ or
$d_{\rm ini} \sim \alpha_{\rm s}^2$, the Sudakov exponent becomes
$const.\,\alpha_{\rm s}\,\log^2(1/\alpha_{\rm s})$. In the limit
$\alpha_{\rm s} \to 0$, these factors get smaller. Thus the Sudakov
exponentials do not change the perturbative power counting and
Eq.~(\ref{dinimin}) is satisfied.

Thus both numerical examples based on Pythia at $\alpha_{\rm s}  =
\alpha_{\rm s}(M_Z)$ and an analysis of the limit $\alpha_{\rm s}  \to 0$
suggest that a choice of $d_{\rm ini}$ in the range $\alpha_{\rm s}^2 <
d_{\rm ini} < \alpha_{\rm s}$ might be useful. When one has a working
program, one will want to revisit these questions with a numerical
analysis using the program.

\section{The Monte Carlo interface function}
\label{sec:moreMCinterface}

In Sec.~\ref{sec:MCinterface}, we introduced a function
$I(\{p,f\}_m;l,k,Y_l)$ that represents the expectation value of the
observable in hadronic states that are generated by a shower Monte Carlo
event generator beginning from initial conditions represented by the
partonic state $\{p,f\}_m$ and the information about the splitting of
parton $l$ contained in the variables $l,k,Y_l$. This function is used for
$\sigma_m^{\rm B + S}$. We also introduced a similar function $\tilde
I(\{p,f\}_n)$ that is used for $\sigma_m^{\rm R + S}$ and $\sigma_m^{\rm
V + S}$ (and may optionally be used for $\sigma_m^{\rm B + S}$ for
$m>m_{\rm NLO}$). For the matching to a NLO calculation, we needed only
certain basic properties of these functions, given in
Eqs.~(\ref{Iproperty1}) and (\ref{Iproperty0}), and for $\tilde
I$, in Eq.~(\ref{tildeIproperty}). 

In this section, we delve a little further into the requirements for the
shower Monte Carlo event generator. There are quite a number of successful
shower algorithms available and it is beyond our scope to specify any
algorithm in detail. However it may be useful to say a little about the
initial conditions for the shower Monte Carlo program. In  particular, we
note that we are handing the Monte Carlo a partially developed shower and
that the Monte Carlo needs to begin at the point where the previous shower
simulation left off.

Consider, then, the function $I(\{p,f\}_m;l,k,Y_l)$ used for
$\sigma_m^{\rm B + S}$. There is an $m$-parton state that can be thought
of as having been created by a showering process with a
hardness cutoff $d_{\rm ini}< d_m(\{p,f\}_m$. After that, one of the
partons (labelled $l$) splits. Suppose for the sake of concreteness that
this dipole splitting is based on the evolution variable of
Eq.~(\ref{ktevollution}),
\begin{equation}
r = s_l\, y z(1-z)\;\;.
\end{equation}
Let us define a transverse momentum $k_\perp$ for this splitting
according to Eq.~(\ref{kperpdef}). With this choice, we have
\begin{equation}
|k_\perp^2| = 2 p_l\cdot p_k\, y z (1-z) = 
\frac{2 p_l\cdot p_k}{s_l}\ r\;\;.
\end{equation}
Note that for a given value of the evolution variable $r$, $|k_\perp^2|$
for the splitting of parton $l$ with the participation of spectator
parton $k$ depends on $l$ and $k$. 

Now, all of the partons are allowed to split, and the one that does is
parton $l$. The others did not split at an evolution variable above the
value $r$. That is, parton $l'$, with the aid of
spectator $k'$, did not split with $r' > r$. Further evolution of these
partons should then be restricted to the range $r'<r$. That is
\begin{equation}
|k_\perp^{\prime 2}| < 
\frac{2 p_{l'}\cdot p_{k'}}{s_{l'}}\ 
\frac{s_l}{2 p_l\cdot p_k}\
|k_\perp^{2}|\;\;.
\label{initialconditionA}
\end{equation}
A restriction like this can be imposed in the chosen shower Monte Carlo
program by using a veto algorithm, as described for instance in
Ref.~\cite{CKKW}. The only problem may be that the Monte Carlo program
may know the index $l'$ of the parton that it is proposing to split, but
may not have a choice for a corresponding spectator parton $k'$. A
sensible choice would be to let $k'$ be one of the final state partons to
which parton $l'$ is color connected (at leading order in $1/N_{\rm c}$)
according to a rule based on the synthetic splitting diagram obtained by
applying the $k_T$-jet algorithm to the $m$-parton state and adding the
one splitting of parton $l$. (For a gluon $l'$ there are two such color
connected partons and one would choose either of them with probability
1/2.) An alternative that avoids selecting a spectator parton is to impose
\begin{equation}
|k_\perp^{\prime 2}| < \max_{k'} \left\{
\frac{2 p_{l'}\cdot p_{k'}}{s_{l'}}\ 
\frac{s_l}{2 p_l\cdot p_k}\
|k_\perp^{2}|\right\}\;\;.
\label{initialconditionB}
\end{equation}
For the splitting of one of the daughters of parton $l$, one may simply
impose
\begin{equation}
|k_\perp^{\prime 2}| < 
|k_\perp^{2}|
\;\;.
\end{equation}
Of course, if the splitting angle is not the evolution variable used in
the shower Monte Carlo program used for secondary showers, then the
program should separately arrange for angular ordering. 

Consider next the function $\tilde I(\{p,f\}_m)$ as used for
$\sigma_m^{\rm V + S}$. The final state, which is the initial state for
the Monte Carlo program, consists of $m$ partons that can be thought
of as having been created by a showering process, with a hardness cutoff
$d_{\rm ini}< d_m(\{p,f\}_m$. There is no further splitting before this
state is passed to the Monte Carlo program. Therefore further splittings
$l \to i + j$ should be generated with a cut $d(p_i,p_j) < d_{\rm ini}$.
The same reasoning applies in the case that one elects to use $\tilde I$
without a dipole splitting for $\sigma_m^{\rm B + S}$ in the case
$m>m_{\rm NLO}$, as described in Sec.~\ref{sec:LOalternative}.

The case of $\tilde I(\{p,f\}_{m+1})$ as used for $\sigma_m^{\rm R + S}$
is similar. Here there are $m+1$ partons with $d_{m+1}(\{p,f\}_{m+1}) <
d_{\rm ini}$. Thus it is $d_{m+1}(\{p,f\}_{m+1})$
rather than $d_{\rm ini}$ that should serve as the upper cutoff for
splittings included in $\tilde I(\{p,f\}_{m+1})$.

\section{Conclusion}
\label{sec:Conclusion}

We have proposed an algorithm for adding showers to next-to-leading
order calculations for $e^+ + e^- \to N \ {\it jets}$. This algorithm
incorporates two kinds of matching, which we might call PS/NLO matching
and $m$-jet/$(m+1)$-jet matching. 

For the $m$-jet/$(m+1)$-jet matching, we make use of the $k_T$-jet
matching scheme of Catani, Krauss, Kuhn, and Webber \cite{CKKW}. We
divide the cross section $\sigma[F]$ corresponding to an observable $F$
into contributions  $\sigma_m[F]$ corresponding to final states with $m$
hard jets.  Here ``hard'' is defined using a jet resolution parameter
$d_{\rm ini}$. In the case that $F$ is an $N$-jet observable that is
infrared-safe in the technical sense and is in fact insensitive to
physics at momentum scales much below $\sqrt s$, we want the dominant
contribution to $\sigma[F]$ to come from $\sigma_m[F]$ with $m = N$.

For PS/NLO matching, we are concerned with the $m$ partons that emerge
from a Born graph that contributes to $\sigma_m[F]$. One of these partons
can split as the first step of a parton shower based on this graph. On
the other hand, there are perturbative corrections to the same graph that
also involve real or virtual splittings of the same $m$ partons. The
MC/NLO matching is needed in order to incorporate the parton shower and
the NLO corrections without double counting. With the parton shower in
place, we not only get $\sigma_m[F]$ correct to NLO, but we also get
sensible answers to more detailed questions about the low momentum scale
structure of the $m$ jets that are used to calculate $\sigma_m[F]$. A
pure NLO calculation cannot (sensibly) answer such questions.

We thus have a good description of hard physics and soft physics and
even both at once. What about physics at an intermediate momentum scale?
Suppose that $F$ is such that $\sigma[F]$ is sensitive to physics at a
scale $K^2$ that is similar to $d_{\rm ini} s$. An example would be the
thrust distribution $d\sigma/dt$ with $K^2 \equiv (1-t)s$ similar to
$d_{\rm ini} s$. In this case neither $\sigma_2[F]$ nor $\sigma_3[F]$
dominates the calculation of  $\sigma[F]$. If we hold $K^2$ fixed and
increase $d_{\rm ini}$ a little, $\sigma_2[F]$ decreases and $\sigma_3[F]$
decreases. If we use the Born level contributions  $\sigma_m^{\rm
B+S}[F]$, the approximations used in $\sigma_2^{\rm B+S}[F]$ and
$\sigma_3^{\rm B+S}[F]$ are so similar that the sum $\sigma_2^{\rm
B+S}[F] + \sigma_3^{\rm B+S}[F]$ is very insensitive to  $d_{\rm ini}$. If
we now include $\sigma_m^{\rm R+S}[F] + \sigma_m^{\rm V+S}[F]$, we improve
the approximations by adding NLO correction terms. However,  the
correction terms $\sigma_2^{\rm R+S}[F] + \sigma_2^{\rm V+S}[F]$ and
$\sigma_3^{\rm R+S}[F] + \sigma_3^{\rm V+S}[F]$ are not matched in the
same sense that the leading terms $\sigma_2^{\rm B+S}[F]$ and
$\sigma_3^{\rm B+S}[F]$ are matched. To do better at matching the
correction terms for different numbers of jets would require
next-to-leading order splitting kernels for the showers, which we do not
have. For this reason, one may expect that adding the correction terms 
$\sigma_m^{\rm R+S}[F] + \sigma_m^{\rm V+S}[F]$ introduces some $d_{\rm
ini}$ dependence into the result. Fortunately, the NLO correction terms
are small, so the added $d_{\rm ini}$ dependence must also be small.

We have been quite specific about the subtraction scheme to use in the
base NLO calculation. We have been deliberately less specific about the
scheme for showering. The hardest splitting of a parton from a Born graph
is to be done using the dipole splitting functions. Softer splittings can
be handled by the reader's favorite shower Monte Carlo method as long as
the initial conditions for this secondary showering are compatible with
the organization of the part of the showering already completed.

It may be useful to compare the algorithm discussed in this paper with
those in other papers. We can compare treatments of the Sudakov factor
$\bm{E}_{l,k}$ in Eq.~(\ref{sigmaBS}), the reweighting factor $W$, and the
factor $I$ that represents further showering.

Consider first the exponent in the Sudakov factor $\bm{E}_{l,k}$ for the
splitting that is to be matched to NLO perturbation theory,
Eq.~(\ref{Elfinal}). In the work of Frixione, Nason, and Webber
\cite{FrixioneWebberI,FrixioneWebberII}, this exponent is the splitting
function of H{\small ERWIG}. Thus one must subtract something from the
NLO calculation that is derived from the order $\alpha_{\rm s}$ expansion
of a H{\small ERWIG} shower. As discussed by Nason \cite{Nasonshowers},
this means that the NLO calculation is to be tuned to the shower Monte
Carlo. Additionally, since the H{\small ERWIG} shower does not precisely
match the perturbative probability for soft wide-angle gluon emission,
one must also provide a special treatment for the soft region. The
algorithm of Ref~\cite{nloshowersI, nloshowersII} is based on the simple
idea that, in a physical gauge, the cut two point function represents
parton splitting, while the virtual two point function represents the
probability not to split. For this reason, the Sudakov exponent is chosen
as the integrand of the virtual two point function in the Coulomb gauge.
There is a separate Sudakov exponential for emission of soft wide-angle
gluons.  This method works, but is not easily adapted to the dipole style
of NLO calculations that most people use. In the present work,
as in the leading order program of Sj\"ostrand and Skands
\cite{SjostrandSkands}, the hardest splitting in a shower gets a special
treatment. Nason \cite{Nasonshowers} has proposed a scheme of this style
that is adapted to an angle ordered shower, as in H{\small ERWIG}. In
this scheme, the Sudakov exponent for the hardest splitting is the ratio
of the exact matrix element $\langle{\cal M}(\{\hat p,\hat f\}_{m+1})|
{\cal M}(\{\hat p,\hat f\}_{m+1})\rangle$ for $m+1$ partons to the matrix
element $\langle{\cal M}(\{p,f\}_m)|{\cal M}(\{p,f\}_m)\rangle$ for $m$
partons, before one of the partons is imagined to have split. The
$m+1$-particle matrix element is to be divided into a sum of terms
associated with individual parton splittings using an algorithm that is
not completely specified in \cite{Nasonshowers}. There is a certain
conceptual simplicity to this choice, but it puts a very complicated
function into the Sudakov exponent.  In this paper, the Sudakov exponent
is the angular average of the splitting function in the dipole
subtraction scheme that is widely used for NLO calculations. With this
choice, the Sudakov exponent is quite simple and at the same time it
incorporates soft wide-angle gluon emissions exactly.

Consider next the reweighting factor $W$, whose main purpose is to provide
a Sudakov suppression factor multiplying the matrix element according to
a synthetic shower that corresponds to the Born level final state
$\{p,f\}_m$. This factor is not present in either of the existing
programs \cite{FrixioneWebberI,FrixioneWebberII} and
\cite{nloshowersI,nloshowersII,nloshowersIII} that add parton showers to
an NLO calculation. The reweighting factor $W$ is simply taken from the
paper of Catani, Krauss, Kuhn and Webber \cite{CKKW} and adapted to the
present case of an NLO calculation. There may be improvements possible.
For instance, L\"onnblad \cite{Lonnblad} and Mrenna and Richardson
\cite{Mrennashowers} have provided more sophisticated algorithms for
calculating a function that plays this role but more closely matches the
Sudakov factor that would actually be generated in a shower Monte Carlo
program.

Consider finally the factor $I$ that represents the expectation value of
the observable after secondary showering and hadronization that is
initiated from the partons produced at the Born level together with one
hardest splitting. We have left this function as something that can be
freely chosen subject only to a few very general conditions. If one wants
a H{\small ERWIG} style angle-ordered shower, the vetoed shower treatment
of Catani, Krauss, Kuhn and Webber \cite{CKKW} is useful. Additionally,
the recent paper of Nason \cite{Nasonshowers} explains in some detail how
an angle ordered-shower can be initiated after a first step consisting of
the hardest splitting rather than the widest angle splitting. If one
would like a $P_T$ ordered shower that properly incorporates coherence,
the recent paper of Sj\"ostrand and Skands \cite{SjostrandSkands}
describes such a treatment, including the feature that emissions are
generated in the order of decreasing overall $P_T$ in the whole event
rather than $P_T$ along a single line. If one would like a shower based
on the Color-Dipole Cascade Model of A{\small RIADNE}, then the work of
L\"onnblad \cite{Lonnblad} is to be recommended. We note that the most
natural choice of shower to follow the initial splitting based on the
Catani-Seymour dipole splitting functions and kinematic definitions would
be a shower in which {\it all} of the splittings are based on this
choice. A construction based on this choice is, however, beyond the scope
of this paper. Instead, we leave the choice of secondary showering scheme
open. We emphasize that the same NLO calculation with the hardest
splitting included could then be coupled to shower Monte Carlo programs
using a variety of splitting styles as long as the subsequent shower is
initiated so as to account for the splitting that has already occurred.

The algorithms covered here include final state showers, but do not allow
for strongly interacting partons in the initial state. The dipole
subtraction scheme of Ref.~\cite{CataniSeymour} covers initial state
partons as well, but we thought it best to leave the case of initial
state showers for a separate publication.

We certainly expect that the algorithm presented here can be improved.  
For instance, we note that the $k_T$-jet matching scheme that we have used
is rather like the phase space slicing method in NLO calculations and we
wonder if it might be replaced by something that is more like the
subtraction method. Additionally, it would be good to incorporate
knowledge of one loop splitting functions into the algorithm so as to
better match the one loop perturbative calculations.

We have not constructed computer code that could demonstrate that the
algorithm presented here works in practice. There may well be
undesirable features that show up when we have real code to test. However,
the algorithms presented here are a refined version of the algorithm
presented in Refs.~\cite{nloshowersI,nloshowersII}. This cruder algorithm
does exist as code and, when coupled appropriately to P{\small YTHIA},
appears to work quite well \cite{nloshowersIII}.

\acknowledgments{We are grateful to S.~Mrenna for advice, particularly
concerning the advisability of dealing with the hardest splitting first.
This work was supported in part the United States Department of Energy
and by the Swiss National Science Foundation (SNF) through grant no.\
200021-101874 and by the Hungarian Scientific Research Fund grants OTKA
T-038240.
}



\begin{thebibliography}{99}

\bibitem{Herwig}
G.~Marchesini, B.~R.~Webber, G.~Abbiendi, I.~G.~Knowles, M.~H.~Seymour
and L.~Stanco,
Comput.\ Phys.\ Commun.\  {\bf 67}, 465 (1992);
G.~Corcella {\it et al.}, JHEP {\bf 0101} 010 (2001) 
[arXiv:hep-ph/0011363].

\bibitem{Pythia}
T.~Sj\"ostrand, Comput.\ Phys.\ Commun.\  {\bf 39} (1986) 347;
T.~Sj\"ostrand, P.~Eden, C.~Friberg, L.~L\"onnblad, G.~Miu, S.~Mrenna and
E.~Norrbin,
Comput.\ Phys.\ Commun.\  {\bf 135}, 238 (2001)
[arXiv:hep-ph/0010017].

\bibitem{Nagy}
Z.~Nagy and Z.~Tr\'ocs\'anyi,
Phys.\ Lett.\ B {\bf 414}, 187 (1997)
[arXiv:hep-ph/9708342];
Phys.\ Rev.\ D {\bf 59}, 014020 (1999)
[Erratum-ibid.\ D {\bf 62}, 099902 (2000)]
[arXiv:hep-ph/9806317];
Phys.\ Rev.\ Lett.\  {\bf 87}, 082001 (2001)
[arXiv:hep-ph/0104315];
Z.~Nagy,
Phys.\ Rev.\ Lett.\  {\bf 88} (2002) 122003
[arXiv:hep-ph/0110315];
Phys.\ Rev.\ D {\bf 68} (2003) 094002
[arXiv:hep-ph/0307268].

\bibitem{MCFM}
J.~M.~Campbell and R.~K.~Ellis,
Phys.\ Rev.\ D {\bf 60} (1999) 113006
[arXiv:hep-ph/9905386];
Phys.\ Rev.\ D {\bf 62} (2000) 114012
[arXiv:hep-ph/0006304];
Phys.\ Rev.\ D {\bf 65} (2002) 113007
[arXiv:hep-ph/0202176].

\bibitem{FrixioneWebberI} 
S.~Frixione and B.~R.~Webber,
JHEP {\bf 0206}, 029 (2002)
[arXiv:hep-ph/0204244].

\bibitem{FrixioneWebberII} 
S.~Frixione, P.~Nason and B.~R.~Webber,
JHEP {\bf 0308}, 007 (2003)
[arXiv:hep-ph/0305252].

\bibitem{FrixioneWebberIII}
S.~Frixione and B.~R.~Webber,
arXiv:hep-ph/0402116.

\bibitem{nloshowersI}
M.~Kr\"amer and D.~E.~Soper,
Phys.\ Rev.\ D {\bf 69}, 054019 (2004)
[arXiv:hep-ph/0306222].

\bibitem{nloshowersII}
D.~E.~Soper,
Phys.\ Rev.\ D {\bf 69}, 054020 (2004)
[arXiv:hep-ph/0306268].

\bibitem{nloshowersIII}
M.~Kr\"amer, S.~Mrenna and D.~E.~Soper
``Next-to-leading order QCD jet production with parton showers
and hadronization,''
in preparation.

\bibitem{CKKW}
S.~Catani, F.~Krauss, R.~Kuhn and B.~R.~Webber,
JHEP {\bf 0111} (2001) 063
[arXiv:hep-ph/0109231];
S.~Mrenna and P.~Richardson,
JHEP {\bf 0405}, 040 (2004)
[arXiv:hep-ph/0312274].

\bibitem{CataniSeymour}
S.~Catani and M.~H.~Seymour,
Nucl.\ Phys.\ B {\bf 485} (1997) 291
[Erratum-ibid.\ B {\bf 510} (1997) 503]
[arXiv:hep-ph/9605323].

\bibitem{kTalgorithm}
Y.L.\ Dokshitzer, in {\it Workshop on jet studies at LEP and HERA},
Durham, 1990, reported in 
W.~J.~Stirling,
J.\ Phys.\ G {\bf 17} (1991) 1567;
S.~Catani, Y.~L.~Dokshitzer, M.~Olsson, G.~Turnock and B.~R.~Webber,
Phys.\ Lett.\ B {\bf 269} (1991) 432;
see also
S.~Bethke, Z.~Kunszt, D.~E.~Soper and W.~J.~Stirling,
Nucl.\ Phys.\ B {\bf 370} (1992) 310
[Erratum-ibid.\ B {\bf 523} (1998) 681].

\bibitem{SjostrandSkands}
T.~Sj\"ostrand and P.~Z.~Skands,
Eur.\ Phys.\ J.\ C {\bf 39} (2005) 129
[arXiv:hep-ph/0408302].

\bibitem{Nasonshowers}
P.~Nason,
JHEP {\bf 0411} (2004) 040
[arXiv:hep-ph/0409146].

\bibitem{ESWbook} 
  See
  R.~K.~Ellis, W.~J.~Stirling and B.~R.~Webber,
  {\it QCD and collider physics},
  Camb.\ Monogr.\ Part.\ Phys.\ Nucl.\ Phys.\ Cosmol.\  {\bf 8} (1996) 1.


\bibitem{Lonnblad}
L.~L\"onnblad,
JHEP {\bf 0205} (2002) 046
[arXiv:hep-ph/0112284].

\bibitem{Mrennashowers}
  S.~Mrenna and P.~Richardson,
  JHEP {\bf 0405} (2004) 040
  [arXiv:hep-ph/0312274].


\end{thebibliography}
\end{document}